\documentclass[aps,floats]{revtex4-2}
\usepackage{amsmath,amssymb}
\usepackage{graphicx,epsfig}
\usepackage[greek,english]{babel}
\usepackage{bbold}

\begin{document}
\bibliographystyle {plain}

\pdfoutput=1
\def\oppropto{\mathop{\propto}} 
\def\opsimeq{\mathop{\simeq}}
\def\opoverderline{\mathop{\overline}}
\def\operarrow{\mathop{\longrightarrow}}
\def\opsim{\mathop{\sim}}

\def\opmin{\mathop{\min}} 
\def\opmax{\mathop{\max}} 
\def\oplim{\mathop{\lim}}

\title{ Journey from the Wilson exact RG towards the Wegner-Morris Fokker-Planck RG  \\
and to the Carosso field-coarsening via Langevin stochastic processes  } 


\author{C\'ecile Monthus}
\affiliation{Universit\'e Paris-Saclay, CNRS, CEA, Institut de Physique Th\'eorique, 91191 Gif-sur-Yvette, France}


\begin{abstract}

Within the Wilson RG of 'incomplete integration' as a function of the effective RG-time $t$, the non-linear differential RG-flow for the energy $E_t[\phi(.)]$ translates for the probability distribution $P_t[\phi(.)] \sim e^{- E_t[\phi(.)]} $ into the linear Fokker-Planck RG-flow associated to independent non-identical Ornstein-Uhlenbeck processes for the Fourier modes. The corresponding Langevin stochastic differential equations for the real-space field $\phi_t(\vec x)$ have been recently interpreted by Carosso as genuine infinitesimal coarsening-transformations that are the analog of spin-blocking, and whose irreversible character is essential to overcome the paradox of the naive description of the Wegner-Morris Continuity-Equation for the RG-flow as a meaningless infinitesimal change of variables in the partition function integral. This interpretation suggests to consider new RG-schemes, in particular the Carosso RG where the Langevin SDE corresponds to the stochastic heat equation also known as the Edwards-Wilkinson dynamics. After a pedestrian self-contained introduction to this stochastic formulation of RG-flows, we focus on the case where the field theory is defined on the large volume $L^d$ with periodic boundary conditions, in order to distinguish between extensive and intensives observables while keeping the translation-invariance. Since the empirical magnetization $m_e \equiv  \frac{1}{L^d} \int_{L^d}  d^d \vec x \ \phi(\vec x) $ is an intensive variable corresponding to the zero-momentum Fourier coefficient of the field, its probability distribution $p_L(m_e)$ can be obtained from the gradual integration over all the other Fourier coefficients associated to non-vanishing-momenta via an appropriate adaptation of the Carosso stochastic RG, in order to obtain the large deviation properties with respect to the volume $L^d$.

\end{abstract}

\maketitle

\section{ Introduction }

In statistical physics, the idea of renormalization is essential to understand 
how some universal macroscopic properties can emerge from many different microscopic models.
The simplest example is the famous Central-Limit-Theorem for
 the sum of a large number $N$ of independent identical random variables,
 that can be directly rephrased in the 
 renormalization language \cite{Jona75,Jona78,JonaReview,calvosum},
 where the Gaussian distribution and the L\'evy stable laws appear as the possible fixed points 
 after an appropriate rescaling involving the system size $N$ and universal critical exponents
($\frac{1}{2}$ for the Gaussian case).
For the Ising model involving spins $S_i=\pm 1$ on a lattice with $N=L^d$ sites in dimension $d$,
the extensive magnetization $M$ associated to the global spin configuration $\{S_{i=1,..,N} \}$
\begin{eqnarray}
M \equiv \sum_{i=1}^{L^d} S_i
\label{extensiveM}
\end{eqnarray}
can be also seen as a sum of $N=L^d$ random variables $S_i$, but the spins $S_i$ are now correlated via their Boltzmann probability distribution (see the reviews \cite{JonaReview,JPBreview,BertinReview} and references therein). 
So besides the high-temperature paramagnetic phase where the magnetization will display the same scaling 
$M^{typ} \propto \sqrt{N} = L^{\frac{d}{2}} $ as in the CLT, there will be in dimension $d >1$ the low-temperature ferromagnetic phase, where the intensive version of Eq. \ref{extensiveM} 
\begin{eqnarray}
m_e \equiv \frac{M}{L^d} = \frac{1}{L^d} \sum_{i=1}^{L^d} S_i 
\ \ \text{ = the empirical magnetization associated to the global configuration $\{S_{i=1,..,N} \}$}
\label{intensivem}
\end{eqnarray}
will converge in the thermodynamic limit $L\to +\infty$ towards two possible opposite finite values $\pm m_{sp}$ corresponding to the symmetry-breaking produced by the correlations between the spins of a given configuration.
The probability distribution $p_L(m_e)$ of the empirical magnetization $m_e$ 
plays an essential role in numerical studies of phase transitions to locate the critical point
 via the famous Binder cumulant method \cite{Binder1,Binder2,BinderReview1985}.
 More recently, the probability distribution $p_L(m_e)$ has been analyzed, at the level of large deviations 
 (see the reviews \cite{oono,ellis,review_touchette} and references therein), in a series of papers 
 \cite{Delamotte2022,Delamotte2024per,Delamotte2024uni,Delamotte2025,Sahu}
 based on the exact functional RG of field theory, in particular to identify the universal and non-universal properties in the critical region \cite{Delamotte2024uni,stella}.
 
 Since the idea of renormalization has been applied in many areas of physics
(high-energy physics, statistical physics, condensed-matter models)
with very different purposes and various technical implementations,
it can be useful to return to a very general level, 
where the goal is to define an RG-flow in a given space of models
as a function of an effective RG-time $t$,
where $t=0$ represents the initial model one is interested in,
while the renormalized models as a function of the growing RG time $t>0$ are effective models 
where more and more possible fluctuations of the initial model are gradually integrated out.
Any RG procedure contains two parts that can be summarized by the two words
 'choose' and 'compute' : (1) the first part consists in choosing for any RG-time $t$
what degrees of freedom of the initial model are still surviving 
and what degrees of freedom have already been integrated out;
(2) the second part consists in writing the corresponding RG-flow for the renormalized model
as a function of the RG-time $t$ and in computing its properties in order to obtain some useful information on the initial model at $t=0$. The success of a given RG procedure for a given type of models is thus usually based on a clever choice of variables, that should be meaningful from the physical point of view, but that should also be smart from the technical point of view, in order to be able to analyze the corresponding RG-flow. 
This explains why there is no such thing as a universal RG procedure that would be 
  appropriate to answer any question on any model, and why so many different RG schemes have been introduced over the years to analyze all kind of issues in all areas of physics.

 In statistical field theory, where the discrete spins $S_i=\pm 1$ living on the lattice
 are replaced by the 'continuous-spin' $ \phi(\vec x)$ living in continuous space $\vec x$,
 the technical possibility to write exact differential functional RG-flows with respect to
 the continuous RG-time $t$, in particular via the Wilson RG-scheme called 'incomplete integration'
 \cite{WilsonKogut,Riedel}, 
 or via the Wegner-Houghton RG-scheme based on a sharp Fourier cut-off \cite{WH}, 
 or via the Polchinski \cite{Polchinski} and Wetterich \cite{Wetterich} RG-schemes based on an additional momentum-dependent Gaussian regulator term,
has attracted a lot of interest (see the various reviews \cite{Aoki,BandB,Wetterich_Review,Polonyi,Pawlowski,Osborn,Rosten,Gies,DelamotteReview,leonie} with different scopes and references therein). 
For statistical physicists working in the area of stochastic processes,
the great outcome is that the Wilson RG-scheme called 'incomplete integration'
described in section 11 of the review by Wilson and Kogut \cite{WilsonKogut}
corresponds to a linear Fokker-Planck Evolution for the renormalized probability distribution, 
even if it is not advertised as such in \cite{WilsonKogut}.
More generally in the literature on exact RG in field theory (see the various reviews \cite{Aoki,BandB,Wetterich_Review,Polonyi,Pawlowski,Osborn,Rosten,Gies,DelamotteReview,leonie}), 
the main focus is instead on the non-linear RG-flows for effective actions or for generating functions, 
with the notable exception of the Wegner-Morris perspective 
\cite{Wegner74,Morris1993,Morris2000,Morris2001,Morris2002}
based on the continuity Equation that is necessary to ensure the conservation of the partition function,
and on the corresponding infinitesimal change of variables in the partition function integral. However, to obtain a meaningful irreversible RG-flow where the RG-time-arrow matters,
as in the Fokker-Planck RG-flow mentioned above,
this field redefinition cannot be a mere change between dummy variables of integration, 
but should be taken very seriously as a physically-meaningful coarsening-transformation for the field.
This step further has been recently developed by Carosso \cite{Carosso,Carosso_conf,Carosso_PhD} :
the Langevin Stochastic Differential equations associated to the Fokker-Planck RG-flow
are interpreted as genuine coarsening-transformations for the field itself
that are the analog of the block-spins in discrete-time RG for lattice spins models,
and where the noise is the technically-convenient way to include the desired irreversibility
associated to the elimination of information as a function of the RG-time-arrow.
In addition, this interpretation has led Carosso \cite{Carosso,Carosso_conf,Carosso_PhD}
to consider different Fokker-Planck RG-flows with respect to the initial Wilson RG-scheme,
where the Langevin Stochastic Differential equation for the field $\phi_t(\vec x)$
involves the Laplacian operator
and corresponds to the stochastic heat equation with additive space-time noise in the mathematical literature (see the lecture notes \cite{StochasticPDE} on the broader area of Stochastic Partial Differential Equations), which is also well-known as the Edwards-Wilkinson dynamics \cite{Stanley,HHZ} in statistical physics, so that the coarsening-transformation of the field in real-space 
is realized by the Gaussian heat-kernel.
It is clear that all these stochastic aspects that appear in field theory exact RG-flows
are of the utmost interest for all statistical physicists familiar with diffusion processes
even without any knowledge about field theory (like me).

The goal of the present paper is thus to give a very pedestrian self-contained introduction into this type of Fokker-Planck RG-flows for readers interested in stochastic processes (see the textbooks \cite{gardiner,vankampen,risken,Pavliotis}), before focusing on a specific application concerning the empirical magnetization $m_e$ of Eq. \ref{intensivem}. While field theory is usually written
 in the infinite space $  \vec x \in {\mathbb R}^d$ with Fourier-transforms involving continuous momenta $\vec q \in {\mathbb R}^d$, we will consider fields defined on the finite volume $\vec x \in [0,L]^d$
 in order to be able to introduce extensive and intensive observables as in lattice models (as in Eqs \ref{extensiveM} and \ref{intensivem}), and in order to analyze the large deviations properties of the empirical magnetization $m_e$ of Eq. \ref{intensivem} with respect to the volume $L^d$. Since the Fourier-series decomposition of the field involves Fourier-coefficients with discrete momenta living on the lattice 
 $\vec k \in {\mathbb Z}^d$, the Fokker-Planck RG-flows for the Fourier coefficients will have the 
 simpler form of a multidimensional Fokker-Planck Equation instead of the functional Fokker-Planck RG-flows that appear within the standard field theory formulation based on continuous momenta.
In addition, to be more pedagogical, all the discussed RG schemes will be explained first 
on the single-variable toy-model already described in section 11 of the review by Wilson and Kogut \cite{WilsonKogut} in order to see more clearly the important ideas and the basic computations.

The paper is organized as follows.
We first describe the statistical properties for the real-space field $\phi(\vec x)$ defined on a volume $L^d$
(Section \ref{sec_Realspace}) and for its Fourier coefficients 
(Section \ref{sec_Fourier}). We then recall
 the main ideas of the Wilson differential RG based on 'incomplete integration' (Section \ref{sec_WK}),
 of the Wegner-Morris perspective where RG is seen as an infinitesimal change of variables 
conserving the partition function (Section \ref{sec_WM}), and of
the Carosso field-coarsening via Langevin stochastic processes (Section \ref{sec_Carosso}).
Finally in Section \ref{sec_RGVolume}, we adapt the Carosso RG-scheme to the finite volume $L^d$
and to the constraint imposing the conservation of
 the empirical magnetization $m_e$ at any RG-time $t$ :
 since the empirical magnetization $m_e \equiv  \frac{1}{L^d} \int_{L^d}  d^d \vec x \ \phi(\vec x) $ corresponds to the zero-momentum Fourier coefficient of the field, its probability distribution $p_L(m_e)$ can be obtained from the gradual elimination of all the information contained in all the other Fourier coefficients associated to non-vanishing-momenta.
Our conclusions are summarized in section \ref{sec_conclusion}.
Three appendices contains complementary discussions and more technical aspects.


\section{ Real-space statistical properties of the field $ \phi( \vec x) $ on the volume $L^d$}

\label{sec_Realspace}

In this section, we recall the standard notations used in statistical field theory
except that the spatial integrations are over the finite volume $\vec x \in [0,L]^d$
instead of the infinite space $  \vec x \in {\mathbb R}^d$. 
We focus on the periodic boundary conditions in the $d$ directions in order to keep the translation invariance and to avoid the discussion of boundary effects.
We then explain the link with the theory of large deviations for the empirical magnetization $m_e$
 with respect to the large volume $L^d$.

\subsection{ Statistical properties of the field $ \phi( \vec x) $ 
on the volume $\vec x \in [0,L]^d$ with periodic B.C. }


\subsubsection{ Probability distribution $  {\cal P}_{t=0}[\phi(.) ] $ of the field $ \phi( \vec x) $ for the 
'continuous-spin' Ising model under study }

To be concrete, 
at the RG-time $t=0$, the model under study is chosen to be the following analog of the lattice Ising model.
The probability distribution ${\cal P}_{t=0}[\phi(.)]$ 
to see the 'continuous-spin' configuration $\phi( \vec x )  $ 
is written in the Boltzmann form involving
 some energy functional ${\cal E}_{t=0}[\phi(.)]$ 
\begin{eqnarray} 
  {\cal P}_0[\phi(.) ]  = \frac{ e^{- {\cal E}_0[\phi(.)]} }{  {\cal Z}_0 }
  \label{ProbaPhiS0}
\end{eqnarray}
where the partition function $ {\cal Z}_0$ ensuring the normalization of probabilities
\begin{eqnarray} 
  {\cal Z}_0  =  \int  {\cal D}\phi(.) e^{- {\cal E}_0[\phi(.)]  }
  \label{Z0}
\end{eqnarray}
 involves the integration 
over the possible field configurations $\phi(.) $ 
with some appropriate measure $\int  {\cal D}\phi(.) $. 
The extensive energy functional ${\cal E}_{t=0}[\phi(.)]$ for the continuous-spin configuration
$\phi(\vec x)$
is written as an integral over the volume $\vec x \in [0,L]^d$ of the form
\begin{eqnarray} 
 {\cal E}_{t=0}[\phi(.)]  =  \int_{L^d}  d^d \vec x \left( U_0[ \phi(\vec x)] 
 +\frac{\Upsilon_0}{2} \left[ \vec \nabla  \phi( \vec x) \right]^2   \right)
  \label{ActionS0}
\end{eqnarray}
The even local potential $ U_0[ \phi]= U_0[-\phi]$ should be minimal at two opposite values 
(for instance $\phi=\pm 1$ in order to mimic the discrete spins $S=\pm 1$)
and should grow sufficiently fast for $\phi \to \pm \infty$.
The gradient-term $\frac{\Upsilon_0}{2} \left[ \vec \nabla  \phi( \vec x) \right]^2 $ is the continuous analog of the lattice nearest-neighbor ferromagnetic interaction.

To clarify the meaning of the partition function of Eq. \ref{Z0},
let us mention the special case of the dimension $d=1$, 
where Eq. \ref{Z0} corresponds
 to the familiar Feynman path-integral for quantum mechanics \cite{feynman}
 with periodic boundary conditions $\phi(x=L)=\phi(x=0) $
\begin{eqnarray} 
\text{ Case $d=1$:  } \ \ 
 {\cal Z}_0  && = \int_{-\infty}^{+\infty} d \phi(0) 
 \int_{\phi(0)}^{\phi(L)=\phi(0)}  {\cal D}\phi(.) e^{-  \int_0^L  dx \left[ 
 \frac{\Upsilon_0}{2} \left( \frac{ d \phi(x)}{dx} \right)^2 + U_0[ \phi( x)] 
   \right]  }
\nonumber \\
&& = \int_{-\infty}^{+\infty} d \phi(0)  \langle \phi(0) \vert e^{- L H } \vert \phi(0) \rangle 
 ={\rm Tr} \left( e^{- L H } \right)  = \sum_{n=0}^{+\infty} e^{ -L e_n }  \opsimeq_{L \to + \infty} e^{ -L e_0 }
  \label{Z1DpathIntegral}
\end{eqnarray}
where $x$ corresponds to time, $\phi(x)$ corresponds to the position of the particle as a function of the time $x \in [0,L]$,
the energy $ {\cal E}_0[\phi(.)]  $ is the Euclidean classical action involving the standard kinetic energy 
$\frac{\Upsilon_0}{2} \left( \frac{ d \phi(x)}{dx} \right)^2 $ with mass $\Upsilon_0$ 
and the scalar potential $U_0(\phi)$, 
while $H$ is the associated quantum Hamiltonian
\begin{eqnarray}
H =  - \frac{1}{2 \Upsilon_0} \frac{ d^2}{d\phi^2}+  U_0[ \phi] 
=  \sum_{n=0}^{+\infty} e_n \vert \psi_n \rangle \langle \psi_n\vert 
\label{hamiltonian}
\end{eqnarray}
with its spectral decomposition involving the discrete eigenvalues $e_n$ and the corresponding 
orthonormalized eigenstates $\vert \psi_n \rangle$ (we assume that the potential $ U_0[ \phi]$ 
grows sufficiently fast for $\phi \to \pm \infty$ to produce only discrete levels), 
so that the groundstate energy $e_0$
will govern the asymptotic behavior of Eq. \ref{Z1DpathIntegral} for large $L$.

More generally in statistical physics, 
the partition function ${\cal Z}_0 $ grows exponentially with respect to the volume $L^d$
\begin{eqnarray} 
  {\cal Z}_0   \opsimeq_{L \to + \infty} e^{ -L^d f }
  \label{Z0expVolume}
\end{eqnarray}
where $f$ is the intensive free-energy,
hence our preference to define models on a large volume $L^d$ (instead of the infinite space usually considered in field theory).


\subsubsection{ Probability distribution $  {\cal P}_t[\phi(.) ] $ of the field $ \phi( \vec x) $ as a function of the RG-time $t \geq 0$ }

In the area of functional RG, the goal is to write
an appropriate differential RG-flow for the probability distribution ${\cal P}_t[\phi(.)]$ 
to see the field configuration $\phi( \vec x )  $ at the RG-time $t$,
or equivalently for the energy functional ${\cal E}_{t}[\phi(.)] $ 
defined by the change of variables
\begin{eqnarray} 
  {\cal P}_t[\phi(.) ] \equiv \frac{ e^{- {\cal E}_t[\phi(.)]} }{  {\cal Z}_0 }
  \label{ProbaPhiS}
\end{eqnarray}
where one imposes that the partition function ensuring the normalization of probabilities is conserved via the RG-flow
\begin{eqnarray} 
  {\cal Z}_t  \equiv  \int  {\cal D}\phi(.) e^{- {\cal E}_t[\phi(.)]  } = {\cal Z}_0 \ \ \text{ conserved at any time $t$ }
  \label{Z}
\end{eqnarray}

Via RG, the energy functional ${\cal E}_{t}[\phi(.)] $ is expected to remain 
an integral over the volume $L^d$ :
the parameters already present in the initial energy functional of Eq. \ref{ActionS0}
 will become RG-time-dependent
\begin{eqnarray} 
 {\cal E}_{t}[\phi(.)]   =  \int_{L^d}  d^d \vec x 
 \left( U_t[ \phi(\vec x)] + \frac{\Upsilon_t }{2} \left[ \vec \nabla  \phi( \vec x) \right]^2  +... \right)
  \label{ActionSt}
\end{eqnarray}
but all the terms involving the field $\phi(\vec x)$ and its derivatives
that are compatible with the symmetries
 are also expected to appear as additional contributions $(...)$.
Note that the power-series expansion of the even local potential $U_t[ \phi] =U_t[-\phi] $
\begin{eqnarray} 
U_t[ \phi] && = \sum_{n=0}^{+\infty} \frac{u_{2n}(t)}{(2n)!}  \phi^{2n}  
  = u_0(t) + \frac{u_{2}(t)}{2}  \phi^{2} (\vec x) + \frac{u_4(t)}{4!}  \phi^{4}  +... 
  \label{Vlocalt}
\end{eqnarray}
already involves an infinite number of couplings $u_{2n}(t)$ that depend on the RG-time $t$,
where the constant part $u_0(t) $ is included in order to be able to satisfy the conservation of the partition function of Eq. \ref{Z}.

In conclusion, the goal of functional RG is to 
write an appropriate differential RG-flow as a function of the RG-time $t$
for the energy functional ${\cal E}_{t}[\phi(.)] $ 
(called the Wilson action $S_t[\phi(.)]$ in field theory)
that will summarize the ordinary RG-flows 
for all the corresponding couplings that will appear in Eq. \ref{ActionSt}.
It is also useful to consider the RG-flows for the probability distribution of Eq. \ref{ProbaPhiS}
and for various generating functions that are described in the next subsections.


\subsubsection{ Generating function ${\cal Z}_t[h(.) ]  $ for the correlations of the field $ \phi( \vec x) $  }

In field theory, it is standard to introduce an arbitrary inhomogeneous magnetic field $h(\vec x)$
(usually called the source $J(\vec x)$) and to
define the generalized partition function $ {\cal Z}_t[h(.) ] $ 
with respect to the zero-field partition function $ {\cal Z}_t[h(.)=0 ] = {\cal Z}_t={\cal Z}_0$ of Eq. \ref{Z} 
\begin{eqnarray} 
 {\cal Z}_t[h(.) ] && \equiv \int  {\cal D}\phi(.) e^{- {\cal E}_t[\phi(.)] +  \int  d^d \vec x h(\vec x) \phi(\vec x) }
   \label{Zh}
\end{eqnarray}
 The ratio
\begin{eqnarray} 
\frac{{\cal Z}_t[h(.) ] }{{\cal Z}_0} && 
= \int  {\cal D}\phi(.) \frac{ e^{- {\cal E}_t[\phi(.)] }}{{\cal Z}_0} e^{  \int_{L^d}  d^d \vec x h(\vec x) \phi(\vec x) } 
\nonumber \\
&& = \sum_{n=0}^{+\infty} \frac{1}{n!} \int d^d \vec x_1 ... \int d^d \vec x_n C^{(n)}_t(\vec x_1,\vec x_2,..,\vec x_n) h(\vec x_1) ... h(\vec x_n)
\nonumber \\
&& = 1 +  \int  d^d \vec x h(\vec x) C^{(1)}_t(\vec x)
+  \frac{1}{2} \int d^d \vec x_1 \int d^d \vec x_2 C^{(2)}_t(\vec x_1,\vec x_2) h(\vec x_1)  h(\vec x_2) +...
   \label{Zhratio}
\end{eqnarray}
can be expanded in the magnetic-field $h(.)$ 
to generate the correlations of the field $\phi( \vec x) $ 
drawn with the probability distribution $  {\cal P}_t[\phi(.) ] $ of Eq. \ref{ProbaPhiS}
\begin{eqnarray} 
C^{(n)}_t(\vec x_1,\vec x_2,..,\vec x_n) && \equiv 
\int  {\cal D}\phi(.) \frac{ e^{- {\cal E}_t[\phi(.)]  } }{{\cal Z}_0} \phi (\vec x_1) ... \phi (\vec x_n)
   \label{Zn}
\end{eqnarray}
The first terms $n=1,2$ correspond to the averaged value $C^{(1)}_t(\vec x) $ 
and to the two-point correlation function $C^{(2)}_t(\vec x_1,\vec x_2) $.


\subsubsection{ Generating function ${\cal W}_t[h(.)]  $ of the connected-correlations of the field $ \phi( \vec x) $  }

The logarithm of the ratio of Eq. \ref{Zhratio}
\begin{eqnarray} 
{\cal W}_t[h(.)] \equiv  \ln \left(\frac{{\cal Z}_t[h(.) ] }{{\cal Z}_0} \right)  
&& 
= \ln \left( 1 +  \int  d^d \vec x h(\vec x) C^{(1)}_t(\vec x)
+  \frac{1}{2} \int d^d \vec x_1 \int d^d \vec x_2 C^{(2)}_t(\vec x_1,\vec x_2) h(\vec x_1)  h(\vec x_2) +...
\right)
\nonumber \\
&& 
=   \int  d^d \vec x h(\vec x) C^{(1)}_t(\vec x)
+  \frac{1}{2} \int d^d \vec x_1 \int d^d \vec x_2 \left[ C^{(2)}_t(\vec x_1,\vec x_2) - C^{(1)}_t(\vec x_1) C^{(1)}_t(\vec x_2)\right] h(\vec x_1)  h(\vec x_2)  
+ ...
\nonumber \\
&& 
\equiv  \sum_{n=1}^{+\infty} \frac{1}{n!} \int d^d \vec x_1 ... \int d^d \vec x_n W^{(n)}_t(\vec x_1,\vec x_2,..,\vec x_n) h(\vec x_1) ... h(\vec x_n)
   \label{Wh}
\end{eqnarray}
is the generating function of the connected correlations $W^{(n)}_t(\vec x_1,\vec x_2,..,\vec x_n) $  of the field 
 with the first terms
\begin{eqnarray} 
W^{(1)}_t(\vec x) && = C^{(1)}_t(\vec x)
\nonumber \\
W^{(2)}_t(\vec x_1,\vec x_2) && =C^{(2)}_t(\vec x_1,\vec x_2) - C^{(1)}_t(\vec x_1) C^{(1)}_t(\vec x_2)
   \label{Wn12}
\end{eqnarray}


\subsubsection{ Legendre transform $\Gamma_t[ \Phi(.)] $ of ${\cal W}_t[h(.)]  $   }

Finally, it is standard to introduce the Legendre transform $\Gamma_t[ \Phi(.)] $ of ${\cal W}_t[h(.)]  $ 
that is usually called the 'generating function of proper vertices' in perturbative field theory 
or the 'effective average action' in non-perturbative field theory
\begin{eqnarray} 
 \Gamma_t[ \Phi(.)]  \equiv \max_{ h(.) } \bigg( \int  d^d \vec x h(\vec x) \Phi(\vec x) - {\cal W}_t[h(.)] \bigg)
   \label{GammaWh}
\end{eqnarray}
so that the variable $ \Phi(\vec x) $ corresponds to the functional derivative of ${\cal W}_t[ h(.) ] $ 
of Eq. \ref{Wh}
with respect to 
the magnetic field $ h(\vec x) $
\begin{eqnarray} 
 \Phi(\vec x) && \equiv \frac{\partial {\cal W}_t[ h(.) ] }{ \partial h(\vec x) } 
 = \frac{\frac{\partial {\cal Z}_t[ h(.) ] }{ \partial h(\vec x) }}{{\cal Z}_t[ h(.)] }  
 =\frac{\int  {\cal D}\phi(.) e^{- {\cal E}_t[\phi(.)] +  \int  d^d \vec x h(\vec x) \phi(\vec x) } \phi(\vec x) }{ \int  {\cal D}\phi(.) e^{- {\cal E}_t[\phi(.)] +  \int  d^d \vec x h(\vec x) \phi(\vec x) } }  
    \label{PhivariableLegendreWdef}
\end{eqnarray}
i.e. to the averaged value of the field $ \phi(\vec x) $
 in the presence of the arbitrary magnetic field $h(.)$,
 while the inverse Legendre transform yields
 \begin{eqnarray} 
h(\vec x) =  \frac{\partial \Gamma_t[ \Phi(.)]  }{ \partial \Phi(\vec x) } 
   \label{hLegendreinverseGamma}
\end{eqnarray}
To see more clearly the relation with the energy functional ${\cal E}_t[\phi(.)] $,
one can use the definitions of Eqs \ref{Zhratio} and \ref{Wh} to write for any $\Phi(\vec x) $
\begin{eqnarray} 
e^{ \displaystyle {\cal W}_t[h(.)]  - \int_{L^d}  d^d \vec x h(\vec x) \Phi(\vec x)}  
&& = \frac{1}{{\cal Z}_0} \int  {\cal D}\phi(.)   
e^{ \displaystyle - {\cal E}_t[\phi(.)] + \int_{L^d}  d^d \vec x h(\vec x) \left[ \phi(\vec x)- \Phi(\vec x) \right] } 
\nonumber \\
&& = \frac{1}{{\cal Z}_0} \int  {\cal D}\zeta(.)   
e^{ \displaystyle - {\cal E}_t[\Phi(.) + \zeta(.)] + \int_{L^d}  d^d \vec x h(\vec x)  \zeta (\vec x)  }
   \label{Zhratiomod}
\end{eqnarray}
The optimization over $h(.)$ then yields using Eqs \ref{GammaWh} and \ref{hLegendreinverseGamma}
\begin{eqnarray} 
e^{ \displaystyle -  \Gamma_t[ \Phi(.)] }  
&& = \frac{1}{{\cal Z}_0} \int  {\cal D}\zeta(.)   
e^{ \displaystyle - {\cal E}_t[\Phi(.) + \zeta(.)] + \int_{L^d}  d^d \vec x   \zeta (\vec x) \frac{\partial \Gamma_t[ \Phi(.)]  }{ \partial \Phi(\vec x) }  }
\nonumber \\
&& = \int  {\cal D}\zeta(.)   P_t [\Phi(.) + \zeta(.) ] 
\ \ e^{ \displaystyle  \int_{L^d}  d^d \vec x   \zeta (\vec x) 
\frac{\partial \Gamma_t[ \Phi(.)]  }{ \partial \Phi(\vec x) }  }
   \label{GammaBackground}
\end{eqnarray}
that can be considered as some kind of generating function of the deviation $\zeta(.)$ 
with respect to the background field $\Phi(.)$, albeit with the implicit conjugated field $h(\vec x) =  \frac{\partial \Gamma_t[ \Phi(.)]  }{ \partial \Phi(\vec x) }  $, that shows that $\Gamma_t[ \Phi(.)] $
would coincide with ${\cal E}_t[\Phi(.) ] $ in the saddle-point approximation for the integral over 
the deviation $\zeta (\vec x) $.


\subsection{ Special case of the uniform magnetic field $h(\vec x)=h$ :  large deviations for the empirical magnetization $m_e$ }

Except for the Gaussian model where the generating function ${\cal W}_t[h(.) ] $ 
can be explicitly written for any function $h(.)$ as recalled in Eq. \ref{Generatinghxgauss}, 
the introduction of an arbitrary inhomogeneous magnetic field $h(.)$
to study spatially-translation-invariant models is actually somewhat strange 
since it breaks the symmetry of translation-invariance 
and it thus complicates very significantly the analysis,
as is well-known from the area of random-field and other disordered systems. 
As a consequence, it is useful to discuss now the special case of the 
uniform magnetic field on the whole volume $\vec x \in [0,L]^d$
\begin{eqnarray} 
h(\vec x)=h
   \label{huniform}
\end{eqnarray}
Then the source-term appearing in Eq. \ref{Zh}
\begin{eqnarray}
 \int_{L^d}  d^d \vec x \ h(\vec x) \phi(\vec x) 
  = h  \int_{L^d}  d^d \vec x \  \phi(\vec x) 
   \equiv h M
\label{sourcemedef}
\end{eqnarray}
involves the extensive magnetization $M=\int_{L^d}  d^d \vec x \  \phi(\vec x) $ of the configuration
that is the continuous analog of Eq. \ref{extensiveM},
while the corresponding intensive variable
representing the empirical magnetization $m_e$ of the 'continuous-spin' configuration $\phi(.)$
is the analog of Eq. \ref{intensivem}
\begin{eqnarray}
m_e  \equiv \frac{1}{L^d} \int_{L^d}  d^d \vec x \ \phi(\vec x) 
\label{medef}
\end{eqnarray}


\subsubsection{ Generating function $w(h)$ for the scaled-cumulants $w_n$ of the empirical magnetization $m_e$ of the initial model at $t=0$}

Plugging the source-term of Eq. \ref{sourcemedef} into 
 the ratio $\frac{{\cal Z}_0[h ] }{{\cal Z}_0} $ of Eq. \ref{Zhratio} or the exponential of ${\cal W}_0(h) $ of Eq. \ref{Wh} associated to the initial model at $t=0$
 yields
\begin{eqnarray} 
\frac{{\cal Z}_0[h ] }{{\cal Z}_0} = e^{{\cal W}_0(h)} && 
= \int  {\cal D}\phi(.) \frac{ e^{- {\cal E}_0[\phi(.)] }}{{\cal Z}_0} e^{  h \int_{L^d}  d^d \vec x  \phi(\vec x) } 
\nonumber \\
&&  =  \int_{-\infty}^{+\infty} d m_e e^{L^d h m_e} 
\int  {\cal D}\phi(.) \frac{ e^{- {\cal E}_0[\phi(.)] }}{{\cal Z}_0} \delta\left(  m_e - \frac{1}{L^d} \int_{L^d}  d^d \vec x \phi(\vec x) \right) 
\nonumber \\
&& \equiv \int_{-\infty}^{+\infty} d m_e e^{L^d h m_e}p_L(m_e)
   \label{Zhratiohomo}
\end{eqnarray}
where
\begin{eqnarray} 
p_L(m_e) 
\equiv \int  {\cal D}\phi(.) \frac{ e^{- {\cal E}[\phi(.)] }}{{\cal Z}_0} \delta\left(  m_e - \frac{1}{L^d} \int_{L^d}  d^d \vec x \phi(\vec x) \right) 
   \label{pme}
\end{eqnarray}
is the probability distribution $p_L(m_e) $
of the empirical magnetization $m_e$.

It is obvious that the generating function of Eq. \ref{Zhratiohomo} 
grows exponentially with respect to the volume $L^d$
\begin{eqnarray} 
\frac{{\cal Z}_0[h ] }{{\cal Z}_0} = e^{{\cal W}_0(h)}= \int_{-\infty}^{+\infty} d m_e e^{L^d h m_e}p_L(m_e)
\opsimeq_{L \to + \infty} e^{L^d w(h) }
   \label{SCGFwhme}
\end{eqnarray}
where the intensive part $w(h) $ of ${\cal W}_0(h) $ is called the scaled-cumulant-generating-function in the theory of large deviations (see the reviews \cite{oono,ellis,review_touchette} and references therein)
\begin{eqnarray} 
w(h) \equiv  \frac{{\cal W}_0[h ] }{L^d}  \equiv \sum_{n=1}^{+\infty} \frac{w_n}{ n!} h^n
= w_1 h + \frac{w_2}{2} h^2+...
   \label{whmeintensive}
\end{eqnarray}
The comparison with the series expansion of ${\cal W}_0[h(.) ] $ of Eq. \ref{Wh}
yields that the scaled-cumulants $w_n $ of $m_e$
are directly
 related to spatial-integrals of the connected correlations $W^{(n)}_0(\vec x_1,\vec x_2,..,\vec x_n) $
\begin{eqnarray} 
w_n = \frac{1}{L^d}  \int d^d \vec x_1 ... \int d^d \vec x_n W^{(n)}_0(\vec x_1,\vec x_2,..,\vec x_n)  
= \int_{L^d} d^d \vec r_1 ... \int_{L^d} d^d \vec r_{n-1} W^{(n)}_0(\vec 0 ,\vec r_1,..,\vec r_{n-1})
  \label{wncorre}
\end{eqnarray}
where we have used the translation-invariance to obtain the last expression.
In particular, the first cumulant
\begin{eqnarray} 
w_1 = \frac{1}{L^d}  \int d^d \vec x  W^{(1)}(\vec x)   = W^{(1)}_0(\vec 0)
  \label{w1corre}
\end{eqnarray}
is the averaged value of the field $W^{(1)}_0(\vec x)   = W^{(1)}_0(\vec 0) $ that does not depend on the position $\vec x$
in this translation-invariant model,
while the second scaled-cumulant 
\begin{eqnarray} 
w_2 = \frac{1}{L^d}  \int_{L^d} d^d \vec x_1 \int_{L^d} d^d \vec x_2 W^{(2)}_0(\vec x_1,\vec x_2) 
= \int_{L^d} d^d \vec r \ W^{(2)}_0(\vec 0,\vec r)
\equiv  \chi_L
  \label{w2corre}
\end{eqnarray}
coincides with the integral of the connected correlation $W^{(2)}_0(\vec 0,\vec r) $ 
over the volume $\vec r \in L^d$
and thus represents the magnetic susceptibility $\chi_L$.


\subsubsection{ Rate function $i(m_e)$ governing the large deviations of the probability $p_L(m_e)$ of the empirical magnetization $m_e$ }

The asymptotic behavior of Eq. \ref{SCGFwhme} for large $L$ means that the probability distribution $p_L(m_e) $
is governed by the large deviations behavior
\begin{eqnarray} 
p_L(m_e)   \opsimeq_{L \to + \infty}   e^{- L^d  i(m_e) }
  \label{rateim}
\end{eqnarray}
where $i(m_e) \geq 0 $ is called the rate-function in the theory of large deviations (see the reviews \cite{oono,ellis,review_touchette} and references therein).
Plugging Eq. \ref{rateim} into Eq. \ref{SCGFwhme}
\begin{eqnarray} 
 e^{{\cal W}_0(h)} = \int_{-\infty}^{+\infty} d m_e e^{L^d h m_e}p_L(m_e)
 \opsimeq_{L \to + \infty} \int_{-\infty}^{+\infty} d m_e e^{L^d \big[ h m_e- i(m_e) \big]}
\opsimeq_{L \to + \infty} e^{L^d w(h) }
   \label{SCGFwhmesaddle}
\end{eqnarray}
yields via the saddle-point evaluation for large $L$ of the integral over $m_e$ 
 that the scaled-cumulant-generating-function $w(h) $ is the Legendre transform of the rate function $i(m_e) $
\begin{eqnarray} 
w(h) = \max_{m_e} \left[ hm_e - i(m_e)  \right] 
  \label{Legendremalonemax}
\end{eqnarray}
i.e. more explicitly
\begin{eqnarray} 
w(h)&& = hm_e - i(m_e)   
 \nonumber \\
 0 && = h -  i' (m_e) 
  \label{Legendremalone}
\end{eqnarray}
with the reciprocal Legendre transform
\begin{eqnarray} 
i(m_e) && = h m_e -  w(h)
 \nonumber \\
 0 && = m_e - w'(h)
  \label{Legendremalonereci}
\end{eqnarray}

So in the theory of large deviations,
 the Legendre transformation between the two intensive functions $i(m_e)$ and $w(h)$
is directly related to the saddle-point evaluation of Eq. \ref{SCGFwhmesaddle}
for large volume $L^d \to + \infty$,
as more generally for all Legendre transformations involving intensive functions in statistical physics.


\subsubsection{ Discussion  }

In summary, the case of the uniform magnetic field $h(\vec x)=h$ 
for the field theory defined on the finite volume $L^d$ 
is useful to analyze the large deviations properties of the empirical magnetization $m_e$.
As already mentioned in the Introduction, these large deviations properties are instrumental 
 in numerical studies to locate the critical point
 via the Binder cumulant method \cite{Binder1,Binder2,BinderReview1985},
 and have attracted a lot of theoretical interest recently \cite{Delamotte2022,Delamotte2024per,Delamotte2024uni,stella,Delamotte2025,Sahu}
(see in particular \cite{Delamotte2024uni} with a very detailed discussion 
for a variety of models and methods, namely perturbative RG in $d=4-\epsilon$, functional RG, hierarchical models, large N limit, and Monte Carlo simulations for the Ising model in dimension $d=3$).
 
Since the empirical magnetization $m_e$ of Eq. \ref{medef}
corresponds to the zero-momentum Fourier-coefficient of the field $\phi (\vec x) $
defined on the volume $L^d$, it is useful in the next section to discuss the statistical properties of the other Fourier-coefficients associated to non-zero momenta.



\section{ Statistical properties of the Fourier coefficients ${\hat \phi}  (\vec k)  $ of the field $\phi(\vec x)  $ }

\label{sec_Fourier}

In this section, we focus on the Fourier-series coefficients ${\hat \phi}  (\vec k) $
 of the real-space field $\phi(\vec x) $ defined on the volume $\vec x \in [0,L]^d$ in order to analyze their statistical properties.

\subsection{ Fourier-series decomposition of the field $\phi(\vec x) $ with Fourier coefficients ${\hat \phi}  (\vec k)   $ living on the lattice $\vec k \in {\mathbb Z}^d $}

\label{subsec_fourierseries}

In the presence of periodic boundary conditions in the $d$ directions of the volume $L^d$, 
the Fourier-series decomposition of the real field $ \phi(\vec x) $
\begin{eqnarray} 
 \phi( \vec x) =  \sum_{\vec k \in {\mathbb Z}^d}   {\hat \phi}  ( \vec k) e^{i  \frac{ 2 \pi }{L} \vec k . \vec x} 
  \label{FourierPerio}
\end{eqnarray}
involves the complex Fourier coefficients ${\hat \phi}  (\vec k)  $ living on the infinite lattice of integers $\vec k \in {\mathbb Z}^d $ (instead of the Fourier-transform involving an integral over the continuous momenta $\vec q \in {\mathbb R}^d$ in field theory usually written in the infinite space $  \vec x \in {\mathbb R}^d$)
\begin{eqnarray} 
{\hat \phi}  (\vec k)  
 = \frac{1}{L^d} \int_{L^d}  d^d \vec x  \phi (\vec x) e^{- i  \frac{ 2 \pi }{L} \vec k . \vec x} 
 = {\hat \phi}^*  (- \vec k) 
  \label{FouriercoefsPerio}
\end{eqnarray}
including the zero-momentum coefficient
\begin{eqnarray} 
{\hat \phi}  (\vec k=\vec 0)   = \frac{1}{L^d} \int_{L^d}  d^d \vec x  \phi (\vec x) \equiv m_e
  \label{FouriercoefsPerioZero}
\end{eqnarray}
that coincides with the empirical magnetization $m_e$ of Eq. \ref{medef}.
For non-vanishing momenta $\vec k \ne \vec 0 $, the Fourier coefficients ${\hat \phi}  (\vec k) $ of Eq. \ref{FouriercoefsPerio}
can be also considered as intensive variables, since they are defined
as spatial-averages of the field $\phi(\vec x)$ over the volume $L^d$
but with some momentum-modulation.
In order to deal with real observables,
it is convenient to define the region $\vec k>0 $ as the region where
 the first non-vanishing coordinate is positive $k_{\mu}>0$
in order to decompose the complex Fourier coefficients satisfying ${\hat \phi}  (\vec k) 
= {\hat \phi}^*  (- \vec k) $ in Eq. \ref{FouriercoefsPerio}
\begin{eqnarray} 
\vec k>\vec 0 : \ \ \ \ \ \ \ {\hat \phi}  (\vec k)  && =  {\hat \phi}^R(\vec k) -i {\hat \phi}^I(\vec k)
\nonumber \\
  {\hat \phi}  ( - \vec k)    &&=  {\hat \phi}^R(\vec k) +i {\hat \phi}^I(\vec k) = {\hat \phi}^*  (\vec k)
  \label{FouriercoefsPeriocab}
\end{eqnarray}
 in terms their real and imaginary parts
\begin{eqnarray} 
{\hat \phi}^R(\vec k)  && =  \frac{1}{L^d} \int_{L^d}  d^d \vec x  \phi (\vec x) \cos \left(  \frac{ 2 \pi }{L} \vec k . \vec x \right) = {\hat \phi}^R(- \vec k)
 \nonumber \\
{\hat \phi}^I(\vec k)&& =  \frac{1}{L^d} \int_{L^d}  d^d \vec x  \phi (\vec x) \sin \left(  \frac{ 2 \pi }{L} \vec k . \vec x \right) 
= - {\hat \phi}^I(- \vec k)
  \label{FouriercoefsPerioab}
\end{eqnarray}
that can be used to rewrite the field $ \phi( \vec x)$ of Eq. \ref{FourierPerio}
\begin{eqnarray} 
 \phi( \vec x) && = {\hat \phi}  (\vec 0) + \sum_{\vec k >0} 
 \left[ \left( {\hat \phi}^R(\vec k) -i {\hat \phi}^I(\vec k)\right)  e^{i  \frac{ 2 \pi }{L} \vec k . \vec x} 
 + \left( {\hat \phi}^R(\vec k) +i {\hat \phi}^I(\vec k)\right)e^{- i  \frac{ 2 \pi }{L} \vec k . \vec x} \right]
 \nonumber \\
 && = m_e + \sum_{\vec k >0} 
 \left[ {\hat \phi}^R(\vec k)  2 \cos \left(  \frac{ 2 \pi }{L} \vec k . \vec x \right) 
 +  {\hat \phi}^I(\vec k) 2 \sin \left(  \frac{ 2 \pi }{L} \vec k . \vec x \right) 
 \right]
  \label{FourierPerioCosSin}
\end{eqnarray}
in terms of the empirical magnetization $m_e={\hat \phi}  (\vec 0) $ of Eq. \ref{FouriercoefsPerioZero}
and in terms of the real Fourier coefficients $[{\hat \phi}^R(\vec k);{\hat \phi}^I(\vec k) ]$ 
associated to the cosine-sine Fourier decomposition of the field $\phi(\vec x)$.

In conclusion, the Fourier series representations of Eq. \ref{FourierPerio} or \ref{FourierPerioCosSin}
is useful to replace the local degrees of freedom $\phi(\vec x)$ on the volume $\vec x \in L^d$
by global intensive variables corresponding to spatial-averages over the volume $L^d$.

At the technical level, the decomposition of Eq. \ref{FourierPerio}
in terms of the complex coefficients ${\hat \phi}  ( \vec k)   $ is usually the most convenient to perform calculations (see examples in Appendix \ref{app_fourier}),
while one can always translate the final results in terms of the real coefficients $[{\hat \phi}^R(\vec k);{\hat \phi}^I(\vec k) ]$, especially when one wishes to discuss their probability distributions more clearly.


\subsection{ Rewriting the energy functional $ {\cal E}[\phi(.)]$ of the field $\phi( \vec x) $ in terms of its Fourier coefficients ${\hat \phi}  (\vec k) $  }

The extensive energy functional $ {\cal E}_t[\phi(.)]  $ of Eq. \ref{ActionSt}
 for the real-space field $\phi( \vec x) $ 
translates into 
\begin{eqnarray} 
{\cal E}_t[\phi(.)]    = L^d E_t[\hat \phi(.)  ]
  \label{ActionSktot}
\end{eqnarray}
where the extensive volume $L^d$ appears directly as prefactor,
while the intensive energy $ E[\hat \phi(.)  ]$ for the intensive Fourier coefficients $\hat \phi(.) $
\begin{eqnarray} 
{\hat E}_t[\hat \phi ]  =  
 \sum_{n=0}^{+\infty}  {\hat E}^{(2n)}_t[\hat \phi(.) ] +  {\hat E}^{Grad}_t[\hat \phi(.)] +...
  \label{Ehatphi}
\end{eqnarray}
will involve in particular the contributions ${\hat E}^{(2n)}_t[\hat \phi(.) ] $ of Eq. \ref{I2n}
associated to the various coefficients $u_{2n}(t)$ of the local potential $U_t[\phi]$ of Eq. \ref{Vlocalt}
and the contribution ${\hat E}^{Grad}[\hat \phi(.)] $ of Eq. \ref{Ederik} associated to the gradient term of coefficient $\Upsilon_t$ (see more detail in Appendix \ref{app_fourier}).


\subsection{ Joint probability distribution $P_t[ m_e;\hat \phi^R(.) ; \hat\phi^I(.)] $ of the Fourier coefficients $[ m_e; \hat \phi^R(.) ; \hat\phi^I(.)]$  }

The Fourier-series representation of Eq. \ref{FourierPerioCosSin}
of the field $\phi(\vec x) $ in terms of its Fourier coefficients $[m_e;\hat \phi^R(.) ; \hat\phi^I(.) ]$
can be considered as a linear change of variables
whose Jacobian is independent of the field. As a consequence,
the measure $\int {\cal D} \phi(.) $ for the real-space field $\phi(\vec x)$ 
can be replaced by the following Fourier-space measure
over the real Fourier coefficients $[m_e;\hat \phi^R(.) ; \hat \phi^I(.) ]$
\begin{eqnarray} 
\int D {\hat \phi} \equiv 
\int_{-\infty}^{+\infty} d m_e \prod_{\vec k >0}  \left[ \int_{-\infty}^{+\infty} d \hat \phi^R(\vec k)
\int_{-\infty}^{+\infty} d \hat \phi^I(\vec k) \right]
  \label{MeasurePhik}
\end{eqnarray}
up to a global multiplicative factor that will disappear when computing averages where the measure appears
both in the numerator and in the denominator as in the ratio of Eq. \ref{Zhratio}.
So the probability ${\cal P}_t[\phi(.)]$ of the real-space field $\phi(\vec x)$ of Eq. \ref{ProbaPhiS}
will be replaced by the probability $P_t[ m_e ; \hat \phi^R(.) ; \hat\phi^I(.)] $ of the Fourier coefficients $[m_e; \hat \phi^R(.) ; \hat\phi^I(.) ]$
\begin{eqnarray} 
  P_t[ m_e;\hat \phi^R(.) ; \hat\phi^I(.)]
  = \frac{ e^{- L^d  {\hat E}_t [ m_e ; \hat \phi^R(.) ; \hat \phi^I(.)] } }  { \hat {\cal Z}_0 }
  \label{ProbaPhik}
\end{eqnarray}
that involves the translation of the energy functional of Eq. \ref{ActionSktot}
discussed in the previous section,
while the normalization of the probabilities of Eq. \ref{ProbaPhik}
 will be ensured by the new partition function $\hat {\cal Z}_t $ computed with the measure of Eq. \ref{MeasurePhik}
\begin{eqnarray} 
  \hat {\cal Z}_t &&  \equiv \bigg( \int_{-\infty}^{+\infty} d m_e \prod_{\vec k >0}  \left[ \int_{-\infty}^{+\infty} d \hat \phi^R(\vec k)
\int_{-\infty}^{+\infty} d \hat \phi^I(\vec k) \right] \bigg) e^{- L^d  {\hat E}_t [ m_e; \hat \phi^R(.) ; \hat \phi^I(.)] }
= \hat {\cal Z}_0
  \label{Zhat}
\end{eqnarray}
that will be conserved by the RG-flow as chosen in Eq. \ref{Z}.

The probability distribution $p_L(m_e,t) $ of the empirical magnetization $m_e $ alone
can be then obtained from the integration of the joint distribution $P_t[m_e; \hat \phi^R(.) ; \hat\phi^I(.)]  $
of Eq. \ref{ProbaPhik}
over all the other Fourier coefficients $ \hat \phi^R(\vec k) ; \hat\phi^I(\vec k)$ associated to $\vec k>0$
\begin{eqnarray} 
p_L(m_e,t) =  
\left( \prod_{\vec k >0}  \left[ \int_{-\infty}^{+\infty} d \hat \phi^R(\vec k)
\int_{-\infty}^{+\infty} d \hat \phi^I(\vec k) \right] \right)
 P_t[m_e;  \hat \phi^R(.) ; \hat\phi^I(.)] 
  \label{pmalone}
\end{eqnarray}


\subsection{ Rewriting the generating functional $ {\cal W}_t [h(.) ]$ involving the magnetic field $h( \vec x) $ in terms of its Fourier coefficients $[ h_e ; \hat h^R(.) ; \hat h^I(.)] $ }

When both the field $\phi(\vec x) $ and the magnetic field $h(\vec x)$ are written in terms of their Fourier coefficients via Eqs \ref{FourierPerioCosSin} and \ref{Fourierh},
the Parseval-Plancherel identity recalled in Eq. \ref{parseval}
yields that the source term involving the scalar product
of the field $\phi(\vec x) $
 and of the magnetic field $h(\vec x)$ translates 
 in terms their intensive Fourier coefficients $[ m_e ; \hat \phi^R(.) ; \hat \phi^I(.)]  $ and $[ h_e ; \hat h^R(.) ; \hat h^I(.)]  $ into
\begin{eqnarray} 
\int_{L^d}  d^d \vec x h(\vec x) \phi(\vec x)
&&  = L^d \sum_{\vec k \in {\mathbb Z}^d} {\hat h}  ( - \vec k)  {\hat \phi}  ( \vec k)
 = L^d \left[ h_e m_e
 + 2 \sum_{\vec k >0} \left(  {\hat h}^R  (  \vec k)  {\hat \phi}^R ( \vec k) + {\hat h}^I  (  \vec k)  {\hat \phi}^I ( \vec k)  \right) \right]
    \label{hphifourier}
\end{eqnarray}
where the extensive volume $L^d$ directly appears as prefactor.

Using Eqs \ref{MeasurePhik}, \ref{ProbaPhik} and \ref{hphifourier}, 
one can translate the real-space generating functions of Eq. \ref{Zhratio} and Eq. \ref{Wh} into
\begin{eqnarray} 
 e^{{\cal W}_t [h(.) ]} = \frac{{\cal Z}_t[h(.) ] }{{\cal Z}_0}
&&= \frac{  \int  {\cal D}\phi(.) \ e^{- {\cal E}_t[\phi(.)] +  \int_{L^d}  d^d \vec x h(\vec x) \phi(\vec x) } }
{ \int  {\cal D}\phi(.) e^{- {\cal E}_0[\phi(.)] } } 
\nonumber \\
&& = \frac{\int D {\hat \phi} \  e^{ \displaystyle L^d \left[ - E_t [ m_e ; \hat \phi^R(.) ; \hat \phi^I(.)]  + h_e m_e
 + 2 \sum_{\vec k >0} \left(  {\hat h}^R  (  \vec k)  {\hat \phi}^R ( \vec k) + {\hat h}^I  (  \vec k)  {\hat \phi}^I ( \vec k)  \right) \right]}  
 }{ \int D {\hat \phi}  e^{- L^d  E_0 [ m_e ;  \hat \phi^R(.) ; \hat \phi^I(.)] } }
 \nonumber \\
 && \equiv \frac{ \hat {\cal Z}_t[ h_e ;  {\hat h}^R (.); {\hat h}^I(.)] }{\hat {\cal Z}_0}
  \equiv e^{{\hat {\cal W}}_t[ h_e ; {\hat h}^R (.); {\hat h}^I(.)]}
  \label{Zhk}
\end{eqnarray}
where the obtained ratio $\frac{ \hat {\cal Z}_t[ h_e ; {\hat h}^R (.); {\hat h}^I(.)] }{\hat {\cal Z}_0} $ or the exponential of ${\hat {\cal W}}_t[ h_e ; {\hat h}^R (.); {\hat h}^I(.)] $ can be used to generate the moments and the correlations of the Fourier coefficients $[m_e;{\hat \phi}^R  (\vec k);{\hat \phi}^I  (\vec k) ] $ that are linearly coupled to the 
 Fourier coefficients $[h_e ; {\hat h}^R (.); {\hat h}^I(.) ]$ of the magnetic field $h(\vec x)$ in the exponential.

 The generating function of the magnetization $m_e$ alone introduced in Eq. \ref{Zhratiohomo}
can be obtained from the generating function of Eq. \ref{Zhk}
with the zero-momentum Fourier coefficient $h_e =h$, 
while all the other Fourier coefficients of the magnetic field vanish $[{\hat h}^R (.) =0; {\hat h}^I(.) =0] $,
so that it corresponds in real-space to the constant magnetic field $h(\vec x)=h$ as expected
\begin{eqnarray} 
\frac{{\cal Z}_t[h ] }{{\cal Z}_0} = e^{{\cal W}_t(h)} = \int_{-\infty}^{+\infty} dm_e e^{L^d h m_e } p_L(m_e,t)   && 
= e^{{\hat {\cal W}}_t[ h_e=h; {\hat h}^R (.) =0; {\hat h}^I(.) =0]}
  \label{Genemalonek}
\end{eqnarray}


\subsection{ Large deviations properties with respect to the large volume $L^d$}

For large $L$, the partition function $\hat {\cal Z}_0 $ of Eq. \ref{Zhat} can be evaluated via the saddle-point method
for the integrals over all the Fourier coefficients $[ m_e;\hat \phi^R(.) ; \hat \phi^I(.)]$
\begin{eqnarray} 
 \hat {\cal Z}_0 
 = \int_{-\infty}^{+\infty} d m_e \prod_{\vec k >0}  \left[ \int_{-\infty}^{+\infty} d \hat \phi^R(\vec k)
\int_{-\infty}^{+\infty} d \hat \phi^I(\vec k) \right] e^{- L^d   {\hat E}_0 [ m_e;\hat \phi^R(.) ; \hat \phi^I(.)] }
 \opsimeq_{L \to + \infty} e^{- L^d {\hat E}_0 }
  \label{hatZlargeL}
\end{eqnarray}
and will be thus dominated by the minimal value 
\begin{eqnarray} 
  {\hat E}_0 = \min_{ \{ m_e ; \hat \phi^R(.) ; \hat \phi^I(.) \}} \bigg( {\hat E} [ m_e ;  \hat \phi^R(.) ; \hat \phi^I(.)] \bigg)
  \label{Emin}
\end{eqnarray}

As a consequence, the joint probability distribution $P_t[ m_e ; \hat \phi^R(.) ; \hat\phi^I(.)] $ of Eq. \ref{ProbaPhik}
will display the following large deviations form with respect to the volume $L^d$
\begin{eqnarray} 
  P_t[ m_e ;  \hat \phi^R(.) ; \hat\phi^I(.)] \opsimeq_{L \to + \infty}  \frac{e^{- L^d  {\hat E}_t [m_e ; \hat \phi^R(.) ; \hat \phi^I(.)] }}{e^{- L^d {\hat E}_0 }} = e^{- L^d  I_t [ m_e ;  \hat \phi^R(.) ; \hat \phi^I(.)] } 
  \label{ProbaPhiklargedev}
\end{eqnarray}
where the rate function 
\begin{eqnarray} 
  I_t [ m_e ; \hat \phi^R(.) ; \hat \phi^I(.)]  =  {\hat E}_t [ m_e ; \hat \phi^R(.) ; \hat \phi^I(.)] - {\hat E}_0
  \label{RatePhi}
\end{eqnarray}
in directly related to the intensive energy ${\hat E} [ m_e ; \hat \phi^R(.) ; \hat \phi^I(.)] $
as a function of the intensive Fourier coefficients $ [ m_e ; \hat \phi^R(.) ; \hat \phi^I(.)] $ 
up to the constant ${\hat E}_0 $.

Plugging the large deviation behavior of Eq. \ref{ProbaPhiklargedev} 
into Eq. \ref{pmalone} for $t=0$
yields
\begin{eqnarray} 
p_L(m_e) && \opsimeq_{L \to + \infty}    
\left( \prod_{\vec k >0}  \left[ \int_{-\infty}^{+\infty} d \hat \phi^R(\vec k)
\int_{-\infty}^{+\infty} d \hat \phi^I(\vec k) \right] \right)
e^{- L^d  I_0 [ m_e ;  \hat \phi^R(.) ; \hat \phi^I(.)] }  
\nonumber \\
&&  \opsimeq_{L \to + \infty} e^{- L^d  i(m_e) }
  \label{pmcontractionFourier}
\end{eqnarray}
so that the rate function $i(m_e) $ introduced in Eq. \ref{rateim}
corresponds to the optimization of the joint rate function $I_0 [ m_e ;  \hat \phi^R(.) ; \hat \phi^I(.)]$ over all the other Fourier coefficients $ \hat \phi_R(\vec k) ; \hat\phi_I(\vec k)$ associated to $\vec k>0$
\begin{eqnarray} 
  i(m_e) = \min_{ \{\hat \phi_R(.) ; \hat\phi_I(.)\}} \left( I_0 [ m_e ;  \hat \phi_R(.) ; \hat\phi_I(.)] \right)
  =  \min_{ \{\hat \phi_R(.) ; \hat\phi_I(.)\}} \left( {\hat E}_t [ m_e ; \hat \phi^R(.) ; \hat \phi^I(.)] \right) 
   - {\hat E}_0 
  \label{imcontraction}
\end{eqnarray}

As in Eq. \ref{hatZlargeL} for $ \hat {\cal Z}_0 $, the generalized partition function $\hat {\cal Z}_t[ m_e ; {\hat h}^R (.); {\hat h}^I(.)] $ of Eq. \ref{Zhk}
can be evaluated for large $L$
via the saddle-point method for the integrals 
over the Fourier coefficients $[ m_e ; {\hat \phi}^R  (\vec k);{\hat \phi}^I  (\vec k) ] $
in order to obtain 
\begin{eqnarray} 
 e^{{\hat {\cal W}_t}[h_e ;  {\hat h}^R (.); {\hat h}^I(.)]} = \frac{ \hat {\cal Z}_t[ h_e ; {\hat h}^R (.); {\hat h}^I(.)] }{\hat {\cal Z}_0}
&& =  \frac{\int D {\hat \phi} \  e^{ \displaystyle L^d \left[ - E_t [ m_e ; \hat \phi^R(.) ; \hat \phi^I(.)]  + h_e m_e
 + 2 \sum_{\vec k >0} \left(  {\hat h}^R  (  \vec k)  {\hat \phi}^R ( \vec k) + {\hat h}^I  (  \vec k)  {\hat \phi}^I ( \vec k)  \right) \right]}  
 }{ \int D {\hat \phi}  e^{- L^d  E_t [ m_e ;  \hat \phi^R(.) ; \hat \phi^I(.)] } }
\nonumber \\
&& \opsimeq_{L \to + \infty} e^{ \displaystyle L^d {\hat W}[ {\hat h}^R (.); {\hat h}^I(.)]}
  \label{saddle}
\end{eqnarray}
where the intensive part ${\hat W}_t[ h_e ; {\hat h}^R (.); {\hat h}^I(.)] $ of ${\hat {\cal W}_t}[h_e ;  {\hat h}^R (.); {\hat h}^I(.)] $
is the Legendre transform of the rate function $I_t [ m_e ;\hat \phi^R(.) ; \hat \phi^I(.)]  
= \big(E_t [ m_e ; \hat \phi^R(.) ; \hat \phi^I(.)] - E_0\big )$ introduced in Eq. \ref{RatePhi}
\begin{eqnarray} 
{\hat W}_t [ h_e ; {\hat h}^R (.); {\hat h}^I(.)] = \max_{\{ m_e ; \hat \phi^R(.) ; \hat \phi^I(.)\}} \bigg(  - I_t [ m_e ; \hat \phi^R(.) ; \hat \phi^I(.)]  
+h_e m_e
 + 2 \sum_{\vec k >0} \left(  {\hat h}^R  (  \vec k)  {\hat \phi}^R ( \vec k) + {\hat h}^I  (  \vec k)  {\hat \phi}^I ( \vec k)  \right)   \bigg)
  \label{Legendremax}
\end{eqnarray}
i.e. more explicitly
\begin{eqnarray} 
{\hat W}_t [ h_e ; {\hat h}^R (.); {\hat h}^I(.)]&& =  - I _t [ m_e ; \hat \phi^R(.) ; \hat \phi^I(.)]  
+ h_e m_e
 + 2 \sum_{\vec k >0} \left(  {\hat h}^R  (  \vec k)  {\hat \phi}^R ( \vec k) + {\hat h}^I  (  \vec k)  {\hat \phi}^I ( \vec k)  \right)   
 \nonumber \\
 h_e && = \frac{ \partial I_t [ m_e ; \hat \phi^R(.) ; \hat \phi^I(.)] }{\partial m_e }
\nonumber \\
2 {\hat h}^R  (  \vec k) && = \frac{ \partial I_t [ m_e ; \hat \phi^R(.) ; \hat \phi^I(.)] }{\partial{\hat \phi}^R  ( \vec k)} 
\nonumber \\
2 {\hat h}^I  (  \vec k) && = \frac{ \partial I_t [ m_e ; \hat \phi^R(.) ; \hat \phi^I(.)] }{\partial{\hat \phi}^I  ( \vec k)} 
  \label{Legendre}
\end{eqnarray}
while the reciprocal Legendre transform reads
\begin{eqnarray} 
I _t [ m_e ; \hat \phi^R(.) ; \hat \phi^I(.)]  && =  - {\hat W}_t [ h_e ; {\hat h}^R (.); {\hat h}^I(.)]
+ h_e m_e
 + 2 \sum_{\vec k >0} \left(  {\hat h}^R  (  \vec k)  {\hat \phi}^R ( \vec k) + {\hat h}^I  (  \vec k)  {\hat \phi}^I ( \vec k)  \right)   
 \nonumber \\
 m_e && = \frac{ \partial {\hat W}_t [ h_e ; {\hat h}^R (.); {\hat h}^I(.)] }{ \partial h_e}
\nonumber \\
2 {\hat \phi}^R  (  \vec k) && = \frac{ \partial {\hat W}_t [ h_e ;{\hat h}^R (.); {\hat h}^I(.)] }{\partial{\hat h}^R  ( \vec k)} 
\nonumber \\
2 {\hat \phi}^I  (  \vec k) && = \frac{ \partial {\hat W}_t [ h_e ; {\hat h}^R (.); {\hat h}^I(.)] }{\partial{\hat h}^I  ( \vec k)} 
  \label{LegendreReci}
\end{eqnarray}

These multidimensional Legendre transformations generalize
the Legendre transformations written in Eqs \ref{Legendremalone} and \ref{Legendremalonereci}
for the single variables $m_e $ and $h_e=h$ alone.


\subsection{ Discussion }

To obtain the probability distribution $p_L(m_e,t=0)$ of the empirical magnetization $m_e$ of the initial model at $t=0$
via Eq. \ref{pmalone}, one needs to integrate
over all the Fourier coefficients
$[ \hat \phi^R(\vec k) ; \hat\phi^I(\vec k)]$ associated to $\vec k>0$ 
but this integration all at once is of course complicated.
So one would like to define some differential RG procedure
 in order to perform this integration more gradually
as a function of some RG-time $t$ while keeping completely frozen the empirical magnetization $m_e$
\begin{eqnarray} 
\text { Goal : } p_L(m_e,t) && =  
\left( \prod_{\vec k >0}  \left[ \int_{-\infty}^{+\infty} d \hat \phi^R(\vec k)
\int_{-\infty}^{+\infty} d \hat \phi^I(\vec k) \right] \right)
 P_t[m_e;  \hat \phi^R(.) ; \hat\phi^I(.)] 
 \nonumber \\
 && = p_L(m_e,t=0) \ \ \text{independent of the RG time $t$}\ \ \ 
  \label{pmalonetfrozen}
\end{eqnarray}
i.e. the RG procedure will only involve the Fourier coefficients $ [\hat \phi^R(\vec k) ; \hat\phi^I(\vec k)]$
 associated to non-zero momenta $\vec k>0$.
 In the three following sections, we discuss what type of differential RG procedures could be appropriate
 before returning to this concrete goal in section \ref{sec_RGVolume}.


\section{ Wilson-Kogut exact differential RG based on 'incomplete integration' }

\label{sec_WK}

In this section, we revisit the Wilson exact differential RG called 'incomplete integration'
as presented in section 11 of the Wilson-Kogut review \cite{WilsonKogut}
 starting with their pedagogical case involving a single variable $\phi$.

\subsection{ Wilson-Kogut differential RG explained for the case of a single variable $\phi$ }

\subsubsection{ Toy model involving a single variable $\phi$ }

\label{subsubsec_single}

To better understand the meaning and the properties of any functional RG,
it is useful, whenever possible, to start with the simpler example of single variable $\phi $, 
where the probability distribution of Eq. \ref{ProbaPhiS}
reduces to 
\begin{eqnarray} 
  P_t (\phi )  = \frac{ e^{- E_t(\phi)} }{  Z_0 }
  \label{p1}
\end{eqnarray}
while the conserved partition function of Eq. \ref{Z} becomes the ordinary one-dimensional integral
\begin{eqnarray} 
  Z_t  =  \int_{-\infty}^{+\infty} d\phi e^{- E_t(\phi)  } = Z_0
  \label{Z1}
\end{eqnarray}
The corresponding generating function of Eq. \ref{Zh} 
\begin{eqnarray} 
 \frac{  Z_t(h) }{Z_0} = \frac{  1 }{Z_0} \int_{-\infty}^{+\infty} d\phi e^{- E_t(\phi) +h \phi } 
= \sum_{n=0}^{+\infty} \frac{h^n}{n!} \int_{-\infty}^{+\infty} d\phi   P_t (\phi ) \phi^n \equiv  \sum_{n=0}^{+\infty} \frac{h^n}{n!} c_n(t)
  \label{Z1h}
\end{eqnarray}
contains the information on the moments $c_t(n)$ of arbitrary order $n$
(instead of the correlations in Eq. \ref{Zh}),
while the generating function of Eq. \ref{Wh} 
\begin{eqnarray} 
W_t(h) = \ln \left( \frac{  Z_t(h) }{Z_0} \right) = \sum_{n=0}^{+\infty} \frac{h^n}{n!} w_n(t)
  \label{W1h}
\end{eqnarray}
contains the information on the cumulants $w_n(t)$ of arbitrary order $n$
(instead of the connected correlations in Eq. \ref{Wh}).

Since there is no space, the energy $ E_t(\phi)$ coincides with the local potential $U_t[\phi]$ 
with its series expansion of Eq. \ref{Vlocalt}  
\begin{eqnarray} 
E_t(\phi) = U_t[ \phi] && = \sum_{n=0}^{+\infty} \frac{u_{2n}(t)}{(2n)!}  \phi^{2n}  
  = u_0(t) + \frac{u_{2}(t)}{2}  \phi^{2}  + \frac{u_4(t)}{4!}  \phi^{4}  +... 
  \label{Vlocalt1}
\end{eqnarray}


\subsubsection{ The notion of 'incomplete integration' via the Fokker-Planck Equation associated to the Ornstein-Uhlenbeck process}

\label{subsec_WK}

In section 11 of the review \cite{WilsonKogut},
the exact differential RG called 'incomplete integration'
is already explained on the simple case of a single variable around their equations 11.1 to 11.5:
in the present subsection, we rephrase their presentation
in terms of notations and notions from the area of stochastic processes
that will play a major role in the remainder of the present paper.

The 'trick of incomplete integration' is introduced 
via their differential equation (11.1) for a function $\psi(.,.) $ :
for us this function will be the unnormalized-version $\rho_t(\phi) $ 
of the probability density $P_t (\phi ) $ of Eq. \ref{p1}
\begin{eqnarray} 
  \rho_t(\phi) \equiv e^{-t E(\phi)}
  \label{rhoEt}
\end{eqnarray}
while we will supplement their equation (11.1)
with the two strictly positive parameters $\omega>0$ and $D>0$ to obtain
\begin{eqnarray} 
\frac{\partial \rho_t (\phi )}{\partial t}  
 = \frac{\partial}{\partial \phi} \left( \omega \phi +D \frac{\partial}{\partial \phi} \right) \rho_t (\phi ) 
 =  \frac{\partial}{\partial \phi} \bigg( \omega \phi \rho_t (\phi )  \bigg) 
 + D \frac{\partial^2 \rho_t (\phi )}{\partial \phi^2}  
  \label{p1FPOU}
\end{eqnarray}
where statistical physicists have the pleasure to recognize the Fokker-Planck Equation associated to the one-dimensional Ornstein-Uhlenbeck process, with the drift $(- \omega \phi )$ and with the diffusion coefficient $D$.
Their solution of Eq. (11.2)  for $\rho_t (\phi ) $ in terms of the initial condition $\rho_0 (\phi_0 ) $ at $t=0$
\begin{eqnarray} 
  \rho_t (\phi )  = \int_{-\infty}^{+\infty} d \phi_0 P(\phi,t \vert \phi_0,0)  \rho_0 (\phi_0 ) 
  \label{p1FPOUsol}
\end{eqnarray}
involves the well-known Gaussian Ornstein-Uhlenbeck 
propagator (called the Green function $G$ in their Eq. 11.3)
\begin{eqnarray} 
P(\phi,t \vert \phi_0,0)
 =\sqrt{ \frac{ \omega}{ 2 \pi D \left( 1- e^{- 2t \omega}\right) }  }
\ \ \  e^{ \displaystyle  -   \frac{\omega \left(  \phi -  e^{-  \omega t } \phi_0 \right)^2}{ 2 D \left( 1- e^{- 2t \omega}\right)} }
  \label{PropagatorOU}
\end{eqnarray}
satisfying the normalization
\begin{eqnarray} 
\int_{-\infty}^{+\infty} d \phi P(\phi,t \vert \phi_0,0) =1
  \label{PropagatorOUnorma}
\end{eqnarray}
The averaged value of $\phi$ at time $t$
\begin{eqnarray} 
 \int_{-\infty}^{+\infty} d \phi \phi P(\phi,t \vert \phi_0,0) = e^{-  \omega t } \phi_0
  \label{PropagatorOUav}
\end{eqnarray}
starts at the initial value $\phi_0$ and decays towards zero 
with the exponential decay $ e^{-  \omega t }$ governed by the parameter $\omega$.
The variance of $\phi$ at time $t$
\begin{eqnarray} 
 \int_{-\infty}^{+\infty} d \phi \bigg( \phi - e^{-  \omega t } \phi_0\bigg)^2 P(\phi,t \vert \phi_0,0) = 
 \frac{  D \left( 1- e^{- 2t \omega}\right)}{\omega}
  \label{PropagatorOUvar}
\end{eqnarray}
starts at the vanishing initial value and grows towards the asymptotic value given by the ratio $\frac{D}{\omega}$ involving the diffusion coefficient $D$ in the numerator.

The notion of 'incomplete integration' can be understood by considering the two following limits for the time $t$:

(i) for $t \to 0$, the propagator $P(\phi,t \vert \phi_0,0) $ reduces to the delta function $\delta(\phi-\phi_0) $
as it should to recover the initial condition $\rho_t(\phi) \to \rho_0(\phi)$ (their Eq 11.4)
\begin{eqnarray} 
P(\phi,t \vert \phi_0,0) && \opsimeq_{t \to 0} \delta(\phi-\phi_0)
\nonumber \\
\rho_t(\phi) && \opsimeq_{t \to 0} \rho_0(\phi)
  \label{PropagatorOUtzero}
\end{eqnarray}

(ii) for $t \to + \infty$, the propagator $P(\phi,t \vert \phi_0,0) $ converges 
towards the following Gaussian steady state for $\phi$ while the initial condition $\phi_0$ completely disappears
\begin{eqnarray} 
P(\phi,t \vert \phi_0,0)
 \opsimeq_{t \to + \infty} \sqrt{ \frac{ \omega}{ 2 \pi D  }  }
\ \ \  e^{ \displaystyle  -   \frac{\omega}{ 2 D } \phi^2} \equiv P_*(\phi)
  \label{PropagatorOUsteady}
\end{eqnarray}
 so that the asymptotic solution $ \rho_t (\phi )$ of Eq. \ref{p1FPOUsol} for $t \to + \infty$ (their Eq 11.5)
\begin{eqnarray} 
  \rho_t (\phi )  \opsimeq_{t \to + \infty} P_*(\phi)  \int_{-\infty}^{+\infty} d \phi_0  \rho_0 (\phi_0 ) = P_*(\phi) Z_0
  \label{p1FPOUsolinfty}
\end{eqnarray}
involves the initial partition function $Z_0$
 besides the normalized Gaussian steady distribution $P_*(\phi) $,
i.e. the only surviving information concerning the initial condition $\rho_0 (\phi_0 ) $ 
is its partition function $Z_0$, while all the other properties of the initial condition $\rho_0 (\phi_0 ) $
have been forgotten and have been replaced by the Gaussian steady distribution $P_*(\phi) $.

In conclusion, the solution $ \rho_t (\phi )$ of Eq. \ref{p1FPOUsol} 
interpolates as a function of the RG-time $t$
between the 'completely unintegrated' function $\rho_0 (\phi ) $ at time $t=0$ (Eq. \ref{PropagatorOUtzero})
and the 'fully integrated' function $Z_0 =\int_{-\infty}^{+\infty} d \phi_0  \rho_0 (\phi_0 ) $ appearing in Eq. \ref{p1FPOUsolinfty} for $t \to + \infty$, hence the notion of 'incomplete integration' for any finite $t$. 
Let us mention that the partition function $Z_t$ of Eq. \ref{Z1}
 is conserved in time as a consequence of Eqs \ref{p1FPOUsol}
and \ref{PropagatorOUnorma}
\begin{eqnarray} 
  Z_t  =  \int_{-\infty}^{+\infty} d\phi e^{- E_t(\phi)  } = \int_{-\infty}^{+\infty} d\phi \rho_t (\phi )
  = \int_{-\infty}^{+\infty} d\phi \int_{-\infty}^{+\infty} d \phi_0 P(\phi,t \vert \phi_0,0)  \rho_0 (\phi_0 ) 
  = \int_{-\infty}^{+\infty} d \phi_0  \rho_0 (\phi_0 ) 
  = Z_0
  \label{Z1WK}
\end{eqnarray}


\subsubsection{  RG-flow for the energy $E_t(\phi) $ containing all the couplings $u_n(t)$ }

The linear Fokker-Planck evolution of Eq. \ref{p1FPOU}
for $ \rho_t(\phi)=e^{- E_t(\phi)}$
translates into the following non-linear RG-flow for the energy $E_t(\phi) $ 
\begin{eqnarray} 
\frac{\partial E_t (\phi )}{\partial t}   = 
D \left[ \frac{\partial^2 E_t (\phi )}{\partial \phi^2 } 
 - \left( \frac{\partial E_t (\phi )}{\partial \phi}\right)^2 
\right] 
+ \omega \left[ \phi \frac{\partial E_t (\phi )}{\partial \phi} -1 \right] 
  \label{WKEnergy}
\end{eqnarray}
where one can plug the series expansion in even powers of $\phi$ of Eq. \ref{Vlocalt}
if one wishes to obtain the corresponding flows for the couplings $u_{2n}(t)$.

Note that the energy $E_t[\phi(.)]$ is called the Wilson action $S_t[\phi]$ in most field theory papers 
but is denoted by $\big(-H_t [.]\big)$ in the section 11 of \cite{WilsonKogut}
that explains the sign difference between Eq. \ref{WKEnergy} and their Eq. 11.8.

The finite-time solution of Eq. \ref{p1FPOUsol} with the propagator of \ref{PropagatorOU}
reads in terms of the energies $E_t(\phi) $ and $E_0(\phi_0)$
\begin{eqnarray} 
 e^{- E_t(\phi)}  = \int_{-\infty}^{+\infty} d \phi_0
 \sqrt{ \frac{ \omega}{ 2 \pi D \left( 1- e^{- 2t \omega}\right) }  }
\ \ \  e^{ \displaystyle  -   \frac{\omega \left(  \phi -  e^{-  \omega t } \phi_0 \right)^2}{ 2 D \left( 1- e^{- 2t \omega}\right)} - E_0(\phi_0)}
  \label{p1FPOUsolE}
\end{eqnarray}


\subsubsection{ Corresponding RG-flow for the generating function $ Z_t(h) $ of the moments $c_n(t)$}

In section 11 of the review \cite{WilsonKogut}, the corresponding generating function of Eq. \ref{Z1h}
is also discussed around Eq. (11.23).
Plugging the solution $\rho_t (\phi ) $ of Eq. \ref{p1FPOUsol}
with the Gaussian propagator of Eq. \ref{PropagatorOU}
into Eq. \ref{Z1h} yields the generating function $Z_t(h)$ at time $t$ 
\begin{eqnarray} 
  Z_t(h) && =  \int_{-\infty}^{+\infty} d\phi e^{- E_t(\phi) +h \phi } = 
  \int_{-\infty}^{+\infty} d\phi  \rho_t (\phi ) e^{ h \phi }
  =  \int_{-\infty}^{+\infty} d\phi   e^{ h \phi } \int_{-\infty}^{+\infty} d \phi_0 P(\phi,t \vert \phi_0,0)  \rho_0 (\phi_0 ) 
  \nonumber \\
  && =\int_{-\infty}^{+\infty} d \phi_0 \rho_0 (\phi_0 ) 
   \int_{-\infty}^{+\infty} d\phi   e^{ h \phi }   P(\phi,t \vert \phi_0,0)  
   \nonumber \\
  && =\int_{-\infty}^{+\infty} d \phi_0 \rho_0 (\phi_0 ) 
     e^{ h e^{-  \omega t } \phi_0 + h^2 \frac{  D \left( 1- e^{- 2t \omega}\right)}{ 2 \omega}}   
     = e^{  h^2 \frac{  D }{2 \omega} \left( 1- e^{- 2t \omega}\right)}   Z_0 \left( h_0 = h e^{-  \omega t } \right)
  \label{Z1hWK}
\end{eqnarray}
in terms of the initial generating function $Z_0(h_0) $. 
As a function of the RG-time $t$, the generating function $Z_t(h)$
interpolates between its initial value $Z_0(h_0=h) $ at time $t=0$
and the asymptotic value
\begin{eqnarray} 
  Z_t(h) \opsimeq_{t \to + \infty}  e^{  h^2 \frac{  D }{2 \omega} }   Z_0 \left( h_0 = 0 \right) =  e^{  h^2 \frac{  D }{2 \omega} }  Z_0
  \label{Z1hWKtlarge}
\end{eqnarray}
associated to the asymptotic solution of Eq. \ref{p1FPOUsolinfty}.

The inversion of Eq. \ref{Z1hWK}
 corresponds to Eq. 11.25 of the review \cite{WilsonKogut} within our present notations
\begin{eqnarray} 
  Z_0(h_0)  =   e^{ - h_0^2 \frac{  D }{2 \omega} \left(  e^{ 2t \omega}-1 \right)}  Z_t(h= h_0 e^{  \omega t })
  \label{Z1h0WK}
\end{eqnarray}

Besides this finite-time correspondence, it is useful to write the differential RG-flow
as obtained from the Fokker-Planck Eq. \ref{p1FPOU} via integration by parts
\begin{eqnarray} 
 \frac{\partial Z_t(h) }{\partial t}    && =  \int_{-\infty}^{+\infty} d\phi e^{h \phi } 
\   \frac{\partial \rho_t (\phi ) }{\partial t} 
 =  \int_{-\infty}^{+\infty} d\phi e^{h \phi } \ \frac{\partial}{\partial \phi} 
 \left( \omega \phi \rho_t (\phi )+D \frac{\partial \rho_t (\phi )}{\partial \phi} \right)  
\nonumber \\
&&  =  -  h  \int_{-\infty}^{+\infty} d\phi e^{h \phi } 
 \left( \omega \phi \rho_t (\phi )+D \frac{\partial \rho_t (\phi )}{\partial \phi} \right) 
 \nonumber \\
&&  =  - \omega h  \frac{\partial}{\partial h}  \int_{-\infty}^{+\infty} d\phi e^{h \phi }  \rho_t (\phi )   
  + D  h^2  \int_{-\infty}^{+\infty} d\phi e^{h \phi }   \rho_t (\phi )
   \nonumber \\
&& =  - \omega h  \frac{\partial  Z_t(h)}{\partial h}     + D  h^2  Z_t(h)
  \label{p1FPOUZ}
\end{eqnarray}


\subsubsection{ Corresponding RG-flow for the generating function $W_t(h)$ of the cumulants $w_n(t)$}

The corresponding RG-flow for $W_t(h)= \ln \left( \frac{Z_t(h)}{Z_0}\right)$ reads
\begin{eqnarray} 
 \frac{\partial W_t(h) }{\partial t}    && = \frac{ \frac{\partial Z_t(h) }{\partial t} }{Z_t(h)}
 =  - \omega h  \frac{\partial  W_t(h)}{\partial h}     + D  h^2  
  \label{p1FPOUW}
\end{eqnarray}
while the finite-time relation of Eq. \ref{Z1hWK}
translates into
\begin{eqnarray} 
  W_t(h) && =   h^2 \frac{  D }{2 \omega} \left( 1- e^{- 2t \omega}\right)
  + W_0 \left( h_0 = h e^{-  \omega t } \right)
  \label{W1hWK}
\end{eqnarray}
where one can plug the series expansion of Eq. \ref{W1h} to obtain that the dynamics
of the cumulants is very simple
\begin{eqnarray} 
w_n(t) = e^{- n \omega t } w_n(0) + \delta_{n,2} \frac{  D }{ \omega} \left( 1- e^{- 2t \omega}\right)
  \label{W1hcumulants}
\end{eqnarray}
All the cumulants of order $n \ne 0$ converge exponentially towards zero as $e^{- n \omega t }  $,
while the second cumulant $n=2$ contains the supplementary second term associated to the 
convergence towards the Gaussian Ornstein-Uhlenbeck steady state of Eq. \ref{PropagatorOUsteady}.


\subsubsection{ Corresponding RG-flow for the Legendre transform $ \Gamma_t(\Phi) $ of $W_t(h)$}

For the two single-variable-functions 
$ \Gamma_t(\Phi) $ and $W_t(h)$,
the Legendre transformations of Eqs \ref{GammaWh} and \ref{PhivariableLegendreWdef}
reduce to
\begin{eqnarray} 
 \Gamma_t( \Phi) && =  \max_{ h } \bigg(  h \Phi - W_t(h) \bigg) = \Phi h_t(\Phi) - W_t(h_t(\Phi))
\nonumber \\ 
 \Phi && = \frac{\partial W_t (h) }{ \partial h } \bigg\vert_{h=h_t(\Phi)}
 \nonumber \\ 
 h_t(\Phi) && = \frac{\partial  \Gamma_t( \Phi) }{ \partial \Phi } 
    \label{PhivariableLegendreWdef1}
\end{eqnarray}

As a consequence, the RG-flow for $\Gamma_t( \Phi) $ can be obtained from the RG-flow of Eq. \ref{p1FPOUW}
concerning $W_t(h)$
\begin{eqnarray} 
\frac{\partial \Gamma_t( \Phi) }{\partial t} && =  -  \frac{\partial W_t(h) }{\partial t}  \bigg\vert_{h=h_t(\Phi)}
+ [\partial_t h_t(\Phi) ] \left( \Phi  - \frac{\partial W_t (h) }{ \partial h } \bigg\vert_{h=h_t(\Phi)} \right)
=  -  \frac{\partial W_t(h) }{\partial t}  \bigg\vert_{h=h_t(\Phi)}
\nonumber \\ 
  && = \left(   \omega h  \frac{\partial  W_t(h)}{\partial h}     - D  h^2 \right) \bigg\vert_{h=h_t(\Phi)}
  = \omega   \Phi  h_t(\Phi)    - D   h_t^2(\Phi)
  \nonumber \\ 
  && = \omega   \Phi \frac{\partial  \Gamma_t( \Phi) }{ \partial \Phi }     - D   \left(\frac{\partial  \Gamma_t( \Phi) }{ \partial \Phi }  \right)^2
    \label{RGFlowGamma1}
\end{eqnarray}

Let us now translate the finite-time relation of Eq. \ref{W1hWK}
concerning $W_t(h) $ for its Legendre transform
\begin{eqnarray} 
\Gamma_t( \Phi) && =  \max_{ h } \bigg[  h \Phi - W_t(h) \bigg]
= \max_{ h } \bigg[  h \Phi - h^2 \frac{  D }{2 \omega} \left( 1- e^{- 2t \omega}\right) 
  - W_0 \left( h_0 = h e^{-  \omega t } \right) \bigg]
  \nonumber \\
&& 
= \max_{ h_0 } \bigg(  h_0 e^{  \omega t } \Phi - h_0^2 e^{2  \omega t }\frac{  D }{2 \omega} \left[ 1- e^{- 2t \omega}\right]    - W_0 ( h_0  ) \bigg)    
  \label{GammatW0}
\end{eqnarray}
where one can plug the definition of $W_0 (h_0) $ in terms of $\Gamma_0(\Phi_0) $
\begin{eqnarray} 
    - W_0 (h_0) =  \min_{ \Phi_0 } \bigg[ - h_0 \Phi_0 + \Gamma_0(\Phi_0) \bigg]
  \label{W0min}
\end{eqnarray}
to obtain
\begin{eqnarray} 
\Gamma_t( \Phi) && 
=\min_{ \Phi_0 } \bigg[ \max_{ h_0 } \bigg(  h_0 e^{  \omega t } \Phi - h_0^2 e^{2  \omega t }\frac{  D }{2 \omega} \left( 1- e^{- 2t \omega}\right)  - h_0 \Phi_0 + \Gamma_0(\Phi_0)   \bigg)    \bigg]
  \nonumber \\
&& =\min_{ \Phi_0 } \bigg[\Gamma_0(\Phi_0)
+  \max_{ h_0 } \bigg(  h_0 (e^{  \omega t } \Phi -\Phi_0) - h_0^2 e^{2  \omega t }\frac{  D }{2 \omega} \left( 1- e^{- 2t \omega}\right)      \bigg)    \bigg]
  \label{Gammatminmax}
\end{eqnarray}
The maximization over $h_0$ of the remaining quadratic function in $h_0$ leads to the optimal value
\begin{eqnarray} 
h_0^{opt}  = \frac{ \left( e^{  \omega t } \Phi - \Phi_0 \right) \omega}{  D  \left( e^{2  \omega t }-1\right) }
      \label{minh0}
\end{eqnarray}
that can be plugged into Eq. \ref{Gammatminmax} to obtain the final result for $\Gamma_t( \Phi)  $
in terms of $\Gamma_{t=0}( \Phi_0) $
\begin{eqnarray} 
\Gamma_t( \Phi)  = 
 \min_{ \Phi_0 } \bigg( \Gamma_0( \Phi_0) + \frac{ \left(  \Phi - \Phi_0 e^{ - \omega t }\right)^2 \omega}
{  2 D  \left( 1- e^{- 2  \omega t }\right)}
 \bigg)
  \label{Gamma1hWK}
\end{eqnarray}


\subsubsection{ Discussion }

\label{subsec_1dflows}

In summary, this single-variable toy-model is very useful to understand the idea of 'incomplete integration,
and to see the relations between the corresponding RG-flows that appear
for the various observables, namely: 

(i) the linear RG-flow of Eq. \ref{p1FPOU} for the unnormalized density $\rho_t(\phi)= e^{- E_t(\phi)}$,
of Eq. \ref{p1FPOUZ} for the generalized partition function $Z_t(h) $ 
and of Eq. \ref{p1FPOUW} for the generating function $W_t(h)= \ln \left( \frac{Z_t(h)}{Z_0}\right)$

(ii) the non-linear RG-flows of Eq. \ref{WKEnergy} for the energy $E_t(\phi)$ (i.e. the Wilson action),
and of Eq. \ref{RGFlowGamma1} for the Legendre transform $\Gamma_t( \Phi) $.

We have also stressed how the explicit Fokker-Planck solution of
Eq. \ref{p1FPOUsol} with the Ornstein-Uhlenbeck propagator of Eq. \ref{PropagatorOU}
gives explicit expressions for all these observables at time $t$ in terms of their counterparts at time $t=0$.


\subsection{ Wilson-Kogut RG scheme for field theory in dimension $d$ }

In section 11 of the review \cite{WilsonKogut}, the previous ideas concerning the single variable $\phi$
are then generalized to the Fourier modes of the field theory :
the total propagator of their Eq. 11.7 corresponds to the product of individual Ornstein-Uhlenbeck
propagators of Eq. \ref{PropagatorOUsteady} involving their single function $\alpha_t(\vec q)$
that can be interpreted as a momentum-dependent-effective-time for the 
Fokker-Planck dynamics of the Fourier mode (their Eq 11.6 is even written with the partial derivative with respect to $\alpha_t(\vec q) $ on the left hand-side instead of the time-derivative).
In our notations of Eq. \ref{PropagatorOUsteady}, this choice corresponds to the equality 
of the two parameters $\omega_t(\vec q) $ and $D_t(\vec q) $
\begin{eqnarray} 
\omega_t(\vec q) = D_t(\vec q) = \frac{\partial \alpha_t(\vec q)}{\partial t}
  \label{WKDq}
\end{eqnarray}
while we will see examples of other choices in further sections and in Appendix \ref{app_timeDepOU}.

Forgetting about the rescaling part of their RG that will not be discussed in the present paper
where our simpler goal is explained around Eq. \ref{pmalonetfrozen},
the choice of their Eq. 11.9 reads
\begin{eqnarray} 
 \alpha_t(\vec q) = (e^{2t}-1) \vec q^{\ 2}
  \label{alphachoice}
\end{eqnarray}
and thus corresponds to
\begin{eqnarray} 
\omega_t(\vec q) = D_t(\vec q) = \frac{\partial \alpha_t(\vec q)}{\partial t} = 2 e^{2t} \vec q^{\ 2} = 2 \left( \frac{\vec q}{ e^{-t} }\right)^2
  \label{WKDqchoice}
\end{eqnarray}
with two important properties :

(i) the growth of $\omega_t(\vec q) $ with respect to the momentum $\vec q$ as $ \vec q^{\ 2}$
means that the exponential decays $e^{-t \omega_t(\vec q) }$ that govern
the dynamics of the averaged values of the Fourier modes 
(as in Eq. \ref{PropagatorOUav}) will damp much more strongly the higher Fourier modes.

(ii) the scaling variable $\left( \frac{\vec q}{ e^{-t} }\right) $ that appears in Eq. \ref{WKDqchoice}
corresponds
to the standard field-theory parametrization of the characteristic
RG-momentum-scale $\Lambda_t=\Lambda_0e^{-t}$ 
in terms of the effective RG-time $t=\ln \frac{\Lambda_0}{\Lambda_t} $ satisfying $dt = - \frac{d\Lambda}{\Lambda} $. In the present paper, we will prefer to remain more flexible in the choice of the RG-time $t$
that parametrizes the RG-flow, since this field-theory standard choice might not be the most natural from the point of view of stochastic processes, and we will indeed discuss other choices in further sections and in Appendix \ref{app_timeDepOU}.


\subsection{ Discussion }

In summary, the Wilson-Kogut RG-scheme does the desired job, 
but contains a certain number of choices that may appear somewhat arbitrary,
for instance the choice of parameters in Eqs \ref{WKDq} and \ref{WKDqchoice} discussed above,
but most importantly, the choice of the Fokker-Planck Eq. \ref{p1FPOU} itself
at the very beginning, i.e. their equation (11.1) accompanied by the minimal technical explanation
'Recall a few facts about differential equations'. 
To better understand what other RG-flows can be chosen,
it is thus important to turn to the Wegner-Morris perspective
discussed in the next section.


\section{ Wegner-Morris RG as an infinitesimal change of variables 
conserving the partition function}

\label{sec_WM}

In this section, we discuss the perspective put forward
by Wegner \cite{Wegner74} and by Morris \cite{Morris1993,Morris2000,Morris2001,Morris2002},
where the possible RG-flows compatible with the conservation of
the partition function are reformulated in terms of the possible field-redefinitions
inside the partition function (see also the discussions in \cite{Caticha} 
and in the reviews \cite{Osborn,Rosten}).


\subsection{ Wegner-Morris differential RG explained for the case of a single variable $\phi$ }

As in the previous section, it is useful to start with the simple case of the single variable $\phi$ of Eq. \ref{p1}.

\subsubsection{ Wegner-Morris continuity Equation for the unnormalized density $\rho_t(\phi) = e^{- E_t(\phi)  } $}

The conservation of the partition function $Z_t$ of Eq. \ref{Z1}
\begin{eqnarray} 
  Z_t  =  \int_{-\infty}^{+\infty} d\phi e^{- E_t(\phi)  } =  \int_{-\infty}^{+\infty} d\phi \rho_t(\phi)  
  = Z_0 \ \ \ \text { for any time $t$}
  \label{Z1conserved}
\end{eqnarray}
can be analyzed between $t$ and $(t+dt)$ via an infinitesimal change of variables 
between $\phi=\phi_t$ and $ {\tilde \phi }=\phi_{t+dt}$
\begin{eqnarray} 
 {\tilde \phi } && = \phi + dt v_t(\phi)
  \label{infinitesimalchangevariables}
\end{eqnarray}
with the corresponding infinitesimal change of measure
\begin{eqnarray} 
 \frac{ d  {\tilde \phi }}{d \phi} && = 1+ dt \frac{ \partial v_t(\phi) }{\partial  \phi}
  \label{infinitesimalchangevariablesjacobien}
\end{eqnarray}
to obtain at the infinitesimal order $dt$
\begin{eqnarray} 
 0&& = Z_{t+dt}  -Z_t =  \int_{-\infty}^{+\infty} d {\tilde \phi } \rho_{t+dt}({\tilde \phi } )- \int_{-\infty}^{+\infty} d\phi \rho_t(\phi) 
 \nonumber \\
 && 
  = \int_{-\infty}^{+\infty} d\phi \   \frac{ d  {\tilde \phi }}{d \phi} \ \ \rho_{t+dt}\big(\phi + dt v(\phi) \big)
    - \int_{-\infty}^{+\infty} d\phi \rho_t(\phi)  
  \nonumber \\
  &&   = \int_{-\infty}^{+\infty} d\phi \ \ \left( 1+ dt \frac{ \partial v_t(\phi) }{\partial  \phi} \right) 
  \bigg[ \rho_t(\phi) + dt \frac{ \partial \rho_t(\phi) }{\partial t} 
  + dt v_t(\phi) \frac{ \partial \rho_t(\phi) }{\partial \phi} \bigg] - \int_{-\infty}^{+\infty} d\phi \rho_t(\phi)  
 \nonumber \\
  &&   =  dt \int_{-\infty}^{+\infty} d\phi \left[\frac{ \partial \rho_t(\phi) }{\partial t} 
  +  v_t(\phi) \frac{ \partial \rho_t(\phi) }{\partial \phi} +\frac{ \partial v_t(\phi) }{\partial  \phi} \rho_t(\phi) \right]
    \label{Z1conserveddt}
\end{eqnarray}
i.e. one obtains the continuity equation for the unnormalized density $\rho_t(\phi) $
\begin{eqnarray} 
\frac{ \partial \rho_t(\phi) }{\partial t} &&  = -    \frac{ \partial  }{\partial \phi} \bigg(  v_t(\phi)  \rho_t(\phi) \bigg)
  \label{Continuity}
\end{eqnarray}
that is associated to the deterministic flow of Eq. \ref{infinitesimalchangevariables}
between $\phi=\phi_t$ and $ {\tilde \phi }=\phi_{t+dt}$
that involves the advective velocity $v_t(\phi) $ (usually denoted by $(- \Psi_t(\phi))$ in the 
Wegner-Morris papers \cite{Wegner74,Morris1993,Morris2000,Morris2001,Morris2002})
\begin{eqnarray} 
\phi_{t+dt} && = \phi_t + dt v_t(\phi_t) 
\nonumber \\
\frac{ d \phi_t}{dt} && = v_t(\phi_t) 
  \label{deterministic}
\end{eqnarray}

In conclusion, the Wegner-Morris continuity Equation \ref{Continuity}
 is useful to encompass all the differential RG-flows 
 that are compatible with the conservation of the partition function $Z_t$,
while the choice of the RG-scheme is contained in the choice of the advective velocity $v_t(\phi)$
appearing in the infinitesimal change of variables of Eq. \ref{infinitesimalchangevariables} 
rewritten as the deterministic-looking flow of Eq. \ref{deterministic}.
An essential issue is then : what are the possible choices for the advective velocity $v_t(\phi)$
in order to obtain a meaningful RG-flow ?


\subsubsection{ Discussion : choice of the velocity $v_t(\phi) $ to include some irreversibility in the RG-flow for the density $\rho_t(\phi) $}

\label{subsec_irreversibility}

Let us first return to the Wilson-Kogut RG scheme 
where the Fokker-Planck evolution of Eq. \ref{p1FPOU}
 corresponds to the Wegner-Morris continuity Eq. \ref{Continuity} where the advective velocity $v_t(\phi) $
 depends on the unnormalized density $\rho_t (\phi )=e^{- E_t (\phi )} $ via
 \begin{eqnarray} 
\text{ Wilson-Kogut RG-scheme : } \ \ \ v_t(\phi)  
&& = -  \omega \phi - D \frac{\partial \ln \rho_t (\phi )}{\partial \phi}
= -  \omega \phi + D \frac{\partial E_t (\phi )}{\partial \phi}
  \label{vWKscheme}
\end{eqnarray}
This example shows that something subtle is happening : 
the advective velocity $v_t(\phi_t) $ that governs the deterministic-looking dynamics of Eq. \ref{deterministic}
actually depends on the solution $\rho_t (\phi ) $ of the Wegner-Morris Continuity Equation itself,
so that the advective-continuity Eq. \ref{Continuity} associated to the velocity $v_t(\phi)$
actually becomes the Fokker-Planck evolution of Eq. \ref{p1FPOU} 
which has of course a completely different physical meaning : 

(i) when the velocity $v_t(\phi)$ of the advective-continuity Eq. \ref{Continuity}
is given independently of the solution $\rho_t (\phi ) $,
 then one can easily reverse the RG-time-arrow 
and consider the dynamics backward in RG-time
 via the inversion of the infinitesimal change of variables of Eq. \ref{infinitesimalchangevariables}
 between $\phi=\phi_t$ and $ {\tilde \phi }=\phi_{t+dt}$
\begin{eqnarray} 
 \phi && = {\tilde \phi } - dt v_t({\tilde \phi })
 \nonumber \\
  \frac{ d  \phi }{d {\tilde \phi }} && = 1- dt \frac{ \partial v_t({\tilde \phi }) }{\partial  {\tilde \phi }}
  \label{infinitesimalchangevariablesinfervion}
\end{eqnarray}
i.e. one can just change the sign of the velocity $v_t({\tilde \phi }) $ to follow the RG-flow backward in time.

(ii) when the velocity $v_t(\phi)$ of the advective-continuity Eq. \ref{Continuity}
depends on the solution $\rho_t (\phi ) $ itself,
then the issue of time-reversal is more subtle.
In the specific case of the Fokker-Planck evolution of Eq. \ref{p1FPOU}, 
we have stressed that the density $\rho_t (\phi )  $
follows the irreversible relaxation towards the steady state of Eq. \ref{p1FPOUsolinfty}
and that the only information surviving asymptotically is the partition function $Z_0= \int d \phi \rho_0(\phi)$,
while all the other properties of the corresponding unnormalized density $\rho_0(\phi)$ inside the partition function have been lost.

More generally, any meaningful RG-flow should contain some mechanism for irreversibility in order to 
be able to obtain that many different microscopic theories flow towards the 
same macroscopic theory at large scale, as in the example of the Central Limit theorem 
recalled at the beginning of the introduction when one considers
sums of independent identical random variables .
In this CLT example, as well as in lattice models involving discrete spins, 
the irreversibility of the RG procedures is obvious 
via the explicit reduction of the number of degrees of freedom.
For instance in the decimation procedure of the one-dimensional Ising model
where half the spins disappear at each RG step, 
it is clear what information is irreversibly lost at each step
and why this information cannot be found deterministically just by returning the RG-time-arrow.
It is actually very interesting to define Inverse-RG-transformations \cite{InverseRGMC,RGinverse_Field,Wavelet,SG,InverseRG_Review,Klinger_InverseRG,Klinger_BayesianRG,Klinger_BayesianRGFlow,InverseRG_Generative}, 
where the goal is to generate via inference the possible pre-images of a given renormalized configuration.
The conclusion of this discussion is that the RG-inverse of a genuine RG-flow cannot be deterministic,
since some information has to be lost as the RG-time grows, either via decimation or via coarse-graining. 
For continuous-time RG-flow involving continuous variables $\phi \in ]-\infty,+ \infty[$, 
the only way to include some irreversibility in the differential deterministic-looking advective flow of Eq. \ref{deterministic} 
is by an appropriate choice of the advective velocity $v_t(\phi)$
that should depend on the solution $\rho_t (\phi )$ of the continuity Equation,
as illustrated by the example of the Wilson-Kogut RG-scheme of Eq. \ref{vWKscheme}.

In conclusion, to obtain a meaningful RG-flow where the RG-time-arrow matters,
the infinitesimal change of Eq. \ref{infinitesimalchangevariables} 
between the seemingly dummy variables of integration in the partition function of Eq \ref{Z1conserveddt} 
cannot be an arbitrary change of variables without meaning, 
but should be taken very seriously as a genuine coarsening-transformation between $\phi=\phi_t$
and $ {\tilde \phi }=\phi_{t+dt}$ as will be discussed in more details in the next section \ref{sec_Carosso}.
But before let us mention how the Wegner-Morris perspective 
is formulated for field theory in dimension $d$.


\subsection{ Wegner-Morris functional continuity Equation for field theory in dimension $d$ }


For a field $\phi_t(\vec x)$ in dimension $d$, the infinitesimal change of variables of Eq. \ref{deterministic}
becomes
\begin{eqnarray} 
\phi_{t+dt}(\vec x) && = \phi_t(\vec x) + dt V_t[\vec x;\phi_t(.)] 
\nonumber \\
\partial_t \phi_t(\vec x) &&=V_t[\vec x,\phi_t(.)]
  \label{deterministicfield}
\end{eqnarray}
where the advective velocity $V_t[\vec x;\phi_t(.)]$ is now 
both a function of the position $\vec x$ and a functional of the field configuration $\phi_t(\vec y)$.
The probability ${\cal P}_t[ \phi (.) ]$ then satisfies
the corresponding 
functional continuity equation 
\begin{eqnarray} 
\partial_t {\cal P}_t[ \phi (.) ] 
= - \int d^d \vec x \frac{ \partial   }{ \partial \phi(\vec x)} \bigg( V_t[\vec x,\phi(.)]  {\cal P}_t[\phi(.) ] \bigg)
  \label{PtdtAdvection}
\end{eqnarray}
that involves the 
functional divergence $\int d^d \vec x \frac{ \partial   }{ \partial \phi(\vec x)} $ of  the advective current $\bigg( V_t[\vec x,\phi(.)]  {\cal P}_t[\phi(.) ] \bigg) $.

In the field-theory literature, the advective velocity $V_t[\vec x;\phi_t(.)]$ is noted $\big(- \Psi_t[\vec x;\phi_t(.)] )$ and is often described as a mere field-redefinition inside the partition function,
while as explained in detail above, the role of the advective velocity $V_t[\vec x;\phi_t(.)]$ 
is to implement some coarsening-transformation
and should depend on the probability ${\cal P}_t[ \phi (.) ]$ itself.
The simplest possibility is when the functional continuity equation of Eq. \ref{PtdtAdvection}
becomes a functional Fokker-Planck Equation (see Eqs \ref{FunctionalFP} and \ref{currentforDiffusion} in Appendix \ref{app_timeDepOU}).


\subsection{ Discussion}

For statistical physicists, Fokker-Planck Equations describe diffusion processes
that can also be described by Langevin Stochastic Differential Equations involving noise,
and it is thus natural to analyze this alternative formulation in the next section.


\section{ Carosso field-coarsening as Langevin stochastic process involving noise} 

\label{sec_Carosso}

In this section, we focus on the more recent developments 
due to Carosso \cite{Carosso,Carosso_conf,Carosso_PhD}, where
the Langevin Stochastic Differential equations associated to the Fokker-Planck RG-flows
are interpreted as genuine coarsening-transformations for the field itself.
Detailed discussions on the Carosso RG, on the comparison with the Polchinski RG,
and on the implementations for field theory defined on the lattice
can be also found in \cite{Cotler_GenerativeDiffusion}.


\subsection{ Langevin process associated to the Fokker-Planck RG-flow for the density $\rho_t(\phi)$
of the single variable $\phi$}

As in the previous sections, let us start with the simple case of the single variable $\phi$,
where the Gaussian Ornstein-Uhlenbeck propagator 
$P(\phi,t \vert \phi_0,0)$ of Eq. \ref{PropagatorOU}
satisfies the Fokker-Planck dynamics of Eq. \ref{p1FPOU}
\begin{eqnarray} 
\partial_t P(\phi,t \vert \phi_0,0)
&&  =  \frac{\partial}{\partial \phi} \bigg( \omega \phi P(\phi,t \vert \phi_0,0)  \bigg) 
 + D \frac{\partial^2 P(\phi,t \vert \phi_0,0)}{\partial \phi^2}  
\nonumber \\
&&  
\equiv {\cal F} P(\phi,t \vert \phi_0,0)
\ \ \ \ \text{with the Fokker-Planck generator } \ {\cal F} \equiv \frac{\partial}{\partial \phi} \left( \omega \phi +D \frac{\partial}{\partial \phi} \right)
  \label{FPgenerator}
\end{eqnarray}
This Fokker-Planck dynamics for the probability $P(\phi,t \vert \phi_0,0) $ to be at $\phi$ at time $t$ 
when starting at $\phi_0$ at time $t=0$ 
is associated to the Ornstein-Uhlenbeck random process $\phi_t$
satisfying the Langevin Stochastic Differential Equation 
\begin{eqnarray} 
\frac{ d \phi_t}{dt}  = - \omega \phi_t + \eta_t
  \label{Langevin1}
\end{eqnarray}
where $ \eta_t$ is a Gaussian white noise 
with vanishing average and with delta-correlations of amplitude $(2D)$
\begin{eqnarray} 
{\mathbb E}\left( \eta_{\tau} \right) && =0
\nonumber \\
{\mathbb E}\left( \eta_{\tau} \eta_{\tau'}\right) 
&& = 2 D  \delta(\tau-\tau') 
  \label{WhiteNoiseRealSpace1}
\end{eqnarray}
The notation $ {\mathbb E}(.)$ has been chosen for averaging over the noise $\eta_t$
in order to avoid any confusion with all other types of averaging considered in the present paper.

The Langevin SDE of Eq. \ref{Langevin1} can be integrated to obtain the solution $\phi_t^{Sol}$
at time $t$ in terms of the initial condition $\phi_{t=0}$ at $t=0$
and in terms on the noise $\nu_{\tau}$ during the time-window $\tau \in [0,t]$
\begin{eqnarray} 
\phi^{Sol}_t = e^{- \omega t} \phi_0 + \int_0^t d\tau e^{- \omega (t-\tau )} \eta_{\tau}
  \label{Langevin1Integrated}
\end{eqnarray}
The Gaussian Ornstein-Uhlenbeck propagator $P(\phi,t \vert \phi_0,0)$ of Eq. \ref{PropagatorOU}
then corresponds to the probability to see the particular value $\phi^{Sol}_t=\phi $ after averaging over the noise $\eta_{.}$
\begin{eqnarray} 
P(\phi,t \vert \phi_0,0)= {\mathbb E}\bigg( \delta(\phi - \phi^{Sol}_t)\bigg) 
  \label{WhiteNoiseAvdelta}
\end{eqnarray}
so that the averaged-value $O^{av}_t $ of an observable $O(\phi)$ computed with the propagator $P(\phi,t \vert \phi_0,0)$
\begin{eqnarray} 
O^{av}_t \equiv \int_{-\infty}^{+\infty} d \phi O(\phi) P(\phi,t \vert \phi_0,0)= {\mathbb E}\bigg( \int_{-\infty}^{+\infty} d \phi O(\phi) \delta(\phi - \phi^{Sol}_t)\bigg) = {\mathbb E}\bigg( O(\phi^{Sol}_t)\bigg)
  \label{Obsav}
\end{eqnarray}
coincides with the averaged-value ${\mathbb E}\bigg( O(\phi^{Sol}_t)\bigg) $ over the noise
of the observable $O(\phi^{Sol}_t) $ evaluated for the Langevin solution of Eq. \ref{Langevin1Integrated}.
In particular, the averaged value and the variance  
of $\phi$ that were computed with the propagator $P(\phi,t \vert \phi_0,0)$ in Eqs \ref{PropagatorOUav} and \ref{PropagatorOUav}
can be recovered from the computations of
the averaged value ${\mathbb E}\bigg( \phi^{Sol}_t \bigg) $ of the Langevin solution $ \phi^{Sol}_t$ of Eq. \ref{Langevin1Integrated} over the noise
\begin{eqnarray} 
{\mathbb E}\bigg( \phi^{Sol}_t \bigg)= e^{- \omega t} \phi_0 = \int_{-\infty}^{+\infty} d \phi \phi P(\phi,t \vert \phi_0,0)
  \label{Langevin1Integratedav}
\end{eqnarray}
and from the computation of the variance of the Langevin solution $\phi^{Sol}_t $
\begin{eqnarray} 
{\mathbb E}\bigg( \left(\phi^{Sol}_t  - e^{- \omega t} \phi_0\right)^2\bigg)
&& ={\mathbb E}\bigg( \int_0^t d\tau e^{- \omega (t-\tau )} \eta_{\tau} \int_0^t d\tau' e^{- \omega (t-\tau' )} \eta_{\tau'}\bigg)
= \int_0^t d\tau e^{- \omega (t-\tau )}  \int_0^t d\tau' e^{- \omega (t-\tau' )} 2 D  \delta(\tau-\tau') 
\nonumber \\
&& = 2D \int_0^t d\tau e^{- 2 \omega (t-\tau )} = \frac{  D \left( 1- e^{- 2t \omega}\right)}{\omega}
 = \int_{-\infty}^{+\infty} d \phi \bigg( \phi - e^{-  \omega t } \phi_0\bigg)^2 P(\phi,t \vert \phi_0,0)
  \label{Langevin1Integratedvar}
\end{eqnarray}

In the area of diffusion processes, it is also standard to write the dynamics of the 
averaged-value $O^{av}_t $ of the observable $O(\phi)$ of Eq. \ref{Obsav}
using the Fokker-Planck Eq \ref{FPgenerator}
satisfied by the propagator $ P(\phi,t \vert \phi_0,0)$
and then making integration by parts 
so that the adjoint ${\cal F}^{\dagger} $ of the Fokker-Planck generator ${\cal F} $ 
of Eq. \ref{FPgenerator}
appears
\begin{eqnarray} 
\partial_t O^{av}_t && = \int_{-\infty}^{+\infty} d \phi O(\phi) \partial_t P(\phi,t \vert \phi_0,0)
= \int_{-\infty}^{+\infty} d \phi O(\phi) \left[  \frac{\partial}{\partial \phi} \bigg( \omega \phi P(\phi,t \vert \phi_0,0)  \bigg) 
 + D \frac{\partial^2 P(\phi,t \vert \phi_0,0)}{\partial \phi^2}  \right]
 \nonumber \\
&& = \int_{-\infty}^{+\infty} d \phi P(\phi,t \vert \phi_0,0)
 \left[ - \omega \phi \frac{\partial O(\phi)}{\partial \phi} 
 + D \frac{\partial^2 O(\phi)}{\partial \phi^2}  \right] 
 \nonumber \\
&& = \int_{-\infty}^{+\infty} d \phi P(\phi,t \vert \phi_0,0) \bigg[ {\cal F}^{\dagger} O(\phi) \bigg]
\ \ \ \ \text{with the adjoint operator } \ {\cal F}^{\dagger} \equiv  \left(-  \omega \phi +D \frac{\partial}{\partial \phi} \right)\frac{\partial}{\partial \phi}
 \nonumber \\
&& \equiv  \bigg[ {\cal F}^{\dagger} O(\phi) \bigg]_t^{av}
  \label{ObsavDyn}
\end{eqnarray}

From the point of view of the physical interpretation,
the writing of the Langevin SDE of Eq. \ref{Langevin1}
suggests to go one step further with respect 
to the previous section concerning the Wegner-Morris perspective, where the coarsening 
was described by the infinitesimal deterministic-looking equation $\frac{ d \phi_t}{dt} =v_t(\phi_t)$ 
of Eq. \ref{deterministic}, but where the advective velocity $v_t(\phi)$ 
had to depend on the solution $\rho_t(\phi) $ itself 
via Eq. \ref{vWKscheme} in order to recover the Fokker-Planck RG-flow of Eq. \ref{p1FPOU}.
With the Langevin SDE of Eq. \ref{Langevin1},
 the coarsening becomes instead a stochastic process involving the white noise $\eta_t$,
 while the propagator $P(\phi,t \vert \phi_0,0) $ satisfying the Fokker-Planck RG-flow
 is recovered only after averaging over the noise via Eq. \ref{WhiteNoiseAvdelta}.
 So the noise represents the information that is still present in the stochastic Langevin trajectories 
 of Eq. \ref{Langevin1Integrated} but that is lost in the noise-averaging of Eq. \ref{WhiteNoiseAvdelta}
 that produces the propagator $P(\phi,t \vert \phi_0,0) $
 satisfying the irreversible RG-flow equation.
This idea of stochastic RG has been put forward by Carosso \cite{Carosso,Carosso_conf,Carosso_PhD}
 for field theory in dimension $d$ with a very nice real-space interpretation as described in the next section.


\subsection{  Carosso coarse-graining for a field $ \phi_t(\vec x)$
based on the stochastic heat equation on the infinite space ${\mathbb R}^d$}

The Carosso stochastic coarse-graining \cite{Carosso,Carosso_conf,Carosso_PhD}
is based on the following Langevin dynamics for the real-space field $\phi_t(\vec x)$ 
as a function of the effective RG-time $t$
 \begin{eqnarray} 
 \partial_t \phi_t( \vec x)  =\Delta \phi_t (\vec x)     + \eta_t(\vec x)
   \label{LangevinCarosso}
\end{eqnarray}
The Laplacian operator $\Delta=\vec \nabla^2$ realizes an infinitesimal local coarse-graining,
while $ \eta_t(\vec x)$ is a Gaussian white noise 
with vanishing average and with delta-correlations both in space and time of amplitude $(2D)$
\begin{eqnarray} 
{\mathbb E}\left( \eta_t(\vec x) \right) && =0
\nonumber \\
{\mathbb E}\left( \eta_t(\vec x) \eta_{\tau}(\vec y)\right) 
&& = 2 D \delta^{(d)} (\vec x- \vec y) \delta(t-\tau) 
  \label{WhiteNoiseRealSpace}
\end{eqnarray}
The Langevin dynamics of Eq. \ref{LangevinCarosso} is well-known 
as the stochastic heat equation with additive noise
in mathematical physics (see the lecture notes \cite{StochasticPDE} on the broader area of Stochastic Partial Differential Equations)
or as the Edwards-Wilkinson dynamics \cite{Stanley,HHZ} in statistical physics.

The Langevin SDE of Eq. \ref{LangevinCarosso} 
can be integrated to obtain the solution $\phi^{Sol}_t( \vec x) $ at time $t$
in terms of the initial field $\phi_{t=0}( \vec y) $ at time $t=0$
and in terms of the space-time noise $\eta_{\tau}( \vec y) $ during the time-window $0 \leq \tau \leq t$
 \begin{eqnarray} 
   \phi_t^{Sol}( \vec x) =    \int d^d \vec y \ \langle \vec x \vert e^{t \Delta } \vert \vec y \rangle \ \phi_0( \vec y)
   + \int_0^t d\tau  \int d^d \vec y \ \langle \vec x \vert e^{(t -\tau) \Delta } \vert \vec y \rangle \ \eta_{\tau}( \vec y) 
   \label{LangevinCarossoIntegInfiniteRealSpace}
\end{eqnarray}
where the heat-kernel $\langle \vec x \vert e^{t \Delta } \vert \vec y \rangle $ 
is well-known when the field is defined on the infinite space $R^d$
 \begin{eqnarray} 
 \langle \vec x \vert e^{t \Delta } \vert \vec y \rangle 
 = \frac{1}{(4 \pi  t)^{\frac{d}{2}}} e^{ - \frac{(\vec x-\vec y)^2}{4  t} } 
 = \int \frac{ d^d \vec q}{ (2 \pi)^{\frac{d}{2} }} e^{- t \vec q \ ^2} e^{i \vec q ( \vec x - \vec y) } 
 \ \ \ \ \ \ \   \text { Heat kernel on the infinite space ${\mathbb R}^d$}
   \label{HeatKernelInfinite}
\end{eqnarray}
but can be also be written for other geometries, 
in particular in lattice field theory using the discrete Laplacian \cite{Carosso,Carosso_conf,Carosso_PhD,Cotler_GenerativeDiffusion}
or for the finite volume $L^d$ that will be considered in the next section.

As in Eq. \ref{WhiteNoiseAvdelta} concerning the single-variable toy-model,
the probability ${\cal P}[\phi(.),t \vert \phi_0(.),0]$ to see the renormalized field configuration $\phi[.]$ at time $t$
given the initial field configuration $\phi_0(.) $ at time $t=0$ can be obtained from the Langevin solutions of Eq. \ref{LangevinCarossoIntegInfiniteRealSpace}
via an average over the noise $\eta_.(.)$ 
\begin{eqnarray} 
{\cal P}[\phi(.),t \vert \phi_0(.),0]= {\mathbb E}\bigg( \prod_{\vec x} \delta(\phi (\vec x) - \phi^{Sol}_t( \vec x))\bigg) 
  \label{WhiteNoiseAvdeltaField}
\end{eqnarray}
so that, as in Eq. \ref{Obsav},
the averaged-value of an observable ${\cal O}[\phi(.)]$ of the renormalized field $\phi[.]$ at time $t$
computed with the propagator ${\cal P}[\phi(.),t \vert \phi_0(.),0] $ 
can be alternatively evaluated from the average-value over the noise of the observable $O[\phi^{Sol}_t(.) ]$ evaluated for the Langevin solution
of Eq. \ref{LangevinCarossoIntegInfiniteRealSpace}
\begin{eqnarray} 
\int {\cal D}\phi(.)  {\cal O}[\phi(.)] {\cal P}[\phi(.),t \vert \phi_0(.),0]
= {\mathbb E}\bigg(  O[\phi^{Sol}_t(.) ]\bigg)
  \label{ObsavField}
\end{eqnarray}

In particular, for a given initial field $\phi_{t=0}( \vec y) $ at time $t=0$,
the averaged value ${\mathbb E}\left(  \phi_t^{Sol}( \vec x) \right) $ of the Langevin solution $\phi_t^{Sol}( \vec x) $ of Eq. \ref{LangevinCarossoIntegInfiniteRealSpace}
over the noise
 \begin{eqnarray} 
 {\mathbb E}\left(  \phi_t^{Sol}( \vec x) \right) 
 = \int d^d \vec y \langle \vec x \vert e^{t \Delta } \vert \vec y \rangle \phi_0( \vec y)
 = \int d^d \vec y \frac{1}{(4 \pi  t)^{\frac{d}{2}}} e^{ - \frac{(\vec x-\vec y)^2}{4  t} }  \phi_0( \vec y)
 \ \ \ \ \ \ \   \text { on the infinite space ${\mathbb R}^d$}
   \label{LangevinCarossoAvHeatKernel}
\end{eqnarray}
corresponds to the spatial-convolution of the given initial field $\phi_{t=0}( \vec y) $  
with the Gaussian heat-kernel $\langle \vec x \vert e^{t \Delta } \vert \vec y \rangle $
that produces some continuous coarse-graining of the initial condition $\phi_0( \vec y) $ 
on a growing scale parametrized by the effective RG-time $t$,
that can be considered as the analog of block-spins in RG procedures concerning lattice spin models.

The connected correlation of the Langevin solution $\phi_t^{Sol}( .) $of Eq \ref{LangevinCarossoIntegInfiniteRealSpace}
between the two positions $\vec x$ and $\vec y$ is independent of the initial field at time $t=0$ and reads
\begin{small}
 \begin{eqnarray} 
&& {\mathbb E}\bigg(    \left[ \phi_t^{Sol}( \vec x) - {\mathbb E}\left(  \phi_t^{Sol}( \vec x) \right) \right] 
\left[ \phi_t^{Sol}( \vec  y) - {\mathbb E}\left(  \phi_t^{Sol}( \vec  y) \right) \right] \bigg)
     =    \int_0^t d\tau  \int d^d \vec z \ \langle \vec x \vert e^{(t -\tau) \Delta } \vert \vec z \rangle \ 
      \int_0^t d\tau'  \int d^d \vec z \ ' \ \langle \vec y \vert e^{(t -\tau') \Delta } \vert \vec z \ ' \rangle \ 
    {\mathbb E}\left(  \eta( \vec z,\tau)\eta( \vec z \ ',\tau') \right)
    \nonumber \\
  &&       =    \int_0^t d\tau  \int d^d \vec z \ \langle \vec x \vert e^{(t -\tau) \Delta } \vert \vec z \rangle \ 
      \int_0^t d\tau'  \int d^d \vec z \ ' \ \langle \vec y \vert e^{(t -\tau') \Delta } \vert \vec z \ ' \rangle \ 
    2 D \delta^{(d)} (\vec z- \vec z \ ') \delta(\tau-\tau') 
       \nonumber \\
  &&       =  2D  \int_0^t d\tau  \int d^d \vec z \ \langle \vec x \vert e^{(t -\tau) \Delta } \vert \vec z \rangle \ 
         \ \langle \vec  y \vert e^{(t -\tau) \Delta } \vert \vec z \rangle \ 
   \label{ConnectedCorreCarossoIntegInfiniteRealSpace}
\end{eqnarray}
\end{small}
where one can use the symmetry between the initial point and the end-point of the heat kernel 
to replace $\langle \vec  y \vert e^{(t -\tau) \Delta } \vert \vec z \rangle =\langle \vec x \vert e^{(t -\tau) \Delta } \vert \vec  y \rangle $ and to perform the integral over $ \vec z$ to obtain
 \begin{eqnarray} 
&& {\mathbb E}\bigg(    \left[ \phi_t^{Sol}( \vec x) - {\mathbb E}\left(  \phi_t^{Sol}( \vec x) \right) \right] 
\left[ \phi_t^{Sol}( \vec  y) - {\mathbb E}\left(  \phi_t^{Sol}( \vec  y) \right) \right] \bigg)
     =  2D  \int_0^t d\tau  \int d^d \vec z \ \langle \vec x \vert e^{(t -\tau) \Delta } \vert \vec z \rangle \ 
         \ \langle \vec z \vert e^{(t -\tau) \Delta } \vert \vec  y \rangle \  
       \nonumber \\
  &&       =   2D  \int_0^t d\tau   \ \langle \vec x \vert e^{2(t -\tau) \Delta } \vert \vec  y \rangle  
  =    2D  \int_0^t d\tau   \ \langle \vec x \vert e^{2 \tau \Delta } \vert \vec  y \rangle
      \nonumber \\
  &&       =   2D  \int_0^t d\tau     \frac{1}{(8 \pi  \tau)^{\frac{d}{2}}} e^{ - \frac{(\vec x-\vec y)^2}{8 \tau} } 
   \ \ \ \ \ \ \   \text { on the infinite space ${\mathbb R}^d$}
    \label{LangevinCarossoIntegInfiniteRealSpaceres}
\end{eqnarray}

In summary, the Carosso stochastic coarse-graining has a clear physical meaning in real-space
both on the infinitesimal time-interval dt with the Laplacian operator $\Delta \phi(\vec x)$ 
governing the non-random part of the Langevin SDE of Eq. \ref{LangevinCarosso},
and on finite-time intervals with the heat-kernel $\langle \vec x \vert e^{t \Delta } \vert \vec y \rangle $ 
governing the average ${\mathbb E}\left(  \phi_t^{Sol}( \vec x) \right) $
 of Eq. \ref{LangevinCarossoAvHeatKernel}.


\subsection{  Carosso RG for the Fourier modes: independent non-identical Ornstein-Uhlenbeck processes }

 When translated towards the Fourier space, the Carosso RG-scheme corresponds to 
 independent non-identical Ornstein-Uhlenbeck processes for the Fourier modes,
 but with different choices for the parameters 
 $\omega_t(\vec q) $ and $D_t(\vec q) $  
 with respect to the Wilson-Kogut choice of Eq. \ref{WKDqchoice} :
 
 (i) the parameters $\omega_t(\vec q) $ are time-independent and reduce to
 \begin{eqnarray} 
\text{ Carosso RG-scheme : } \ \ \omega(\vec q) = \vec q^{\ 2}  \text { on the infinite space ${\mathbb R}^d$}
  \label{CarossochoiceOmega}
\end{eqnarray}
while they are given more generally in other geometries  by the opposite Laplacian eigenvalues,
in particular for lattice field theory \cite{Carosso,Carosso_conf,Carosso_PhD,Cotler_GenerativeDiffusion}
or on the finite volume $L^d$ as will be discussed in the next section.

(ii) the parameters $D_t(\vec q) $ are also time-independent
 \begin{eqnarray} 
\text{ Carosso RG-scheme : } \ \ D_t(\vec q) = D(\vec q)
  \label{CarossochoiceD}
\end{eqnarray}
 but may depend on the momentum $\vec q$ as one wishes \cite{Carosso,Carosso_conf,Carosso_PhD,Cotler_GenerativeDiffusion}, 
 even if above we have focused on the constant
 value $D$ in the real-space noise correlations of Eq. \ref{WhiteNoiseRealSpace}
in order to simplify the discussion and to recover exactly the 
the stochastic heat equation of mathematical physics 
also known as the Edwards-Wilkinson dynamics in statistical physics.

When the parameters $\omega(\vec q) $ and $ D(\vec q) $ are both time-independent, 
the whole Fokker-Planck generator is time-independent,
which is the most standard case considered in the area of stochastic processes,
with the many corresponding simplifications.


\subsection{ Conclusions on the stochastic formulation of exact RG-flows }

\label{subsec_conclusionStochasticRG}

\subsubsection{ Comparison of the three perspectives described in the three last sections } 

Let us first summarize the similarities and the differences between the perspectives 
described the three last sections:

$\bullet$ in section \ref{sec_WK}, we have explained how the Wilson RG-scheme of 'incomplete integration'
corresponds to a Fokker-Planck Evolution for the renormalized probability distribution ${\cal P}_t[ \phi (.) ]$, 
and more precisely to independent Ornstein-Uhlenbeck processes for the Fourier modes.
However it should be stressed that, for reasons that are not clear to us, Wilson and Kogut have chosen
to completely avoid the names 'Fokker-Planck' or 'Ornstein-Uhlenbeck' or even any other terminology related to
stochastic processes, and have preferred to present their procedure as a 'trick' based on 'a differential equation'. 

$\bullet$ in section \ref{sec_WM}, we have described the Wegner-Morris perspective, 
where the possible RG-procedures are seen as infinitesimal changes of variables conserving the partition function. As a consequence, the continuity equation satisfied by the renormalized probability distribution ${\cal P}_t[ \phi (.) ]$ involves some advective velocity $V_t[\vec x;\phi_t(.)] $ that reflects the precise RG-scheme. However, as emphasized in subsection \ref{subsec_irreversibility}, some mechanism for irreversibility
is needed to have a genuine RG-flow where the RG-time-arrow matters,
so that the advective velocity $V_t[\vec x;\phi_t(.)] $ 
should actually depend also on the solution ${\cal P}_t[ \phi (.) ] $ itself,
the simplest possibility being when the Wegner-Morris
continuity equation corresponds to a Fokker-Planck equation involving both advective and diffusive contributions, as in the previous Wilson-Kogut example.

$\bullet$ in the present section \ref{sec_Carosso}, we have explained how Carosso
has chosen to fully acknowledge the stochastic character of the Fokker-Planck RG-flows  
for the probability distribution ${\cal P}_t[ \phi (.) ]$
and has writen the corresponding Langevin Stochastic Differential Equations for the field $\phi_t(\vec x)$ involving the space-time noise $\eta_{\tau}(\vec y)$, whose physical meaning is discussed 
in more details in the next subsection.


\subsubsection{ Discussion on the role and on the physical meaning of the noise in the Langevin SDE of Carosso stochastic RG} 

\label{subsec_noise}

$\bullet$ As explained by Carosso \cite{Carosso,Carosso_conf,Carosso_PhD},
the role of the noise in the stochastic RG can be understood
from the comparison with the block-spin-RG concerning discrete spins models :

(i) In discrete spins models, the information contained 
in the initial configuration of the $N$ spins can be decomposed  
into the information still surviving in the smaller number $N_b$ of block-spins 
and into the information that has been lost when the other degrees of freedom 
have been integrated out.

(ii) Similarly in the Langevin formulation of the stochastic RG, 
the information contained in the initial field $\phi_{t=0}(.)$ can be decomposed
into the information still surviving in the renormalized field $\phi(.)$ at the RG-time $t$, 
that represents the analog of the block-spins,
and into the information that is lost when one averages over the noise in Eq. \ref{WhiteNoiseAvdeltaField} 
to obtain the propagator ${\cal P}[\phi(.),t \vert \phi_0(.),0]$ 
between the initial fields $\phi_0(.) $ at time $t=0$ and the renormalized field $\phi(.) $ at time $t$.
So at the level of the Langevin SDE associated to the infinitesimal time-interval $[t,t+dt]$,
the noise $\eta_t(.)$ at time $t$ plays the role of the degrees of freedom that have to be integrated-out 
during this elementary RG-step to obtain the new renormalized field at time $(t+dt)$.

 $\bullet$ In his PhD-manuscript \cite{Carosso_PhD}, 
 Carosso gives, in addition, a very suggestive and profound analogy with the usual Brownian motion:  
 
 (a) In the very famous theory of the 'physical' Brownian motion by Albert Einstein in 1905 
 and by Paul Langevin in 1908, 
 the mesoscopic Brownian particle immersed in the fluid
 receives a very large number of kicks per unit time
 from the much smaller microscopic molecules of the fluid,
 so that besides the averaged effect producing the viscous force,
 the remaining global random contribution can be modeled mathematically
 as a white noise in the Langevin SDE, 
 where the diffusion coefficient $ D \propto \frac{T}{{\cal N}_A}$ 
 is computed to be proportional to the temperature $T$ of the fluid
 and to be inversely proportional to the Avogadro number ${\cal N}_A \simeq 6.10^{23}$,
 as checked experimentally by Jean Perrin in 1909 in order to 'prove' the existence of atoms
 and in order to obtain the first experimental measure of the Avogadro number ${\cal N}_A$.
 
 (b) Similarly during the infinitesimal RG-time-interval $[t,t+dt]$ of the stochastic RG for the field theory, 
 one can have in mind that, 
 besides the averaged effect of the coarsening procedure that damps 
 infinitesimally all the Fourier modes according to the momentum-dependent parameters $\omega_t(\vec q)$, each Fourier-mode is also 'kicked' by all the other Fourier-modes 
 that are simultaneously infinitesimally integrated over,
 so that the noise produced by this very large number of infinitesimal kicks 
 can be considered as a direct consequence of
  the interactions that should exist between Fourier modes of different momenta in all non-gaussian models.
 This physical interpretation of the noise in terms of the interactions existing between the Fourier modes
 suggests that it would be very important in the future to develop this argument further at the quantitative level,
 at least in some special cases,
 in order to 'derive' the statistical properties of the noise, and to 'compute' the diffusion coefficients $D_t(\vec q)$, instead of making assumptions about them a priori.


\subsubsection{ Advantages of the stochastic formulation of exact RG-flows } 

As stressed by Carosso \cite{Carosso,Carosso_conf,Carosso_PhD},
the stochastic formulation of exact RG-flows
has a certain number of advantages, both conceptual and computational:

$\bullet$ At the conceptual level, the first important point
is that the Markovian character of the stochastic process, 
where the future depends only on the present state,
and where the Chapman-Kolmogorov Equation describes 
how two successive steps can be composed via the integration over the intermediate state,
corresponds exactly to the desired properties for RG procedures built from the iteration of elementary RG-steps.
For diffusion processes, this decomposition into infinitesimal RG-steps 
leads to the path-integral representation of the propagator 
${\cal P}[\phi(.),t \vert \phi_0(.),0] $ over the RG-time-window $[0,t]$.

The second important point is that the desired irreversibility of the RG-flow is contained :

(i) in the diffusive-part of the Fokker-Planck Equation, that shows that the advective-form
of the Wegner-Morris Continuity Equation cannot be interpreted as a meaningless field-redefinition
in the partition function integral, as discussed in detail in subsection \ref{subsec_irreversibility};

(ii) in the noise of the Carosso Langevin SDE, that represents the information
that needs to be integrated over to produce an appropriate RG-step, as discussed
in detail in the previous subsection \ref{subsec_noise}.

So at the conceptual level, the stochastic formulation, that may seem strange at first sight,
turns out to fit perfectly into the desired RG-framework,
while it offers completely new ways of thinking. In particular, the general framework of Fokker-Planck Equations
and Langevin SDE suggests to consider the cases of more general drifts and more general diffusion coefficients beyond the specific examples of Ornstein-Uhlenbeck processes chosen by Wilson and by Carosso:

(a) the drift term $\left[ -\omega_t(\vec q) \phi_t(\vec q) \right]$ that is linear with respect to the field $\phi_t(.)$ 
in the Ornstein-Uhlenbeck SDE can be replaced by drifts that are non-linear with respect to the field;

(b) the noise $\eta_.(.)$ that appears additively in the Ornstein-Uhlenbeck SDE, i.e. the 
diffusion coefficients $D_t(\vec q)$ do not depend on the field $\phi_t(.)$,
could instead appear multiplicatively in the SDE, 
so that the diffusion coefficients of the Fokker-Planck RG-flow could
also depend on the field.

$\bullet$ At the computational level, the main advantage put forward by Carosso 
\cite{Carosso,Carosso_conf,Carosso_PhD}
 is a new type of Monte Carlo RG in lattice field simulation,
 where the dynamics of observables along the RG-flow 
 is studied from the Langevin trajectories 
 using the property discussed in Eq. \ref{ObsavField}, with the application
 to the measure of anomalous dimensions of scaling operators close to critical points.
Beyond this specific important example, the correspondence with diffusion processes means 
more generally that one can 
directly import all the knowledge and all the toolbox of analytical/numerical techniques
 that have been developed 
over the years to analyze Fokker-Planck flows and Langevin SDE.


\section{ Adaptation of the Carosso RG to the finite volume $L^d$ with $m_e$ frozen }

\label{sec_RGVolume}

\subsection{ Returning to the finite volume $L^d$ with the RG-goal of Eq. \ref{pmalonetfrozen} }

In this section, we return
to the task explained at the end of section \ref{sec_Fourier} around Eq. \ref{pmalonetfrozen},
i.e. we return to the framework described in the two sections \ref{sec_Realspace} and \ref{sec_Fourier},
 where the field $\phi_t(\vec x)$ is defined on the finite volume $L^d$ with periodic boundary conditions,
with the Fourier-series decomposition of Eq. \ref{FouriercoefsPerioab} and \ref{FourierPerioCosSin} at any time $t$
\begin{eqnarray} 
 \phi_t( \vec x)  && = m_e + \sum_{\vec k >0} 
 \left[ {\hat \phi}_t^R(\vec k)  2 \cos \left(  \frac{ 2 \pi }{L} \vec k . \vec x \right) 
 +  {\hat \phi}_t^I(\vec k) 2 \sin \left(  \frac{ 2 \pi }{L} \vec k . \vec x \right) 
 \right]
 \nonumber \\
 {\hat \phi}_t^R(\vec k)  && =  \frac{1}{L^d} \int_{L^d}  d^d \vec x  \phi_t (\vec x) \cos \left(  \frac{ 2 \pi }{L} \vec k . \vec x \right) 
 \nonumber \\
{\hat \phi}_t^I(\vec k)&& =  \frac{1}{L^d} \int_{L^d}  d^d \vec x  \phi_t (\vec x) \sin \left(  \frac{ 2 \pi }{L} \vec k . \vec x \right) 
  \label{FourierPerioCosSint}
\end{eqnarray}
where the empirical magnetization $m_e$ corresponding to the zero-momentum Fourier coefficient
remains frozen
\begin{eqnarray} 
 \frac{1}{L^d} \int_{L^d}  d^d \vec x  \phi_t (\vec x) = m_e \ \ \text{for any RG-time $t$}  
  \label{mefrozen}
\end{eqnarray}
in order to pursue our goal explained around Eq. \ref{pmalonetfrozen} : the probability distribution $p_L(m_e)$
of the empirical magnetization $m_e$ of the initial model at $t=0$ will be conserved during the RG-flow at any time $t$ and will be the only information that we wish to keep about the initial condition
(instead of the partition function $Z_0$ for the single-variable toy-model as discussed after Eq. \ref{p1FPOUsolinfty}).

It is clear that there is still a lot of freedom in the choice of the RG-scheme,
i.e. in the precise way one wishes to integrate over the non-zero-momentum Fourier-modes 
as a function of the effective RG-time $t$. Even within the framework of stochastic RG based
on independent Ornstein-Uhlenbeck processes for the Fourier modes, 
one can make different choices for the parameters
$\omega_t(\vec q) $ and $D_t(\vec q) $,
as already discussed with the Wilson-Kogut choice of Eq. \ref{WKDq} and with the Carosso choice of
Eqs \ref{CarossochoiceOmega} and \ref{CarossochoiceD} for the case
where the momentum $\vec q$ is a continuous variable.

In the present section, we have chosen 
to adapt the Carosso RG described in section \ref{sec_Carosso} to the finite volume $L^d$ and
to the constraint of Eq. \ref{mefrozen},
while the generalization to Ornstein-Uhlenbeck processes with time-dependent parameters
$\omega_t(\vec q) $ and $D_t(\vec q) $
is described in Appendix \ref{app_timeDepOU}.


\subsection{ Adaptation of the Carosso Langevin real-space SDE to the volume $L^d$ and to the conservation of $m_e$ }

The Carosso stochastic field-coarsening described in the previous section for the infinite space $\vec x \in {\mathbb R}^d$ can be adapted  
to our present purposes via the two following modifications:

(1) The stochastic heat equation of Eq. \ref{LangevinCarosso},
and the corresponding integrated form of Eq. \ref{LangevinCarossoIntegInfiniteRealSpace}
for the solution that will be denoted with the simpler notation $\phi^{Sol}_t( \vec x) \equiv \phi_t( \vec x) $ from now on,
are now valid on the finite volume $\vec x \in L^d$ with periodic boundary conditions
 for the field itself and for the noise $\eta_t( \vec x) $
\begin{eqnarray} 
   \phi_t( \vec x) =    \int_{L^d} d^d \vec y \ \langle \vec x \vert e^{t \Delta } \vert \vec y \rangle \ \phi_0( \vec y)
   + \int_0^t d\tau  \int_{L^d} d^d \vec y \ \langle \vec x \vert e^{(t -\tau) \Delta } \vert \vec y \rangle \ \eta_{\tau}( \vec y) 
   \label{LangevinCarossoIntegFiniteRealSpace}
\end{eqnarray}
So the heat-kernel $\langle \vec x \vert e^{t \Delta } \vert \vec y \rangle $ 
 given by Eq. \ref{HeatKernelInfinite} on the infinite space ${\mathbb R}^d$
is now replaced by the following heat-kernel on the finite volume $L^d$ with periodic boundary conditions
 \begin{eqnarray} 
\text{ Heat kernel on $L^d$ : } \ \ \ \  \langle \vec x \vert e^{t \Delta } \vert \vec y \rangle 
&& =  \frac{1}{L^d}  \sum_{\vec k \in {\mathbb Z}^d} e^{- t \omega_L(\vec k) } 
e^{i  \frac{ 2 \pi }{L} \vec k . (\vec x- \vec y)   }
\nonumber \\
&& = \frac{1}{L^d} \left[1 + \sum_{\vec k >0} e^{- t \omega_L(\vec k) } 
2 \cos \left(  \frac{ 2 \pi }{L} \vec k . (\vec x- \vec y) \right)  \right]
   \label{HeatKernelFinite}
\end{eqnarray}
where we have introduced the notation
 \begin{eqnarray} 
 \omega_L(\vec k)  \equiv  \frac{ 4 \pi^2 }{L^2} \vec k^2
   \label{omega}
\end{eqnarray}
for the opposite eigenvalues of the Laplacian $\Delta$ that govern the relaxation towards
the uniform density $\frac{1}{L^d}$ for $t \to + \infty$ in Eq. \ref{HeatKernelFinite}.

(2) In order to impose that the empirical magnetization $m_e$ remains frozen in time (Eq. \ref{mefrozen}),
we will add the following global spatial constraints on the noise $\eta_t(\vec x) $ at any time $t$
\begin{eqnarray} 
\int_{L^d} d^d \vec  x \ \eta_t(\vec x) =0  \ \ \text{for any time $t$}  
  \label{WhiteNoiseRealSpaceConstraint}
\end{eqnarray}
As a consequence, the standard gaussian space-time white noise of Eq. \ref{WhiteNoiseRealSpace}
will be slightly changed into
the following joint probability for all the noises on the time-interval $[0,T]$ 
\begin{eqnarray} 
{\mathbb P} \left[ \eta_. (.)  \right] = e^{ \displaystyle - \int_0^T  dt \int_{L^d} d^d \vec x \ \frac{\eta_t^2(\vec x)}{4D}  }
\prod_{0 \leq t \leq T} \delta \left( \int_{L^d} d^d \vec x \eta_t(\vec x) \right)
  \label{WhiteNoisesRealSpaceDistri}
\end{eqnarray}
with the consequences for the real-space two-point correlations
discussed in Eq. \ref{NoiseCorreAvecContrainte}.

It is clear that these two modifications will be both simpler when translated towards the Fourier space,
as described in the next subsection.


\subsection{ Corresponding Langevin SDE for the Fourier coefficients 
$[{\hat \phi}^R_t  (\vec k);{\hat \phi}^I_t  (\vec k)]$   }

\subsubsection{ Probability distribution of the Fourier coefficients $[{\hat \eta}^R_t(\vec k),{\hat \eta}^I_t(\vec k)] $ of the real-space noise $\eta_t( \vec x)$ }

For the real-space noise $\eta_t(\vec x) $ of Eq. \ref{WhiteNoisesRealSpaceDistri}, 
the Fourier decomposition analog to Eq. \ref{FourierPerioCosSin}
only involves the Fourier coefficients associated to $ \vec k >0$ (since the zero-momentum Fourier coefficient vanishes at any time as a consequence of the constraints of Eq. \ref{WhiteNoisesRealSpaceDistri})
 \begin{eqnarray} 
\eta_t(\vec x) =    \sum_{\vec k >0} 
 \left[    {\hat \eta}^R_t (\vec k)   2 \cos \left(  \frac{ 2 \pi }{L} \vec k . \vec x \right) 
 +    {\hat \eta}^I_t (\vec k)  2 \sin \left(  \frac{ 2 \pi }{L} \vec k . \vec x \right) 
 \right] 
   \label{xi}
\end{eqnarray}
with 
 \begin{eqnarray} 
  {\hat \eta}^R_t(\vec k)  && =  \frac{1}{L^d} \int_{L^d}  d^d \vec x  \eta_t(\vec x) \cos \left(  \frac{ 2 \pi }{L} \vec k . \vec x \right) 
 \nonumber \\
 {\hat \eta}^I_t(\vec k)&& =  \frac{1}{L^d} \int_{L^d}  d^d \vec x  \eta_t (\vec x) \sin \left(  \frac{ 2 \pi }{L} \vec k . \vec x \right) 
  \label{xiinverse}
\end{eqnarray}
The Parseval-Plancherel identity analogous to Eq. \ref{hphifourier}
\begin{eqnarray} 
 \int_{L^d} d^d \vec x \eta_t^2(\vec x)= 
2 L^d    \sum_{\vec k >0} \left( [ {\hat \eta}^R_t  (  \vec k)]^2   +  [ {\hat \eta}^I_t  (  \vec k)]^2 \right) 
  \label{ParsevalNoise}
\end{eqnarray}
then yields that the measure of Eq. \ref{WhiteNoisesRealSpaceDistri} for the real-space noise $\eta_t(\vec x) $
translates for its Fourier coefficients ${\hat \eta}^{\alpha=R,I}_t (\vec k) $ into the fully factorized form
\begin{eqnarray} 
{\mathbb P} \left[ {\hat \eta}^._{.} (.)  \right] = e^{ \displaystyle - \frac{L^d}{2 D} \int_0^T  dt 
 \sum_{ \vec k>0} \left[  [ {\hat \eta}^R_t  (  \vec k)]^2   +  [ {\hat \eta}^I_t  (  \vec k)]^2 \right] }
= \prod_{\vec k >0} \prod_{\alpha=R,I} 
e^{ \displaystyle - \frac{L^d}{2 D} \int_0^T  dt 
[{\hat \eta}^{\alpha}_t  (  \vec k)]^2    }
  \label{WhiteNoisesFourierDistri}
\end{eqnarray}
In conclusion, the Fourier coefficients  ${\hat \eta}^{\alpha=R,I}_t (\vec k) $ 
are independent identical white noises with vanishing averages and delta-correlations
\begin{eqnarray} 
{\mathbb E}\left( {\hat \eta}^{\alpha}_t (\vec k)  \right) && =0
\nonumber \\
{\mathbb E}\left( {\hat \eta}^{\alpha}_t (\vec k) {\hat \eta}^{\alpha'}_{\tau} (\vec q) \right) 
&& = \frac{D}{L^d}  \delta_{\vec k, \vec q} \delta_{\alpha,\alpha'} \delta(t-\tau) 
\equiv 2 D_L \delta_{\vec k, \vec q} \delta_{\alpha,\alpha'} \delta(t-\tau) 
  \label{WhiteNoisesFourier}
\end{eqnarray}
where the amplitude $(2 D_L)$ scales as the inverse of the volume $L^d$
\begin{eqnarray} 
D_L \equiv \frac{D}{2 L^d}  
  \label{DL}
\end{eqnarray}
as a consequence of the presence of $L^d$ in the Parseval-Plancherel identity of Eq. \ref{ParsevalNoise},
since the Fourier coefficients  ${\hat \eta}^{\alpha=R,I}_t (\vec k) $ are intensive variables
defined as spatial-averages of the real-space noise over the volume $L^d$ in Eq. \ref{xiinverse}.
  The scaling of Eq. \ref{DL} will have of course very important consequences in the following subsections.


\subsubsection{ Langevin SDE for the Fourier coefficients $[{\hat \phi}^R_t(\vec k);{\hat \phi}^I_t(\vec k)] $ of the real-space field $\phi_t( \vec x)$ }

The Fourier coefficients of Eq. \ref{FourierPerioCosSin}
satisfy the following Langevin Equations obtained from Eq. \ref{LangevinCarosso}
 \begin{eqnarray} 
\partial_t  {\hat \phi}^R_t(\vec k)  && =  \frac{1}{L^d} \int_{L^d}  d^d \vec y  \bigg(\Delta \phi_t (\vec x)     + \eta_t(\vec x) \bigg) \cos \left(  \frac{ 2 \pi }{L} \vec k . \vec y \right) 
= - \frac{ 4 \pi^2 }{L^2} \vec k^2 {\hat \phi}^R_t(\vec k)+ {\hat \eta}^R_t(\vec k)
\equiv - \omega_L(\vec k){\hat \phi}^R_t(\vec k)+ {\hat \eta}^R_t(\vec k)
 \nonumber \\
\partial_t {\hat \phi}^I_t(\vec k)&& =  \frac{1}{L^d} \int_{L^d}  d^d \vec y  \bigg(\Delta \phi_t (\vec x)     + \eta_t(\vec x) \bigg) \sin \left(  \frac{ 2 \pi }{L} \vec k . \vec y \right) 
  = - \frac{ 4 \pi^2 }{L^2} \vec k^2 {\hat \phi}^I_t(\vec k)+ {\hat \eta}^I_t(\vec k)
  \equiv - \omega_L(\vec k){\hat \phi}^I_t(\vec k)+ {\hat \eta}^I_t(\vec k)
  \label{OUFourier}
\end{eqnarray}
where one recognizes the Fourier coefficients of the noise of Eq \ref{xiinverse},
while the two integrations by parts involving the Laplacian $\Delta$ in the presence of periodic boundary conditions
have produced the Fourier eigenvalues $\left( - \omega_L(\vec k) \right)$ of the Laplacian $\Delta$ with the notation of Eq. \ref{omega}.

In conclusion, the Fourier coefficients ${\hat \phi}^{\alpha}_t  (\vec k) $ with 
 $\vec k>0$ and $\alpha=R,I$ are independent non-identical Ornstein-Uhlenbeck 
 processes satisfying the Langevin dynamics
\begin{eqnarray} 
\partial_t {\hat \phi}^{\alpha}_t  (\vec k) && = - \omega_L(\vec k) {\hat \phi}^{\alpha}_t  (\vec k)  
+   {\hat \eta}^{\alpha}_t (\vec k) \ \ \ \text{for $\vec k>0$ and $\alpha=R,I$} \ \ 
  \label{LangevinOU}
\end{eqnarray}
with the noise properties of Eqs \ref{WhiteNoisesFourierDistri} and \ref {WhiteNoisesFourier}.
Note that all the following calculations will be written with the general notation $\omega_L(\vec k) $
and can be thus applied to other choices than the Laplacian eigenvalues of Eq. \ref{omega}
if one wishes.
The case of Eq. \ref{LangevinOU} that will be analyzed in the remainder of this section
is generalized to the case of time-dependent parameters $[\omega_L(\vec k,t);D_L(\vec k,t) ] $
  in Appendix \ref{app_timeDepOU} in order to include the case of the Wilson-Kogut choice of Eq. \ref{WKDqchoice}.


\subsection{ Statistical properties of the single Fourier coefficient ${\hat \phi}^{\alpha}_t  (\vec k)$ 
as an Orstein-Uhlenbeck process of parameters $[\omega_L(\vec k);D_L= \frac{D}{2 L^d}]$   }

Since the single Fourier coefficient ${\hat \phi}^{\alpha}_t  (\vec k)$ 
is an Orstein-Uhlenbeck process satisfying the Langevin SDE of Eq. \ref{LangevinOU}
of parameters $[\omega_L(\vec k);D_L= \frac{D}{2 L^d}]$,
one can directly translate all the results recalled in the previous sections
for the one-dimensional Ornstein-Uhlenbeck process of parameters $[\omega,D]$ :

(i) The Ornstein-Uhlenbeck propagator 
${\mathbb p}_t^{[k]} [{\hat \phi}^{\alpha}  (\vec k) \vert {\hat \phi}^{\alpha}_0  (\vec k)] $
for the single Fourier coefficient ${\hat \phi}^{\alpha}  (\vec k) $ 
translated from Eq. \ref{PropagatorOU}
\begin{eqnarray} 
{\mathbb p}_t^{[k]} [{\hat \phi}^{\alpha}  (\vec k) \vert {\hat \phi}^{\alpha}_0  (\vec k)]
 =\sqrt{ \frac{L^d \omega_L(\vec k)}{  \pi D \left[ 1- e^{- 2t \omega_L(\vec k)}\right] }  }
 e^{\displaystyle  - L^d \omega_L(\vec k) \frac{\left[ {\hat \phi}^{\alpha}  (\vec k) -  e^{- t \omega_L(\vec k) } {\hat \phi}^{\alpha}_0  (\vec k) \right]^2}{ D \left[ 1- e^{- 2t \omega_L(\vec k)}\right]}}
  \label{SingleModePropagatorOU}
\end{eqnarray}
involves the averaged value analog to Eq. \ref{Langevin1Integratedav}
\begin{eqnarray} 
{\mathbb E}\left( {\hat \phi}^{\alpha}_t  (\vec k) \right)  
\equiv \int_{-\infty}^{+\infty} d{\hat \phi}^{\alpha}  (\vec k)  {\hat \phi}^{\alpha}  (\vec k)  {\mathbb p}_t^{[k]} [{\hat \phi}^{\alpha}  (\vec k) \vert {\hat \phi}^{\alpha}_0  (\vec k)]
= e^{- t \omega_L(\vec k) } {\hat \phi}^{\alpha}_0  (\vec k)
  \label{OUaverage}
\end{eqnarray}
and the variance analog to Eq. \ref{Langevin1Integratedvar}
\begin{eqnarray} 
{\mathbb E}\left(  \left[ {\hat \phi}^{\alpha}_t(\vec k) - e^{- t \omega_L(\vec k) } {\hat \phi}^{\alpha}_0  (\vec k) \right]^2 \right)  
&&  =  \frac{D}{L^d}  \left( \frac{1- e^{- 2t \omega_L(\vec k)}}{  2 \omega_L(\vec k)} \right)
  \label{OUvariance}
\end{eqnarray}
while it satisfies the Fokker-Planck dynamics analog to Eq. \ref{p1FPOU}
\begin{eqnarray} 
\partial_t {\mathbb p}_t^{[k]} [{\hat \phi}^{\alpha}  (\vec k) \vert {\hat \phi}^{\alpha}_0  (\vec k)]
 = \frac{\partial}{\partial {\hat \phi}^{\alpha} (\vec k)} \left[ \omega_L(\vec k) {\hat \phi}^{\alpha}  (\vec k)
 +\frac{D}{2 L^d} \frac{\partial}{\partial {\hat \phi}^{\alpha} (\vec k)} \right] 
 {\mathbb p}_t^{[k]} [{\hat \phi}^{\alpha}  (\vec k) \vert {\hat \phi}^{\alpha}_0  (\vec k)]
  \label{FPOU}
\end{eqnarray}

(ii) The corresponding generating function for the Fourier coefficient 
${\hat \phi}^{\alpha}_t  (\vec k) $ 
for the given initial condition ${\hat \phi}^{\alpha}_0  (\vec k) $ reads 
\begin{eqnarray} 
{\mathbb E}\left( e^{\lambda {\hat \phi}^{\alpha}_t  (\vec k)} \vert {\hat \phi}^{\alpha}_0  (\vec k) \right)  
&& \equiv \int_{-\infty}^{+\infty} d{\hat \phi}^{\alpha}  (\vec k) e^{\lambda {\hat \phi}^{\alpha}  (\vec k)}  {\mathbb p}_t^{[k]} [{\hat \phi}^{\alpha}  (\vec k) \vert {\hat \phi}^{\alpha}_0  (\vec k)]
 \nonumber \\
&& = e^{ \displaystyle \lambda e^{- t \omega_L(\vec k) } {\hat \phi}^{\alpha}_0  (\vec k)
+ \lambda^2  \frac{D}{2 L^d}  \left( \frac{1- e^{- 2t \omega_L(\vec k)}}{  2 \omega_L(\vec k)} \right)} 
  \label{SingleModePropagatorOUgenerating}
\end{eqnarray}


\subsection{ Joint statistical properties of all the Fourier coefficients ${\hat \phi}^{\alpha}_t  (\vec k) $ for a given initial condition at $t=0$   }

Since the Fourier coefficients ${\hat \phi}^{\alpha}_t  (\vec k) $ for $\vec k>0 $ and $\alpha=R,I $ are independent non-identical Ornstein-Uhlenbeck processes,
their global propagator reduces to the product of the individual propagators 
${\mathbb p}_t^{[k]} [{\hat \phi}^{\alpha}  (\vec k) \vert {\hat \phi}^{\alpha}_0  (\vec k)]$
 of Eq. \ref{SingleModePropagatorOU} 
\begin{eqnarray} 
&&  {\mathbb P}_t[{\hat \phi}^{.}  (.) \vert {\hat \phi}^._0 (.)]
  = \prod_{\vec k>0} \prod_{\alpha=R,I}
{\mathbb p}_t^{[k]} [{\hat \phi}^{\alpha}  (\vec k) \vert {\hat \phi}^{\alpha}_0  (\vec k)]
 \nonumber \\
 &&  = \left[ \prod_{\vec k>0} 
 \frac{L^d \omega_L(\vec k)}{  \pi D \left[ 1- e^{- 2t \omega_L(\vec k)}\right]   }
 \right]
e^{\displaystyle  - L^d \sum_{\vec k>0} \sum_{\alpha=R,I} \omega_L(\vec k) \frac{\left[ {\hat \phi}^{\alpha}  (\vec k) -  e^{- t \omega_L(\vec k) } {\hat \phi}^{\alpha}_0  (\vec k) \right]^2}{ D \left[ 1- e^{- 2t \omega_L(\vec k)}\right]}} 
  \label{Propagatorfactorization}
\end{eqnarray}

Using the Parseval-Plancherel identity of Eq. \ref{hphifourier},
the generating function of the field $\phi_t(\vec x)$ for a given initial condition $ \phi_{t=0}(\vec y)$ 
can be computed in Fourier space from the product of the individual
generating functions of Eq. \ref{SingleModePropagatorOUgenerating}
for the Fourier modes with the replacement 
$\lambda \to L^d 2 h^{\alpha}  (\vec k)$ to obtain 
\begin{eqnarray} 
&& {\mathbb E} \left(   e^{ \displaystyle \int_{L^d} d^d \vec x h(\vec x) \phi_t(\vec x) } \bigg\vert \phi_0(. ) \right)
 =  {\mathbb E} \left(   e^{ \displaystyle L^d \left[ h_e  m_e
 + 2 \sum_{\vec k >0} \left(  {\hat h}^R  (  \vec k)  {\hat \phi}^R ( \vec k) + {\hat h}^I  (  \vec k)  {\hat \phi}^I ( \vec k)  \right)\right] } \bigg\vert {\hat \phi}^._0 (.) \right)
 \nonumber \\
 && = e^{ L^d h_e m_e} 
 \prod_{\vec k>0} \prod_{\alpha=R,I}
\int_{-\infty}^{+\infty} d{\hat \phi}^{\alpha}  (\vec k) e^{L^d 2 h^{\alpha}  (\vec k) {\hat \phi}^{\alpha}  (\vec k)}  {\mathbb p}_t^{[k]} [{\hat \phi}^{\alpha}  (\vec k) \vert {\hat \phi}^{\alpha}_0  (\vec k)]
 \nonumber \\
 && \equiv e^{ \displaystyle L^d \left( h_e m_e
 +  \sum_{\vec k>0} \sum_{\alpha=R,I}
 \left[  2 h^{\alpha}  (\vec k)e^{- t \omega_L(\vec k) } {\hat \phi}^{\alpha}_0  (\vec k)
+   [h^{\alpha}  (\vec k) ]^2 D \left( \frac{1- e^{- 2t \omega_L(\vec k)}}{   \omega_L(\vec k)} \right) 
\right] \right)} 
  \label{OUgeneratingtotdef}
\end{eqnarray}

In summary, for any given initial field configuration $\phi_{t=0}(\vec x) $ on the volume $L^d$,
we have characterized in Fourier space
the statistical properties of the field $\phi_{t}(\vec x) $ at time $t$
obtained via the Carosso stochastic coarse-graining adapted to the finite volume $L^d$ and to
the conservation of the empirical magnetization $m_e$ at any time $t$.
In the next subsection, we take into account the statistical 
properties of the initial condition $\phi_{t=0}(\vec x) $.


\subsection{ Joint distribution $P_t[m_e;{\hat \phi}^{.}  (.) ] $ 
of the Fourier coefficients at time $t$ 
for the initial distribution $P_0[ m_e;\hat \phi^._0(.) ]$  }

When the initial Fourier coefficients ${\hat \phi}^{\alpha=R,I}_0  (\vec k) $ are correlated
via their joint distribution $P_{t=0}[ m_e;\hat \phi^R(.) ; \hat\phi^I(.)] $ of Eq. \ref{ProbaPhik}
\begin{eqnarray} 
P_0\left[ m_e;{\hat \phi}^._0  (.) \right]
= \frac{ e^{-   L^d {\hat  E}_0 \left[ m_e;{\hat \phi}^._0 (.)  \right]} }
  { \hat {\cal Z}_0 }
  \label{ProbaPhikInitialTime}
\end{eqnarray}
then the joint probability distribution $P_t \left[m_e; {\hat \phi}^.  (.) \right] $ 
of the Fourier coefficients at time $t$ 
is given by the convolution of the factorized propagator of Eq. \ref{Propagatorfactorization}
with the correlated initial condition of Eq. \ref{ProbaPhikInitialTime}
\begin{eqnarray} 
&& P_t \left[ m_e; {\hat \phi}^.  (.) \right]  
 =  \left[ \prod_{\vec k >0} \prod_{\alpha=R,I}  \int_{-\infty}^{+\infty} d \hat \phi^{\alpha}_0(\vec k) \right]
 {\mathbb P}_t[{\hat \phi}^{.}  (.) \vert {\hat \phi}^._0 (.)] \ \ 
   P_{0}\left[ m_e;{\hat \phi}^._0  (.) \right]
  \label{SoluOUt}  \\
&&  =\frac{1}{\hat {\cal Z}_0} \left[ \prod_{\vec k >0} \prod_{\alpha=R,I}  \int_{-\infty}^{+\infty} d \hat \phi^{\alpha}_0(\vec k) 
\sqrt{ \frac{L^d \omega_L(\vec k)}{  \pi D \left[ 1- e^{- 2t \omega_L(\vec k)}\right]   } }
 \right]
 e^{\displaystyle  - L^d  \sum_{\vec k>0} \sum_{\alpha=R,I} \omega_L(\vec k) \frac{\left[ {\hat \phi}^{\alpha}  (\vec k) -  e^{- t \omega_L(\vec k) } {\hat \phi}^{\alpha}_0  (\vec k) \right]^2}{ D \left[ 1- e^{- 2t \omega_L(\vec k)}\right] }   }
  P_{0}\left[ m_e;{\hat \phi}^._0  (.) \right]
\nonumber
\end{eqnarray}

The distribution $p_L[m_e] $ of the empirical magnetization $m_e$ of the initial condition
$ P_{0}\left[ m_e;{\hat \phi}_R  (.,0);{\hat \phi}_I  (.,0) \right]$ is conserved by construction
and 
can be obtained at any time from the integration of the joint distribution $  P_t \left[ m_e; {\hat \phi}^.  (.) \right]$
over all the initial Fourier coefficients $\vec k>0$ 
\begin{eqnarray} 
p_L[m_e]
  =\left[ \prod_{\vec k >0} \prod_{\alpha=R,I}  \int_{-\infty}^{+\infty} d \hat \phi^{\alpha}(\vec k) \right]
P_t \left[ m_e; {\hat \phi}^.  (.) \right] 
=    \left[ \prod_{\vec k >0} \prod_{\alpha=R,I}  \int_{-\infty}^{+\infty} d \hat \phi^{\alpha}_0(\vec k) \right]
 P_{0}\left[ m_e;{\hat \phi}^._0  (.) \right] 
  \label{pmezerotime}
\end{eqnarray}

For large time $t=+\infty$, the joint probability distribution of Eq. \ref{SoluOUt}
will converge towards the fully factorized steady state
\begin{eqnarray} 
P_t \left[m_e;{\hat \phi}^.  (.) \right]  
&&  \opsimeq_{t \to + \infty} 
p_L[m_e]
 \prod_{\vec k >0} \prod_{\alpha=R,I}  
  \left[ \sqrt{ \frac{L^d \omega_L(\vec k)}{  \pi D    } }
  e^{\displaystyle  - L^d    
 \frac{\omega_L(\vec k) }{ D  }  \left[{\hat \phi}^{\alpha}  (\vec k)  \right]^2   }
 \right]
  \label{SoluOUtinfty}
\end{eqnarray}
that is the analog of Eq. \ref{p1FPOUsolinfty} : here the only surviving information 
concerning the initial condition is the probability distribution $p_L[m_e] $
(instead of partition function $Z_0$ in Eq. \ref{p1FPOUsolinfty}),
while all the Fourier coefficients for $\vec k>0$ and $\alpha=R,I$ have converged towards 
their Gaussian steady states and have no memory of their initial values.

On infinitesimal time-intervals, the joint probability $P_t \left[m_e;{\hat \phi}^.  (.) \right]  $ evolves according
to the multidimensional Fokker-Planck dynamics associated
to the Langevin stochastic differential Equations \ref{LangevinOU} for the Fourier coefficients
\begin{eqnarray} 
 \frac{ \partial P_t \left[m_e;{\hat \phi}^.  (.) \right] }{\partial t}
  =  \sum_{\vec k>0} \sum_{\alpha=R,I}
 \frac{ \partial }{ \partial {\hat \phi}^{\alpha}  (\vec k)} 
 \left[   \omega_L(\vec k) {\hat \phi}^{\alpha}  (\vec k) P_t \left[m_e;{\hat \phi}^.  (.) \right]
 + \frac{D}{2 L^d}  \frac{ \partial P_t \left[m_e;{\hat \phi}^.  (.) \right]}{ \partial {\hat \phi}^{\alpha}  (\vec k)}
\right]  
  \label{FP}
\end{eqnarray}
where one recognizes the individual Fokker-Planck generators of Eq. \ref{FPOU}.
As discussed in Appendix \ref{app_timeDepOU},
 the translation in real-space involves instead
the functional Fokker-Planck Equation of the form of Eq. \ref{FunctionalFP}
with the kernels given in Eqs \ref{KernelOmegaOppositeLaplacian} and \ref{NoiseCorreAvecContrainte}.


\subsection{ RG-flow for the intensive energy ${\hat E}_t[m_e; \hat \phi^.(.)  ] $ of the intensive Fourier-coefficients 
$[m_e; \hat \phi^.(.)  ] $ at time $t$ }

Plugging the form of Eq. \ref{ProbaPhik} 
into the Fokker-Planck Eq. \ref{FP} 
\begin{eqnarray} 
 \frac{ \partial  }{\partial t}\left( e^{\displaystyle  - L^d  {\hat  E}_t \left[ m_e;{\hat \phi}^._.  (.) \right]  } \right)
 =  \sum_{\vec k>0} \sum_{\alpha=R,I}
 \frac{ \partial }{ \partial {\hat \phi}^{\alpha}  (\vec k)} 
 \left[   \omega_L(\vec k) {\hat \phi}^{\alpha}  (\vec k) 
 + \frac{D}{2 L^d} \frac{ \partial }{ \partial {\hat \phi}^{\alpha}  (\vec k)}
\right] e^{\displaystyle  - L^d  {\hat  E}_t \left[ m_e;{\hat \phi}^._.  (.) \right]  } 
  \label{PlugInFP}
\end{eqnarray}
yields  the following RG-flow for the intensive energy 
$ {\hat  E}_t \left[ m_e;{\hat \phi}^._.  (.) \right]$ of the intensive Fourier coefficients 
\begin{eqnarray} 
 \frac{ \partial {\hat  E}_t \left[ m_e;{\hat \phi}^._.  (.) \right] }{\partial t}  
&& =  \sum_{\vec k>0} \sum_{\alpha=R,I}
 \left[   \omega_L(\vec k) {\hat \phi}^{\alpha}  (\vec k) \frac{ \partial  {\hat  E}_t \left[ m_e;{\hat \phi}^._.  (.) \right] }{ \partial {\hat \phi}^{\alpha}  (\vec k)}
 - \frac{D}{2 } 
 \left( \frac{ \partial  {\hat  E}_t \left[ m_e;{\hat \phi}^._.  (.) \right] }{ \partial {\hat \phi}^{\alpha}  (\vec k)} \right)^2
\right] 
\nonumber \\
&& + \frac{1}{L^d} \sum_{\vec k>0} \sum_{\alpha=R,I} \left[ \frac{D}{2} \frac{ \partial^2  {\hat  E}_t \left[ m_e;{\hat \phi}^._.  (.) \right] }{ \partial {\hat \phi}^{\alpha}  (\vec k) \partial {\hat \phi}^{\alpha}  (\vec k)} -  \omega_L(\vec k) \right]
  \label{DynamicsIntensiveE}
\end{eqnarray}
which is the direct analog of Eq. \ref{WKEnergy} concerning the single-variable toy-model,
but with the two replacements $E_t \to L^d {\hat  E}_t $ and $D \to  \frac{D}{2 L^d} $.
As a consequence, the contributions of the second line (involving the second-derivatives terms and the constant terms) are only of order $ O \left( \frac{1}{L^d} \right)$ with respect to the finite contributions
of the first line, so that one obtains at leading order for large $L$
the simplified RG-flow
\begin{eqnarray} 
 \frac{ \partial {\hat  E}_t \left[ m_e;{\hat \phi}^._.  (.) \right] }{\partial t}  
&& =  \sum_{\vec k>0} \sum_{\alpha=R,I}
 \left[   \omega_L(\vec k) {\hat \phi}^{\alpha}  (\vec k) \frac{ \partial  {\hat  E}_t \left[ m_e;{\hat \phi}^._.  (.) \right] }{ \partial {\hat \phi}^{\alpha}  (\vec k)}
 - \frac{D}{2 } 
 \left( \frac{ \partial  {\hat  E}_t \left[ m_e;{\hat \phi}^._.  (.) \right] }{ \partial {\hat \phi}^{\alpha}  (\vec k)} \right)^2
\right] 
+ O \left( \frac{1}{L^d} \right)
  \label{DynamicsIntensiveEleading}
\end{eqnarray}

Plugging the form of Eq. \ref{ProbaPhik} 
both for the time $t$ and for the time $t=0$ 
into Eq. \ref{SoluOUt}
yields
\begin{eqnarray} 
&& e^{\displaystyle  - L^d  {\hat  E}_t \left[ m_e;{\hat \phi}^._.  (.) \right]  }
  \label{PlugInSoluOUt}  \\
&&  = \left[ \prod_{\vec k >0} \prod_{\alpha=R,I}  \int_{-\infty}^{+\infty} d \hat \phi^{\alpha}_0(\vec k) 
\sqrt{ \frac{L^d \omega_L(\vec k)}{  \pi D \left[ 1- e^{- 2t \omega_L(\vec k)}\right]   } }
 \right]
 e^{\displaystyle  - L^d \left( \sum_{\vec k>0} \sum_{\alpha=R,I} \omega_L(\vec k) \frac{\left[ {\hat \phi}^{\alpha}  (\vec k) -  e^{- t \omega_L(\vec k) } {\hat \phi}^{\alpha}_0  (\vec k) \right]^2}{ D \left[ 1- e^{- 2t \omega_L(\vec k)}\right] } 
  + {\hat  E}_0 \left[ m_e;{\hat \phi}^._0  (.) \right] \right) }
\nonumber
\end{eqnarray}
which is the direct analog of Eq. \ref{p1FPOUsolE} concerning the single-variable toy-model.
For large $L \to + \infty$,
 the saddle-point evaluation of the integrals 
 over the initial coefficients $\hat \phi^{\alpha}_0(\vec k) $
yields that the intensive energy ${\hat  E}_t \left[ m_e;{\hat \phi}^.  (.)  \right] $
at time $t$ corresponds to the optimization of the function in the exponential
with respect to all the initial Fourier coefficients $\phi^{\alpha=R,I}_0(\vec k) $
\begin{eqnarray} 
 {\hat  E}_t \left[ m_e;{\hat \phi}^.  (.)  \right] 
 && = \min_{ \{\phi^._0(.)\} } \left(  \sum_{\vec k>0} \sum_{\alpha=R,I} \omega_L(\vec k) \frac{\left[ {\hat \phi}^{\alpha}  (\vec k) -  e^{- t \omega_L(\vec k) } {\hat \phi}^{\alpha}_0  (\vec k) \right]^2}{ D \left[ 1- e^{- 2t \omega_L(\vec k)}\right] } 
  + {\hat  E}_0 \left[ m_e;{\hat \phi}^._0  (.) \right] \right)
  \label{SoluOUtlargeLsaddlemin}
\end{eqnarray}

For $t \to + \infty$, the variables $\{\phi^._0(.)\} $ disappear from the first contribution
and one recovers the intensive energy of the asymptotic steady state of Eq. \ref{SoluOUtinfty}
\begin{eqnarray} 
 {\hat  E}_{t=+\infty} \left[ m_e;{\hat \phi}^.  (.)  \right] 
 && = \min_{ \{\phi^._0(.)\} } \left(  \sum_{\vec k>0} \sum_{\alpha=R,I} \omega_L(\vec k) \frac{\left[ {\hat \phi}^{\alpha}  (\vec k)  \right]^2}{ D  }   + {\hat  E}_0 \left[ m_e;{\hat \phi}^.  (.)  \right] \right)
 \nonumber \\
&& = \sum_{\vec k>0} \sum_{\alpha=R,I} \omega_L(\vec k) \frac{\left[ {\hat \phi}^{\alpha}  (\vec k)  \right]^2}{ D  }   +
\min_{ \{\phi^._0(.)\} } \left(   {\hat  E}_0 \left[ m_e;{\hat \phi}^._0  (.) \right] \right) 
  \label{SoluOUtlargeLsaddlemintinfty}
\end{eqnarray}
where the last term will be directly related to the rate function $i(m_e)$ of the empirical magnetization $m_e$ introduced in Eq. \ref{rateim}  up to the constant $ {\hat E}_0 $ (see Eq. \ref{imcontraction}).


\subsection{RG-flow for the generating function $ {\hat {\cal W}}_t[h_e, {\hat h}^.(.) ] $ 
involving the Fourier coefficients $  [h_e, {\hat h}^.(.) ]$ of the magnetic field}

Using Eq. \ref{OUgeneratingtotdef} concerning the generating function for a given initial field $\phi_0(\vec x)$
with Fourier coefficients $\hat \phi^{\alpha}_0(\vec k) $,
one obtains that the generating function associated to the probability distribution $P_t [ m_e,{\hat \phi}^.(.)] $ of Eq. \ref{SoluOUt}
reads
\begin{small}
\begin{eqnarray} 
&& e^{{\cal W}_t[h(.) ]}  \equiv \int {\cal D} \phi(.)  e^{ \displaystyle\int_{L^d} d^d \vec x h(\vec x) \phi(\vec x) } {\cal P}_t [ \phi(.)]
= \int {\cal D} {\hat \phi}^{.}(.)  e^{ \displaystyle L^d \left[ h_e  m_e
 + 2 \sum_{\vec k >0} \sum_{\alpha=R,I}  {\hat h}^{\alpha}  (  \vec k)  {\hat \phi}^{\alpha} ( \vec k) \right] } P_t [ m_e,{\hat \phi}^.(.)]
 \nonumber \\
 && = 
 \left[ \prod_{\vec k >0} \prod_{\alpha=R,I}  \int_{-\infty}^{+\infty} d \hat \phi^{\alpha}_0(\vec k) \right]
 P_0 [ m_e,{\hat \phi}^._0(.)]
 e^{ \displaystyle L^d \left( h_e m_e
 +  \sum_{\vec k>0} \sum_{\alpha=R,I}
 \left[  2 h^{\alpha}  (\vec k)e^{- t \omega_L(\vec k) } {\hat \phi}^{\alpha}_0  (\vec k)
+   [h^{\alpha}  (\vec k) ]^2 D \left( \frac{1- e^{- 2t \omega_L(\vec k)}}{   \omega_L(\vec k)} \right) 
\right] \right)} 
 \nonumber \\
 && = 
 e^{ \displaystyle L^d  \sum_{\vec k>0} \sum_{\alpha=R,I}
  [h^{\alpha}  (\vec k) ]^2 D \left( \frac{1- e^{- 2t \omega_L(\vec k)}}{   \omega_L(\vec k)} \right) 
}
 \left[ \prod_{\vec k >0} \prod_{\alpha=R,I}  \int_{-\infty}^{+\infty} d \hat \phi^{\alpha}_0(\vec k) \right]
 P_0 [ m_e,{\hat \phi}^._0(.)]
 e^{ \displaystyle L^d \left( h_e m_e
 +  \sum_{\vec k>0} \sum_{\alpha=R,I}
  2 h^{\alpha}  (\vec k)e^{- t \omega_L(\vec k) } {\hat \phi}^{\alpha}_0  (\vec k)
 \right)} 
 \nonumber \\
&& = e^{ \displaystyle L^d   \sum_{\vec k>0} \sum_{\alpha=R,I}
   [h^{\alpha}  (\vec k)  ]^2 D \left( \frac{1- e^{- 2t \omega_L(\vec k)}}{   \omega_L(\vec k)} \right) }
\times e^{ \displaystyle  {\hat {\cal W}}_0[h_e, {\hat h}^{\alpha}_0(\vec k) = h^{\alpha}  (\vec k) e^{- t \omega_L(\vec k) }]}
  \equiv e^{ \displaystyle {\hat {\cal W}}_t[h_e, {\hat h}^.(.) ]} 
  \label{OUgeneratingtotdeft}
\end{eqnarray}
\end{small}
leading to the final result
that involves the generating function ${\hat {\cal W}}_0 $ at time $t=0$ with the  
Fourier coefficients ${\hat h}^{\alpha}_0(\vec k) = h^{\alpha}  (\vec k) e^{- t \omega_L(\vec k) } $
\begin{eqnarray} 
{\hat {\cal W}}_t[h_e, {\hat h}^.(.) ] =
 L^d   \sum_{\vec k>0} \sum_{\alpha=R,I}
   [h^{\alpha}  (\vec k)  ]^2 D \left( \frac{1- e^{- 2t \omega_L(\vec k)}}{   \omega_L(\vec k)} \right) 
+  {\hat {\cal W}}_0[h_e, {\hat h}^{\alpha}_0(\vec k) = h^{\alpha}  (\vec k) e^{- t \omega_L(\vec k) }]
  \label{WOUgeneratingtotdeft}
\end{eqnarray}
where one recognizes the direct analog of the one-dimensional result of Eq. \ref{W1hWK}.

For $t=+\infty$, the asymptotic value 
\begin{eqnarray} 
 {\hat {\cal W}}_{t=+\infty}[h_e, {\hat h}^.(.) ] 
 && =  L^d   \sum_{\vec k>0} \sum_{\alpha=R,I}
   [h^{\alpha}  (\vec k)  ]^2 \frac{D}{   \omega_L(\vec k)} 
   +  {\hat {\cal W}}_0[h_e , {\hat h}^{\alpha}_0(\vec k   = 0]
   \nonumber \\
 && =  L^d   \sum_{\vec k>0} \sum_{\alpha=R,I}
   [h^{\alpha}  (\vec k)  ]^2 \frac{D}{   \omega_L(\vec k)} 
   + L^d w(h_e)  
  \label{OUgeneratingtottinfty}
\end{eqnarray}
involves the generating function associated to the steady state of Eq. \ref{SoluOUtinfty}
for the Fourier coefficients associated to momenta $\vec k>0$ and
$w(h_e) $  is the scaled-cumulant-generating-function of the empirical magnetization $m_e$
discussed in Eqs \ref{SCGFwhme} and \ref{Genemalonek}.

As in the one-dimensional calculation of Eq. \ref{p1FPOUZ},
the Fokker-Planck Equation of Eq. \ref{FP}
for the probability density $P_t \left[m_e;{\hat \phi}^.  (.) \right] $
can be translated via integrations by parts
into the following Linear RG-flow 
\begin{eqnarray} 
\frac{ \partial  }{\partial t} e^{ \displaystyle {\hat {\cal W}}_t[h_e, {\hat h}^.(.) ]}  
&&  = \int {\cal D} {\hat \phi}^{.}(.)  e^{ \displaystyle L^d \left[ h_e  m_e
 + 2 \sum_{\vec k >0} \sum_{\alpha=R,I}  {\hat h}^{\alpha}  (  \vec k)  {\hat \phi}^{\alpha} ( \vec k) \right] } \frac{ \partial  }{\partial t} P_t [ m_e,{\hat \phi}^.(.)]
 \nonumber \\
 &&   =   \sum_{\vec k>0} \sum_{\alpha=R,I} 
 \left[ 2 D L^d [ {\hat h}^{\alpha}  (  \vec k) ]^2-    \omega_L(\vec k) {\hat h}^{\alpha}  (  \vec k) \frac{\partial}{\partial {\hat h}^{\alpha}  (\vec k)} 
 \right]
 e^{ {\hat {\cal W}}_t[h_e, {\hat h}^.(.) ]} 
  \label{dynOUgenerating}
\end{eqnarray}
which is the analog of Eq. \ref{p1FPOUZ} concerning the single-variable toy-model
with the two rescaling $h \to 2 L^d {\hat h}^{\alpha}  (  \vec k) $ and $D \to  \frac{D}{2 L^d} $.

The RG-flow for ${\hat {\cal W}}_t[h_e, {\hat h}^.(.) ]  $ itself is the analog of Eq. \ref{p1FPOUW}
\begin{eqnarray} 
&&  \partial_t   {\hat {\cal W}}_t[h_e, {\hat h}^.(.) ]     =   \sum_{\vec k>0} \sum_{\alpha=R,I} 
 \left[ 2 D L^d [ {\hat h}^{\alpha}  (  \vec k) ]^2 
 -    \omega_L(\vec k) {\hat h}^{\alpha}  (  \vec k) 
  \frac{\partial {\hat {\cal W}}_t[h_e, {\hat h}^.(.) ]}{\partial {\hat h}^{\alpha}  (\vec k)} 
 \right] 
  \label{dynOUgeneratingW}
\end{eqnarray}

Let us now translate the above results for the intensive part $ {\hat W}_t[h_e, {\hat h}^.(.)]$ of ${\hat {\cal W}}_t[h_e, {\hat h}^.(.) ] $ introduced in Eq. \ref{saddle} :
Eq. \ref{WOUgeneratingtotdeft} leads to
\begin{eqnarray} 
{\hat W}_t[h_e, {\hat h}^.(.)]
= \frac{ {\hat {\cal W}}_t[h_e, {\hat h}^.(.) ] }{ L^d  }
=  \sum_{\vec k>0} \sum_{\alpha=R,I}
   [h^{\alpha}  (\vec k)]^2 D \left( \frac{1- e^{- 2t \omega_L(\vec k)}}{   \omega_L(\vec k)} \right)   
   + {\hat W}_0[h_e, {\hat h}^{\alpha}_0(\vec k) = {\hat h}^{\alpha}  (\vec k) e^{- t \omega_L(\vec k) }]
  \label{OUgeneratingtotintensive}
\end{eqnarray}
while the dynamics of Eq. \ref{dynOUgeneratingW}
yields 
\begin{eqnarray} 
&&  \partial_t     {\hat W}_t[h_e, {\hat h}^.(.)]   =   \sum_{\vec k>0} \sum_{\alpha=R,I} 
 \left[ 2 D  [ {\hat h}^{\alpha}  (  \vec k)  ]^2
 -    \omega_L(\vec k) {\hat h}^{\alpha}  (  \vec k) 
  \frac{\partial   {\hat W}_t[h_e, {\hat h}^.(.)]}{\partial {\hat h}^{\alpha}  (\vec k)} 
 \right] 
  \label{dynOUgeneratingWintensive}
\end{eqnarray}


\subsection{ Discussion}

Since the rate function $I_t [ m_e ;\hat \phi^R(.) ; \hat \phi^I(.)]  
= \big(E_t [ m_e ; \hat \phi^R(.) ; \hat \phi^I(.)] - E_0\big )$ introduced in Eq. \ref{RatePhi}
that coincides with the intensive energy $E_t [ m_e ; \hat \phi^R(.) ; \hat \phi^I(.)] $ up to the constant $E_0$
is the Legendre transform of the intensive part ${\hat W}_t[ h_e ; {\hat h}^R (.); {\hat h}^I(.)] $ of ${\hat {\cal W}_t}[h_e ;  {\hat h}^R (.); {\hat h}^I(.)] $ (see Eqs \ref{Legendremax} and \ref{LegendreReci}),
one understands why the simplified RG-flow of Eq. \ref{DynamicsIntensiveEleading}
for the intensive energy $ {\hat  E}_t \left[ m_e;{\hat \phi}^._.  (.) \right] $
is actually the analog of Eq. \ref{RGFlowGamma1} concerning the Laplace transform $\Gamma_t(.)$ of the single variable toy-model,
and why the finite-time solution of Eq. \ref{SoluOUtlargeLsaddlemin}
is also the analog of Eq. \ref{Gamma1hWK} concerning the Laplace transform $\Gamma_t(.)$ of the single variable toy-model.


\section{ Conclusion}

\label{sec_conclusion}

The two main goals of the present paper can be summarized as follows:

(i) the goal of the three sections \ref{sec_WK}, \ref{sec_WM} and \ref{sec_Carosso}
was to give a pedagogical introduction to the stochastic formulation of exact RG-flows
for statistical physicists interested in stochastic processes,
with the final detailed discussion given in subsection \ref{subsec_conclusionStochasticRG}.

(ii) the goal of the three other sections \ref{sec_Realspace}, \ref{sec_Fourier} and \ref{sec_WK}
was to focus on the case where the field theory is defined on the finite volume $L^d$
in order to study the
empirical magnetization $m_e \equiv  \frac{1}{L^d} \int_{L^d}  d^d \vec x \ \phi(\vec x) $
corresponding to the zero-momentum Fourier coefficient of the field,
so that its probability distribution $p_L(m_e)$ can be obtained from the gradual integration over all the other Fourier coefficients associated to non-vanishing-momenta via an appropriate adaptation of the Carosso
stochastic RG.

\vskip 0.3cm

Among many other possible interesting directions for the future, let us mention the following topics :

\vskip 0.3cm

$\bullet$ We have stressed how some exact RG-flows translate into a linear Fokker-Planck evolution
for the probability $P_t[\phi(.)]  $ of the field $\phi(.) $ at the RG-time $t$,
and we have discussed in detail the two examples of  
the Wilson-Kogut RG-scheme and of the Carosso RG-scheme.
This linearity alone is actually already a very remarkable property with respect 
to many other RG procedures written in other contexts
that lead to non-linear RG-flows for the renormalized probability itself.
This point suggests to reconsider all these other contexts to see whether 
a change of perspective could lead instead to linear RG-flows for probabilities
that are much easier to analyze.

\vskip 0.3cm

$\bullet$ The Carosso interpretation of the associated Langevin Stochastic Differential 
equation for the real-space field $\phi_t(\vec x)$ as genuine coarsening-transformations 
that are the analog of the block-spins in discrete-time RG for lattice spins models
is very suggestive and had led Carosso to consider new RG-schemes,
where the Langevin SDE for the field $\phi_t(\vec x)$
corresponds to the stochastic heat equation in the mathematical literature, 
which is also well-known as the Edwards-Wilkinson dynamics in statistical physics, 
so that the coarsening-transformation of the real-space field  
is realized by the Gaussian heat-kernel.
This example where the RG-flow turns out to coincide
 with a standard stochastic dynamics that has been much studied in the past for other purposes,
suggests that other well-known Markov dynamics 
 could be also re-interpreted as appropriate RG-flows in some contexts,
 i.e. that some models could be 'naturally' renormalized via some existing Markov dynamics
 without the need to design ad-hoc RG-schemes that would never appear outside the RG literature.

\vskip 0.3cm

$\bullet$ Many other fascinating links between field theory exact RG procedures and stochastic processes 
have also been analyzed recently,  
in particular in the area of Inverse-RG-transformation related to statistical inference \cite{Klinger_InverseRG,Klinger_BayesianRG,Klinger_BayesianRGFlow,InverseRG_Generative},
in the context of generative diffusion models  
\cite{Cotler_GenerativeDiffusion,InverseRG_Generative,Generative_HigherCumulants,Generative_StochasticQuantization}
and more generally in the context of Machine Learning \cite{QFTMachineLearning},
while the exact RG has also been rephrased as an optimal transport gradient flow for the relative entropy \cite{Cotler_OptimalTransport}.

In conclusion, we feel that all these new perspectives on RG
based on the core concepts of statistical physics (probability, information, entropy, irreversibility, inference, optimization) and on stochastic processes with respect to the RG-time 
(Fokker-Planck evolutions and Langevin SDE involving noise)
open up whole new avenues for future research, both for models without disorder considered
in the present paper,
and for disordered models, where exact RG procedures have also been studied,
either within field theory (see the reviews \cite{Pierre,RFReview} and references therein)
or outside field theory based on the pioneering work of Daniel Fisher \cite{DanielFisher} (see the reviews \cite{Ferenc} and references therein), while the real-space coarse-graining is useful to understand
the emergence of multifractal properties at random critical points \cite{c_nancy,c_cascade}.


\appendix

\section{ Translations between the real-space volume $\vec x \in L^d$ 
and the Fourier lattice $\vec k \in {\mathbb Z}^d $  }

\label{app_fourier}

\subsection{ Fourier-series decomposition of spatial-functions like the field $\phi(\vec x) $
 or the magnetic field $h(\vec x)$  }

The Fourier-series decomposition of the field $\phi(\vec r) $ explained 
in detail in subsection \ref{subsec_fourierseries} of the main text,
 can be applied to any other spatial-function,
in particular to the magnetic field $h(\vec x)$
\begin{eqnarray} 
 h( \vec x) && =  \sum_{\vec k \in {\mathbb Z}^d}   {\hat h}  ( \vec k) e^{i  \frac{ 2 \pi }{L} \vec k . \vec x}  
 \nonumber \\
&& = h_e  + \sum_{\vec k >0} 
 \left[ {\hat h}^R(\vec k)  2 \cos \left(  \frac{ 2 \pi }{L} \vec k . \vec x \right) 
 +  {\hat h}^I(\vec k) 2 \sin \left(  \frac{ 2 \pi }{L} \vec k . \vec x \right) 
 \right]
  \label{Fourierh}
\end{eqnarray}
with the complex Fourier coefficients
\begin{eqnarray} 
{\hat h}  (\vec k)  
 = \frac{1}{L^d} \int_{L^d}  d^d \vec x  h (\vec x) e^{- i  \frac{ 2 \pi }{L} \vec k . \vec x} 
 = {\hat h}^*  (- \vec k) 
  \label{FouriercoefsPerioh}
\end{eqnarray}
with the special case of the zero-momentum coefficient that corresponds to the empirical magnetic field $h_e$
\begin{eqnarray} 
{\hat h}  (\vec k=\vec 0)   = \frac{1}{L^d} \int_{L^d}  d^d \vec x  h (\vec x) \equiv h_e
  \label{FouriercoefsPerioZeroh}
\end{eqnarray}
while the real Fourier coefficients for $\vec k >0$ analogous to Eq. \ref{FouriercoefsPerioab}
read
\begin{eqnarray} 
{\hat h}^R(\vec k)  && =  \frac{1}{L^d} \int_{L^d}  d^d \vec x  h (\vec x) \cos \left(  \frac{ 2 \pi }{L} \vec k . \vec x \right)
 \nonumber \\
{\hat h}^I(\vec k)&& =  \frac{1}{L^d} \int_{L^d}  d^d \vec x  h (\vec x) \sin \left(  \frac{ 2 \pi }{L} \vec k . \vec x \right) 
  \label{FouriercoefsPerioabh}
\end{eqnarray}


\subsection{ Parseval-Plancherel identity for the scalar product of two spatial-functions   }
 
 The Parseval-Plancherel identity for the scalar product of two spatial-functions
 like the field $\phi(\vec x) $
 and the magnetic field $h(\vec x)$ 
\begin{eqnarray} 
\int_{L^d}  d^d \vec x h(\vec x) \phi(\vec x)
&& =\sum_{\vec q \in {\mathbb Z}^d}   {\hat h}  ( \vec q) \sum_{\vec k \in {\mathbb Z}^d}   {\hat \phi}  ( \vec k)   
 \int_{L^d}  d^d \vec x e^{i  \frac{ 2 \pi }{L} (\vec k+\vec q ). \vec x} 
 =\sum_{\vec q \in {\mathbb Z}^d}   {\hat h}  ( \vec q) \sum_{\vec k \in {\mathbb Z}^d}   {\hat \phi}  ( \vec k)   
 L^d \delta_{ \vec k+\vec q , \vec 0} 
 \nonumber \\
 && = L^d \sum_{\vec k \in {\mathbb Z}^d} {\hat h}  ( - \vec k)  {\hat \phi}  ( \vec k)
 \nonumber \\
 && = L^d \left[ h_e m_e
 + 2 \sum_{\vec k >0} \left(  {\hat h}^R  (  \vec k)  {\hat \phi}^R ( \vec k) + {\hat h}^I  (  \vec k)  {\hat \phi}^I ( \vec k)  \right) \right]
    \label{parseval}
\end{eqnarray}
involves the extensive volume $L^d$ as prefactor, since the Fourier coefficients of Eqs \ref{FouriercoefsPerioh}
and \ref{FouriercoefsPerioabh}
are intensive variables.


\subsection{ Energy associated to the local potential $U_t[\phi]$ in terms of the Fourier coefficients   }

In the energy functional of Eq. \ref{ActionSt}, the contribution associated
 to the local potential $U_t[\phi]$ of Eq. \ref{Vlocalt}
\begin{eqnarray} 
{\cal E}^{U}_t[\phi(.)] && \equiv \int_{L^d}  d^d \vec x  U_t[ \phi(\vec x)]  
 = \sum_{n=0}^{+\infty}  \frac{u_{2n}(t)}{(2n)!}\int_{L^d}  d^d \vec x \phi^{2n} (\vec x)
 \nonumber \\
 && = \sum_{n=0}^{+\infty} \frac{u_{2n}}{(2n)!}  \sum_{\vec k_1 \in {\mathbb Z}^d}   {\hat \phi}  ( \vec k_1)   ...
 \sum_{\vec k_{2n} \in {\mathbb Z}^d}   {\hat \phi}  ( \vec k_{2n})   
 \int_{L^d}  d^d \vec x e^{i  \frac{ 2 \pi }{L} (\vec k_1+...+\vec k_{2n} ). \vec x} 
\nonumber \\
&&  = L^d \sum_{n=0}^{+\infty} \frac{u_{2n}}{(2n)!}  \sum_{\vec k_1 \in {\mathbb Z}^d}   {\hat \phi}  ( \vec k_1)   ...
 \sum_{\vec k_{2n} \in {\mathbb Z}^d}   {\hat \phi}  ( \vec k_{2n}) \delta_{\vec k_1+...+\vec k_{2n},\vec 0}
 \equiv L^d \sum_{n=0}^{+\infty}  {\hat E}^{(2n)}_t[\hat \phi(.) ]
  \label{ActionpotSk}
\end{eqnarray}
involves the extensive volume $L^d$ as prefactor,
while the intensive functions ${\hat E}^{(2n)}_t[\hat \phi(.) ] $ of the intensive Fourier coefficients
associated to the various orders $(2n)$ of the local potential $U_t[.]$ read
\begin{eqnarray} 
{\hat E}^{(2n)}_t[\hat \phi(.) ]
  \equiv \frac{u_{2n}}{(2n)!} \sum_{\vec k_1 \in {\mathbb Z}^d}   {\hat \phi}  ( \vec k_1)   ...
 \sum_{\vec k_{2n} \in {\mathbb Z}^d}   {\hat \phi}  ( \vec k_{2n}) \delta_{\vec k_1+...+\vec k_{2n},\vec 0}
  \label{I2n}
\end{eqnarray}

In particular, the quadratic term $2n=2$ corresponding to a special case of the 
Parseval-Plancherel identity of Eq. \ref{parseval}  
\begin{eqnarray} 
  {\hat E}^{(2)}_t[\hat \phi(.)  ] 
&&  = \frac{u_2(t)}{2} \sum_{\vec k \in {\mathbb Z}^d}   {\hat \phi}  ( \vec k)      {\hat \phi}  ( - \vec k) 
 =\frac{u_2(t)}{2}  \left[ m_e^2 + 2 \sum_{\vec k >0}  \left( [{\hat \phi}^R  ( \vec k)]^2 + [{\hat \phi}^I  ( \vec k)]^2\right) \right] \equiv E^{(2)}_t[\hat \phi^R(.) ; \hat \phi^I(.)] 
  \label{I2}
\end{eqnarray}
is diagonal for the real Fourier modes $[\hat \phi^R(.) ; \hat \phi^I(.)] $,
while the quartic term $2n=4$ 
\begin{eqnarray} 
 {\hat E}^{(4)}_t[\hat \phi(.)  ] 
&& = \frac{u_4 (t)}{4!} \sum_{\vec k_1 \in {\mathbb Z}^d}  \sum_{\vec k_2 \in {\mathbb Z}^d} \sum_{\vec k_3 \in {\mathbb Z}^d} 
 {\hat \phi}  ( \vec k_1)   {\hat \phi}  ( \vec k_2)   {\hat \phi}  ( \vec k_3)   {\hat \phi}  ( - \vec k_1- \vec k_2 -\vec k_3)   
  \label{I4}
\end{eqnarray}
will produce interactions between the real Fourier modes $[\hat \phi^R(.) ; \hat \phi^I(.)] $ associated to different momenta.


\subsection{ Fourier-series decomposition involving derivatives the field $\phi(\vec x) $  }

The Fourier-series of the derivatives of the field $\phi( \vec x) $ can be directly obtained from Eq. \ref{FourierPerio}
\begin{eqnarray} 
 \frac{ \partial \phi( \vec x) }{\partial x_{\mu} } && 
 =  \sum_{\vec k \in {\mathbb Z}^d}   {\hat \phi}  ( \vec k) \left( i  \frac{ 2 \pi }{L} k_{\mu} \right) e^{i  \frac{ 2 \pi }{L} \vec k . \vec x} 
  \label{FourierPerioDeri}
\end{eqnarray}
with the simple consequence for the Laplacian $\Delta$
\begin{eqnarray} 
\Delta \phi( \vec x) \equiv \sum_{\mu=1}^d \frac{ \partial^2 \phi( \vec x) }{\partial x_{\mu}^2}  && 
 = - \sum_{\vec k \in {\mathbb Z}^d}   {\hat \phi}  ( \vec k) \left(   \frac{ 4 \pi^2 }{L^2} \sum_{\mu=1}^d k_{\mu}^2 \right) e^{i  \frac{ 2 \pi }{L} \vec k . \vec x} = - \sum_{\vec k \in {\mathbb Z}^d}   {\hat \phi}  ( \vec k) \left(   \frac{ 4 \pi^2 }{L^2} \vec k^2 \right) e^{i  \frac{ 2 \pi }{L} \vec k . \vec x} 
  \label{FourierPerioLaplacian}
\end{eqnarray}
So
the gradient term in the energy functional of Eq. \ref{ActionSt} translates into
\begin{eqnarray} 
 {\cal E}^{Grad}_t[\phi(.)] &&  \equiv  \frac{\Upsilon_t}{2} \int_{L^d}  d^d \vec x  \left[ \vec \nabla  \phi( \vec x) \right]^2 
 = \frac{\Upsilon_t}{2} \int_{L^d}  d^d \vec x \phi( \vec x) (-\Delta ) \phi( \vec x)
\nonumber \\
&&  = L^d \frac{ \Upsilon_t }{2}\sum_{\vec k \in {\mathbb Z}^d} {\hat \phi}  ( - \vec k)  {\hat \phi}  ( \vec k) \left(   \frac{ 4 \pi^2 }{L^2} \vec k^2 \right)
 \equiv L^d {\hat E}^{Grad}_t[\hat \phi(.)]
  \label{ActionSderik}
\end{eqnarray}
where the extensive volume $L^d$ appears as prefactor, 
while the intensive function $E^{Grad}[\hat \phi(.)] $ of the intensive Fourier coefficients $\hat \phi(.) $
\begin{eqnarray} 
 {\hat E}_t^{Grad}[\hat \phi(.)] &&  = \frac{\Upsilon_t }{2}\sum_{\vec k \in {\mathbb Z}^d} {\hat \phi}  ( - \vec k)  {\hat \phi}  ( \vec k) \left(   \frac{ 4 \pi^2 }{L^2} \vec k^2 \right)
\nonumber \\
 && = \Upsilon_t \sum_{\vec k >0} \left[ [{\hat \phi}^R (  \vec k) ]^2+ [{\hat \phi}^I  ( \vec k)]^2 \right]  \left(   \frac{ 4 \pi^2 }{L^2} \vec k^2 \right)
 \equiv {\hat E}^{Grad}[\hat \phi^R(.) ; \hat \phi^I(.)] 
  \label{Ederik}
\end{eqnarray}
is diagonal in terms of the real Fourier modes $[\hat \phi^R(.) ; \hat \phi^I(.)] $.


\section{ Simplest example of the Gaussian model on the volume $L^d$  }

\label{app_gauss}

In this Appendix, we use the simple example of the Gaussian model
to illustrate some notations introduced in the main text. 


\subsection{ Real-space properties to illustrate the notations of section \ref{sec_Realspace}  }

Let us recall the simplest example of the Gaussian model where  
the local potential $U[\phi ] $ of Eq. \ref{Vlocalt} reduces to the quadratic term with coefficient $u_2>0$, 
so that the energy functional of Eq. \ref{ActionS0}
is quadratic
\begin{eqnarray} 
 {\cal E}^{Quadratic}[\phi (.)]  
&&  = \int_{L^d}  d^d \vec x \left( \frac{1}{2} \left[ \vec \nabla  \phi ( \vec x) \right]^2 +  \frac{u_2}{2} \phi ^2(\vec x)  \right)
\nonumber \\
&&  = \frac{1}{2}  \int_{L^d}  d^d \vec x \phi (\vec x) \left(  -\Delta  +  u_2 \right) \phi (\vec x)  
\equiv \frac{1}{2}  \int_{L^d}  d^d \vec x \int_{L^d}  d^d \vec y \phi (\vec x) Q(\vec x, \vec y) \phi (\vec y)  
  \label{calEquadratic}
\end{eqnarray}
where the kernel $ Q(\vec x, \vec y) $ involves the Laplacian operator $\Delta$
\begin{eqnarray} 
 Q(\vec x, \vec y) =\left(  -\Delta  +  u_2 \right) \delta^{(d)} (\vec x - \vec y)
  \label{KernelK}
\end{eqnarray}

The probability distribution of Eq. \ref{ProbaPhiS}
\begin{eqnarray} 
  {\cal P}^{Gauss}[\phi (.) ] && = \frac{ e^{- {\cal E}^{Quadratic}[\phi (.)]} }{  {\cal Z}^{Quadratic} }
  \label{ProbaPhiSGauss}
\end{eqnarray}
can be considered as the continuous version of multivariate Gaussian distributions where the 2-point connected correlation 
\begin{eqnarray} 
W^{(2)Gauss}(\vec x_1, \vec x_2) =C^{(2)}(\vec x_1,\vec x_2) 
=  \langle \phi  (\vec x_1)  \phi  (\vec x_2) \rangle 
   \label{W2gauss}
   \end{eqnarray}
is given by the inverse of the kernel $Q(\vec x, \vec y) $ introduced in Eq. \ref{KernelK}
\begin{eqnarray} 
\delta^{(d)} (\vec x - \vec z) =\int_{L^d}  d^d \vec y Q(\vec x, \vec y) W^{(2)Gauss}(\vec y, \vec z)
= \left(  -\Delta  +  u_2 \right) W^{(2)Gauss}(\vec x, \vec z)
  \label{screenedPoisson}
\end{eqnarray}
This screened Poisson equation can be diagonalized in Fourier space (see Eq. \ref{W2GaussFourier}),
but within the real-space point of view described in the present subsection,
the important point is that  
the ratio of Eq. \ref{Zhratio}
corresponding to the exponential of ${\cal W}[h(.) ] $ of Eq. \ref{Wh}
can be explicitly computed as a function of the arbitrary inhomogeneous magnetic field $h(\vec x)$
\begin{eqnarray} 
 e^{{\cal W}^{Gauss}[h(.) ]} && 
= \frac{\int  {\cal D}\phi (.) e^{- {\cal E}^{Quadratic}[\phi (.)] +  \int_{L^d}  d^d \vec x h(\vec x) \phi (\vec x) } }
{\int  {\cal D}\phi (.) e^{- {\cal E}^{Quadratic}[\phi (.)]} } 
\nonumber \\
&& = e^{ \displaystyle \frac{1}{2} \int d^d \vec x_1 \int d^d \vec x_2 W^{(2)Gauss}(\vec x_1,\vec x_2) h(\vec x_1)  h(\vec x_2)} 
   \label{Generatinghxgauss}
\end{eqnarray}
in terms of the 2-point connected correlation $W^{(2)Gauss}(\vec x_1,\vec x_2) $ of Eq. \ref{W2gauss},
while all the higher connected correlations vanish $W^{(n>2)Gauss}(\vec x_1,\vec x_2,..,\vec x_n) =0$.


\subsection{ Fourier-space properties to illustrate the notations of section \ref{sec_Fourier}  }

\subsubsection{ Independent Gaussian distributions for the Fourier coefficients}

For the Gaussian model of Eq. \ref{calEquadratic}
where the only non-vanishing coefficient is $u_2$,
the intensive energy of Eq. \ref{Ehatphi} in terms the intensive Fourier coefficients reduces to
\begin{eqnarray} 
{\hat E}^{Quadratic} [ m_e ; \hat \phi^R(.) ; \hat \phi^I(.)]  = \frac{u_2 }{2}m_e^2
 +\sum_{\vec k > 0} \left( [{\hat \phi}^R  (  \vec k)]^2 + [{\hat \phi}^I  ( \vec k)]^2 \right) \left(   \frac{ 4 \pi^2 }{L^2} \vec k^2+u_2 \right)
  \label{EQuadratic}
\end{eqnarray}
The probability of Eq. \ref{ProbaPhik} is then fully factorized into independent Gaussian distributions for the intensive Fourier coefficients
$[ m_e ;  \hat \phi^R(.) ; \hat\phi^I(.)] $
\begin{eqnarray} 
  P^{Gauss}[m_e ;  \hat \phi^R(.) ; \hat\phi^I(.)] 
&&  = \frac{ e^{- L^d  {\hat E}^{Quadratic} [m_e ;  \hat \phi^R(.) ; \hat \phi^I(.)] } }
  { \displaystyle \int_{-\infty}^{+\infty} d m_e  \left( \prod_{\vec k >0}  \left[ \int_{-\infty}^{+\infty} d \hat \phi^R(\vec k)
\int_{-\infty}^{+\infty} d \hat \phi^I(\vec k) \right] \right) e^{- L^d  {\hat E}^{Quadratic} [ m_e ; \hat \phi^R(.) ; \hat \phi^I(.)] }}
  \nonumber \\
&&  = \left( \sqrt{ \frac{ u_2 L^d}{ 2 \pi } }  e^{- L^d \frac{u_2 }{2} m_e^2 } \right)
  \prod_{\vec k > 0} \left[ \frac{ L^d \left(   \frac{ 4 \pi^2 }{L^2} \vec k^2+u_2 \right) }{\pi } 
  e^{- L^d \left(   \frac{ 4 \pi^2 }{L^2} \vec k^2+u_2 \right)  \left( [{\hat \phi}^R  (  \vec k)]^2 + [{\hat \phi}^I  ( \vec k)]^2 \right)} \right]
  \label{ProbaPhikGauss}
\end{eqnarray}

Using the Fourier series of Eq. \ref{FourierPerioCosSin}, the real-space connected 2-point correlation 
$W^{(2)}(\vec x, \vec y) $ reads
\begin{eqnarray} 
W^{(2)Gauss}(\vec x, \vec y) && = \langle  \phi( \vec x)   \phi( \vec y) \rangle
 \nonumber \\
 && = \langle m_e^2 \rangle + 4 \sum_{\vec k >0}  \
 \left[\langle [{\hat \phi}^R(\vec k) ]^2\rangle  
 \cos \left(  \frac{ 2 \pi }{L} \vec k . \vec x \right) \cos \left(  \frac{ 2 \pi }{L} \vec k . \vec y \right)
 +  \langle [{\hat \phi}^I(\vec k)]^2 \rangle  
 \sin \left(  \frac{ 2 \pi }{L} \vec k . \vec x \right)   \sin \left(  \frac{ 2 \pi }{L} \vec k . \vec y \right)
  \right]
 \nonumber \\
 && = \frac{1}{ L^d u_2 } + 4 \sum_{\vec k >0}  \frac{1}{2 L^d \left(   \frac{ 4 \pi^2 }{L^2} \vec k^2+u_2 \right)}
 \left[  
 \cos \left(  \frac{ 2 \pi }{L} \vec k . \vec x \right) \cos \left(  \frac{ 2 \pi }{L} \vec k . \vec y \right) 
 +   \sin \left(  \frac{ 2 \pi }{L} \vec k . \vec x \right)    \sin \left(  \frac{ 2 \pi }{L} \vec k . \vec y \right)  \right]
   \nonumber \\
 && = \frac{1}{ L^d u_2 } + \sum_{\vec k >0}  \frac{1}{ L^d \left(   \frac{ 4 \pi^2 }{L^2} \vec k^2+u_2 \right)}
    2 \cos \left(  \frac{ 2 \pi }{L} \vec k . (\vec x - \vec y) \right)
      \nonumber \\
 && = \frac{1}{ L^d u_2 } + \sum_{\vec k >0}  \frac{1}{ L^d \left(   \frac{ 4 \pi^2 }{L^2} \vec k^2+u_2 \right)}
    \left( e^{i  \frac{ 2 \pi }{L} \vec k . (\vec x - \vec y) } +  e^{-i  \frac{ 2 \pi }{L} \vec k . (\vec x - \vec y) } \right)
     \nonumber \\ && 
 =  \sum_{\vec k \in {\mathbb Z}^d }  \frac{1}{ L^d \left(   \frac{ 4 \pi^2 }{L^2} \vec k^2+u_2 \right)}
     e^{i  \frac{ 2 \pi }{L} \vec k . (\vec x - \vec y) }  
  \label{W2GaussFourier}
\end{eqnarray}
that satisfies the screened Poisson Eq. \ref{screenedPoisson} as it should
\begin{eqnarray} 
\left(  -\Delta  +  u_2 \right) W^{(2)Gauss}(\vec x, \vec y)
&& = \sum_{\vec k \in {\mathbb Z}^d }  \frac{1}{ L^d \left(   \frac{ 4 \pi^2 }{L^2} \vec k^2+u_2 \right)}
   \left(  -\Delta  +  u_2 \right)  e^{i  \frac{ 2 \pi }{L} \vec k . (\vec x - \vec y) } 
\nonumber \\
&&    = \frac{1}{ L^d } \sum_{\vec k \in {\mathbb Z}^d }    e^{i  \frac{ 2 \pi }{L} \vec k . (\vec x - \vec y) } 
=\delta^{(d)} (\vec x - \vec y) 
  \label{screenedPoissonFourier}
\end{eqnarray}


\subsubsection{ Corresponding generating functions }

Since the joint distribution $P^{Gauss}[ m_e ; \hat \phi^R(.) ; \hat\phi^I(.)]  $ of Eq. \ref{ProbaPhikGauss}
is factorized into independent Gaussian distributions for the Fourier coefficients
$[ m_e;\hat \phi^R(.) ; \hat\phi^I(.)] $,
the generating function of Eq. \ref{Zhk} involving the Fourier coefficients $ [ h_e; {\hat h}^R (.); {\hat h}^I(.)]$ of the magnetic field $h(\vec x)$
reduces to
\begin{eqnarray} 
 e^{{\hat {\cal W}}^{Gauss}[ h_e; {\hat h}^R (.); {\hat h}^I(.)]}
&& = \int D {\hat \phi}^R(.) \int D {\hat \phi}^I(.)  P^{Gauss}[ \hat \phi^R(.) ; \hat\phi^I(.)] 
e^{ \displaystyle L^d \left[  {\hat h}^R  (  \vec 0)  {\hat \phi}^R  ( \vec 0)
 + 2 \sum_{\vec k >0} \left(  {\hat h}^R  (  \vec k)  {\hat \phi}^R ( \vec k) + {\hat h}^I  (  \vec k)  {\hat \phi}^I ( \vec k)  \right) \right]}
 \nonumber \\
 && = e^{\displaystyle L^d \left[ \frac{ h_e^2}{ 2  u_2}   
 +  \sum_{\vec k >0} \frac{ [{\hat h}^R  (  \vec k) ]^2+ [{\hat h}^I  ( \vec k) ]^2 }{ \left(   \frac{ 4 \pi^2 }{L^2} \vec k^2+u_2 \right)}  \right] } 
  \label{Zhkgauss}
\end{eqnarray}
When plugging the 2-point correlation $W^{(2)Gauss}(\vec x, \vec y) $ of Eq. \ref{W2GaussFourier}
into the real-space expression of the generating function ${\cal W}^{Gauss}[h(.) ] $ Eq. \ref{Generatinghxgauss}
\begin{eqnarray} 
 {\cal W}^{Gauss}[h(.) ] && =\frac{1}{2} \int d^d \vec x_1 \int d^d \vec x_2 W^{(2)Gauss}(\vec x_1,\vec x_2) h(\vec x_1)  h(\vec x_2)
 \nonumber \\
&& = \frac{1}{2} \int d^d \vec x_1 \int d^d \vec x_2  h(\vec x_1)  h(\vec x_2)
 \sum_{\vec k \in {\mathbb Z}^d }  \frac{1}{ L^d \left(   \frac{ 4 \pi^2 }{L^2} \vec k^2+u_2 \right)}
     e^{i  \frac{ 2 \pi }{L} \vec k . (\vec x - \vec y) } 
     \nonumber \\
     && =  \frac{L^d}{2} 
 \sum_{\vec k \in {\mathbb Z}^d }  \frac{{\hat h}(\vec k) {\hat h}(- \vec k)}{  \left(   \frac{ 4 \pi^2 }{L^2} \vec k^2+u_2 \right)}
 =  L^d \left[ \frac{{\hat h}^2(\vec 0) }{ 2 u_2 }
 +  \sum_{\vec k >0 }  \frac{[{\hat h}^R(\vec k)]^2+ [{\hat h}^I( \vec k)]^2}{  \left(   \frac{ 4 \pi^2 }{L^2} \vec k^2+u_2 \right)}\right] = {\hat {\cal W}}^{Gauss}[  {\hat h}^R (.); {\hat h}^I(.)]
   \label{Generatinghxgaussfourier}
\end{eqnarray}
one obtains indeed the same expression as ${\hat {\cal W}}^{Gauss}[  {\hat h}^R (.); {\hat h}^I(.)] $ of Eq. \ref{Zhkgauss} as it should.


\section{ Generalization to RG-schemes involving time-dependent Ornstein-Uhlenbeck processes } 

\label{app_timeDepOU}

In this Appendix, we describe how the RG-scheme described in section \ref{sec_RGVolume}
can be generalized to the case where the parameters of the Ornstein-Uhlenbeck processes
depend on the RG-time $t$, in order to include the special case of the Wilson-Kogut choice of Eq. \ref{WKDq}.


\subsection{ Fourier coefficients 
${\hat \phi}_t^{\alpha=R,I}  (\vec k>0)  $ : OU-processes 
with time-dependent parameters $[\omega_L(\vec k,t);D_L(\vec k,t) ] $   }

The OU-process of Eq. \ref{LangevinOU} with time-independent positive parameters $[\omega_L(\vec k)>0;D_L(\vec k)>0 ] $ can be generalized to the case of time-dependent positive parameters $[\omega_L(\vec k,t)>0;D_L(\vec k,t)>0 ] $
as follows :
the Fourier coefficient ${\hat \phi}^{\alpha}_t  (\vec k) $ with 
 $\vec k>0$ and $\alpha=R,I$ satisfies the Langevin SDE
\begin{eqnarray} 
\partial_t {\hat \phi}^{\alpha}_t  (\vec k) && = - \omega_L(\vec k,t) {\hat \phi}^{\alpha}_t  (\vec k)  
+   {\hat \eta}^{\alpha}_t (\vec k) \ \ \ \text{for $\vec k>0$ and $\alpha=R,I$} \ \ 
  \label{LangevinOUtimedep}
\end{eqnarray}
where the joint distribution of Fourier coefficients ${\hat \eta}^{\alpha}_t (\vec k) $ of the noises 
\begin{eqnarray} 
{\mathbb P} \left[ {\hat \eta}^._{.} (.)  \right] = e^{ \displaystyle -  \int_0^T  dt 
 \sum_{ \vec k>0} \frac{  [{\hat \eta}_t^R(\vec k)]^2 +[{\hat \eta}_t^I(\vec k)]^2 }{ 4  D_L(\vec k,t) } }
  \label{NoisesFourierTimeDep}
\end{eqnarray}
correspond to vanishing averages and to delta-correlations of amplitude $ 2  D_L(\vec k,t)$
\begin{eqnarray} 
{\mathbb E}\left( {\hat \eta}^{\alpha}_t (\vec k)  \right) && =0
\nonumber \\
{\mathbb E}\left( {\hat \eta}^{\alpha}_t (\vec k) {\hat \eta}^{\alpha'}_{\tau} (\vec q) \right) 
&& = 2  D_L(\vec k,t)  \delta_{\vec k, \vec q} \delta_{\alpha,\alpha'} \delta(t-\tau) 
  \label{AvCorreNoisesFourierTimeDep}
\end{eqnarray}

\subsection{ Statistical properties of the Fourier coefficient ${\hat \phi}^{\alpha}_t  (\vec k) $ at time $t$ 
for a given initial condition ${\hat \phi}^{\alpha}_0  (\vec k) $ }

For a given initial condition ${\hat \phi}^{\alpha}_{t=0}  (\vec k) $ at $t=0$,
the Langevin SDE of Eq. \ref{LangevinOUtimedep} for the Fourier coefficient ${\hat \phi}^{\alpha}_t  (\vec k) $
can be integrated to obtain the solution 
\begin{eqnarray} 
{\hat \phi}^{\alpha}_t  (\vec k) ={\hat \phi}^{\alpha}_0  (\vec k) e^{- \int_0^t d\tau \omega_L(\vec k,\tau) } 
+  \int_0^t d\tau e^{- \int_{\tau}^t ds \omega_s(\vec k) } {\hat \eta}^{\alpha}_{\tau} (\vec k)
  \label{IntegLangevinTimeDep}
\end{eqnarray}
So its averaged value over the noise generalizing Eq. \ref{OUaverage} reads
\begin{eqnarray} 
{\hat \phi}^{\alpha (av)}_t  (\vec k) \equiv {\mathbb E}\left( {\hat \phi}^{\alpha}_t  (\vec k) \right) 
= {\hat \phi}^{\alpha}_0  (\vec k) e^{- \int_0^t d\tau \omega_L(\vec k,\tau) } 
  \label{OUaverageTimeDep}
\end{eqnarray}
while its variance generalizing Eq. \ref{OUvariance} is given by
\begin{eqnarray} 
\Sigma_t(\vec k) \equiv {\mathbb E}  \left(  \left[ {\hat \phi}^{\alpha}_t(\vec k) - {\mathbb E}\left( {\hat \phi}^{\alpha}_t  (\vec k) \right) \right]^2 \right)  
&& =    \int_0^t d\tau_1 e^{- \int_{\tau_1}^t \omega_L(\vec k,s_1) } 
\int_0^t d\tau_2 e^{- \int_{\tau_2}^t \omega_L(\vec k,s_2) }
{\mathbb E}\left( {\hat \eta}^{\alpha}_{\tau_1} (\vec k) {\hat \eta}^{\alpha}_{\tau_2} (\vec k) \right)
\nonumber \\
&& =   \int_0^t d\tau_1 e^{- \int_{\tau_1}^t \omega_L(\vec k,s_1) } 
\int_0^t d\tau_2 e^{- \int_{\tau_2}^t \omega_L(\vec k,s_2) }
2 D_{\tau_1}(\vec k)   \delta(\tau_2-\tau_1) 
\nonumber \\
&& = 2  \int_0^t d\tau D_L(\vec k,\tau) e^{- 2 \int_{\tau}^t ds \omega_{s}(\vec k) } 
  \label{OUvarianceTimeDep}
\end{eqnarray}
with their dynamics 
\begin{eqnarray} 
\partial_t {\hat \phi}^{\alpha (av)}_t  (\vec k) && = -\omega_L(\vec k,t) {\hat \phi}^{\alpha (av)}_t  (\vec k) 
\nonumber \\
\partial_t\Sigma_t(\vec k) && = - 2 \omega_L(\vec k,t) \Sigma_t(\vec k)+ 2  D_L(\vec k,t)  
  \label{dynOUaverageVarianceTimeDep}
\end{eqnarray}

Since the Fourier coefficient ${\hat \phi}^{\alpha}_t  (\vec k) $ of Eq. \ref{IntegLangevinTimeDep}
depends linearly on the noise
with the gaussian measure of Eq. \ref{NoisesFourierTimeDep},
it is also a Gaussian process, so that the averaged value and the variance computed in Eqs \ref{OUaverageTimeDep} and \ref{OUvarianceTimeDep}
are sufficient to write:

(i) the generating function generalizing Eq. \ref{SingleModePropagatorOUgenerating}
\begin{eqnarray} 
 {\mathbb E}\left( e^{\lambda {\hat \phi}^{\alpha}_t  (\vec k)} \vert {\hat \phi}^{\alpha}_t  (\vec k,0) \right)  
&&  = e^{ \displaystyle \lambda {\hat \phi}^{\alpha (av)}_t  (\vec k)
+  \frac{ \lambda^2}{2 } \Sigma_t(\vec k) } 
\nonumber \\
&& = e^{ \displaystyle \lambda {\hat \phi}^{\alpha}_0  (\vec k) e^{- \int_0^t d\tau \omega_L(\vec k,\tau) } 
+   \lambda^2  \int_0^t d\tau D_L(\vec k,\tau) e^{- 2 \int_{\tau}^t ds \omega_{s}(\vec k) }  } 
  \label{SingleModePropagatorOUgeneratingTimeDep}
\end{eqnarray}

(ii) the finite-time propagator generalizing Eq. \ref{SingleModePropagatorOU}
\begin{eqnarray} 
{\mathbb p}_t^{[k]} [{\hat \phi}^{\alpha}  (\vec k) \vert {\hat \phi}^{\alpha}_0  (\vec k)]
&& = \frac{1}{ \sqrt{2 \pi \Sigma_t(\vec k) } }  
\  e^{\displaystyle -  \frac{\left[ {\hat \phi}^{\alpha}  (\vec k) -  {\hat \phi}^{\alpha (av)}_t  (\vec k) \right]^2}{ 2 \Sigma_t(\vec k)}}
\nonumber \\
&&  =\sqrt{ \frac{1 }{ 4 \pi      \int_0^t d\tau D_L(\vec k,\tau)e^{- 2 \int_{\tau}^t ds \omega_{s}(\vec k) }   }  }
\  e^{\displaystyle -   \frac{\left[ {\hat \phi}^{\alpha}  (\vec k) - {\hat \phi}^{\alpha}_0  (\vec k) e^{- \int_0^t d\tau \omega_L(\vec k,\tau) }  \right]^2}
 { 4     \int_0^t d\tau D_L(\vec k,\tau)e^{- 2 \int_{\tau}^t ds \omega_{s}(\vec k) }}}
  \label{SingleModePropagatorOUTimeDep}
\end{eqnarray}
satisfying the Fokker-Planck dynamics with time-dependent coefficients with respect to Eq. \ref{FPOU}
\begin{eqnarray} 
  \partial_t {\mathbb p}_t^{[k]} [{\hat \phi}^{\alpha}  (\vec k) \vert {\hat \phi}^{\alpha}_0  (\vec k)]
 = \frac{ \partial}{\partial {\hat \phi}^{\alpha}  (\vec k)} 
 \left[  \omega_L(\vec k,t) {\hat \phi}^{\alpha}  (\vec k){\mathbb p}_t^{[k]} [{\hat \phi}^{\alpha}  (\vec k) \vert {\hat \phi}^{\alpha}_0  (\vec k)] 
 + D_L(\vec k,t)  \frac{ \partial {\mathbb p}_t^{[k]} [{\hat \phi}^{\alpha}  (\vec k) \vert {\hat \phi}^{\alpha}_0  (\vec k)]}{\partial {\hat \phi}^{\alpha}  (\vec k)}
 \right] 
  \label{FPOUTimeDep}
\end{eqnarray}

Here the convergence of Eq. \ref{SingleModePropagatorOUTimeDep} for $t \to + \infty$
towards the Gaussian asymptotic state for any initial condition ${\hat \phi}^{\alpha}_0  (\vec k) $
\begin{eqnarray} 
{\mathbb p}_t^{[k]} [{\hat \phi}^{\alpha}_t  (\vec k) \vert {\hat \phi}^{\alpha}_0  (\vec k)]
&& \opsimeq_{t \to + \infty} \frac{1}{ \sqrt{2 \pi \Sigma_{\infty}(\vec k) } }  
 e^{ -  \frac{\left[ {\hat \phi}^{\alpha}  (\vec k)  \right]^2}{ 2 \Sigma_{\infty}(\vec k)}}
  \label{SingleModePropagatorOUTimeDepSteady}
\end{eqnarray}
requires the divergence of the integral governing the relaxation of the averaged value in Eq. \ref{OUaverageTimeDep}
\begin{eqnarray} 
\lim_{t \to + \infty} \left( \int_0^t d\tau \omega_L(\vec k,\tau) \right) = +\infty 
  \label{OUaverageTimeDepRelaxation}
\end{eqnarray}
and the convergence of the integral corresponding to the asymptotic variance in Eq. \ref{OUvarianceTimeDep}
\begin{eqnarray} 
\Sigma_{\infty}(\vec k) = 2  \lim_{t \to + \infty} \left( \int_0^{t} d\tau D_L(\vec k,\tau) e^{- 2 \int_{\tau}^t ds \omega_{s}(\vec k) } \right) <+\infty
  \label{OUvarianceTimeDepinfty}
\end{eqnarray}

In order to decide what choices of the time-dependent parameters $[\omega_L(\vec k,t);D_L(\vec k,t) ] $
might be interesting for these OU-processes in Fourier space,
it is useful in the next subsections to analyze the real-space interpretation.


\subsection{ Langevin SDE : from the Fourier-space towards the real-space  }

\subsubsection{ Statistical properties of the real-space noise $\eta_t(\vec x)  $ }

The real-space noise $\eta_t(\vec x) $
as reconstructed from the linear combination of
its Fourier coefficients $[{\hat \eta}^R_t(\vec k),{\hat \eta}^I_t(\vec k) ] $ 
via Eq. \ref{FourierPerioCosSin} using ${\hat \eta}_R(\vec 0,t)= 0 $
will be also a Gaussian process with vanishing average
\begin{eqnarray} 
{\mathbb E}\left( \eta_t(\vec x)  \right) 
&& = 0
  \label{WhiteNoisesRealSpace}
\end{eqnarray}
while its correlations can be evaluated using Eq. \ref{AvCorreNoisesFourierTimeDep}
 \begin{eqnarray} 
&& {\mathbb E}\bigg(     \eta_t( \vec x)  \eta_{\tau}( \vec y)  \bigg)
\nonumber \\
&&      = 4 \sum_{\vec k >0} \sum_{\vec q >0}   {\mathbb E}\bigg(   
 \left[    {\hat \eta}^R_t (\vec k)    \cos \left(  \frac{ 2 \pi }{L} \vec k . \vec x \right) 
 +    {\hat \eta}^I_t (\vec k)   \sin \left(  \frac{ 2 \pi }{L} \vec k . \vec x \right)  \right] 
 \left[    {\hat \eta}^R_{\tau} (\vec q)    \cos \left(  \frac{ 2 \pi }{L} \vec q . \vec y \right) 
 +    {\hat \eta}^I_{\tau} (\vec q)   \sin \left(  \frac{ 2 \pi }{L} \vec q . \vec y \right)  \right] 
   \bigg)
       \nonumber \\ 
  && = 4 \sum_{\vec k >0}\sum_{\vec q >0} 
\left[ {\mathbb E}\left(    {\hat \eta}^R_t (\vec k)  {\hat \eta}^R_{\tau} (\vec q) \right)
    \cos \left(  \frac{ 2 \pi }{L} \vec k . \vec x \right) \cos \left(  \frac{ 2 \pi }{L} \vec q . \vec y \right) 
 +   {\mathbb E}\left(    {\hat \eta}^I_t (\vec k)  {\hat \eta}^I_{\tau} (\vec q) \right) \sin \left(  \frac{ 2 \pi }{L} \vec k . \vec x \right)   \sin \left(  \frac{ 2 \pi }{L} \vec q . \vec y \right) 
 \right]  
    \nonumber \\
&&      = 4 \sum_{\vec k >0}  2 D_L(\vec k,t)   \delta(t-\tau)
 \left[     \cos \left(  \frac{ 2 \pi }{L} \vec k . \vec x \right)  \cos \left(  \frac{ 2 \pi }{L} \vec k . \vec y \right) 
 +    \sin \left(  \frac{ 2 \pi }{L} \vec k . \vec x \right)    \sin \left(  \frac{ 2 \pi }{L} \vec k . \vec y \right)  \right] 
   \bigg) 
     \nonumber \\
 && = \delta(t-\tau)  8 \sum_{\vec k >0}  D_L(\vec k,t)    \cos \left(  \frac{ 2 \pi }{L} \vec k . (\vec x -\vec y) \right)  
 \equiv  \delta(t-\tau) 2 {\cal D}_L(\vec x- \vec y,t)
  \label{GeneralNoisesRealSpace}
\end{eqnarray}
so the correlations of the real-space noise are still 'delta-correlated-in-time' 
but are governed by the spatial-translation-invariant kernel
\begin{eqnarray} 
{\cal D}_L(\vec x- \vec y,t) \equiv 4 \sum_{\vec k >0}  D_L(\vec k,t)    \cos \left(  \frac{ 2 \pi }{L} \vec k . (\vec x -\vec y) \right)  
  \label{Dkernel}
\end{eqnarray}


\subsubsection{ Langevin SDE for the real-space field $\phi_t( \vec x) $  }

The real-space field $ \phi_t( \vec x) $ 
as reconstructed from
the linear combination of
 the Fourier coefficients $[{\hat \phi}^R(\vec k),{\hat \phi}^I_t(\vec k) ] $ 
via Eq. \ref{FourierPerioCosSint} 
satisfies the following Langevin Equation obtained from Eqs \ref{LangevinOUtimedep}
that involves the real-space noise $\eta_t(\vec x) $ discussed above
\begin{small}
 \begin{eqnarray} 
\partial_t \phi_t( \vec x) &&   
= \sum_{\vec k >0} 
 \left[ \left( \partial_t {\hat \phi}^R_t(\vec k) \right) 2 \cos \left(  \frac{ 2 \pi }{L} \vec k . \vec x \right) 
 + \left( \partial_t {\hat \phi}^I_t(\vec k)  \right) 2 \sin \left(  \frac{ 2 \pi }{L} \vec k . \vec x \right) 
 \right] 
\nonumber \\ &&
= -  2 \sum_{\vec k >0}  \omega_L(\vec k,t)
 \left[  {\hat \phi}^R_t  (\vec k)     \cos \left(  \frac{ 2 \pi }{L} \vec k . \vec x \right) 
 +   {\hat \phi}^I_t  (\vec k)    \sin \left(  \frac{ 2 \pi }{L} \vec k . \vec x \right) 
 \right] 
   + \sum_{\vec k >0} 
 \left[   {\hat \eta}^R_t (\vec k)  2 \cos \left(  \frac{ 2 \pi }{L} \vec k . \vec x \right) 
 +   {\hat \eta}^I_t (\vec k) 2 \sin \left(  \frac{ 2 \pi }{L} \vec k . \vec x \right) 
 \right] 
 \nonumber \\
 &&   = -  2 \sum_{\vec k >0}  \omega_L(\vec k,t )\frac{1}{L^d} \int_{[0,L]^d}  d^d \vec y  \phi_t (\vec y)
 \left[   \cos \left(  \frac{ 2 \pi }{L} \vec k . \vec y \right)    \cos \left(  \frac{ 2 \pi }{L} \vec k . \vec x \right) 
 +   \sin \left(  \frac{ 2 \pi }{L} \vec k . \vec y \right)    \sin \left(  \frac{ 2 \pi }{L} \vec k . \vec x \right) 
 \right] 
   + \eta_t(\vec x)
 \nonumber \\
 && =- \int_{[0,L]^d}  d^d \vec y  \phi_t (\vec y) \frac{1}{L^d} \sum_{\vec k >0}  \omega_L(\vec k,t)
2   \cos \left(  \frac{ 2 \pi }{L} \vec k . (\vec x-\vec y ) \right)    
 + \eta_t(\vec x)  \nonumber \\
 && \equiv  - \int_{[0,L]^d}  d^d \vec y  \Omega_L(\vec x-\vec y,t)  \phi_t (\vec y)     + \eta_t(\vec x)
  \label{LangevinRealspacet}
\end{eqnarray}
\end{small}
So the first term of this Langevin SDE involves the translation-invariant kernel 
 \begin{eqnarray} 
 \Omega_L(\vec x-\vec y,t) \equiv   \frac{1}{L^d} \sum_{\vec k >0}  \omega_L(\vec k)
2   \cos \left(  \frac{ 2 \pi }{L} \vec k . (\vec x-\vec y) \right)    
   \label{KernelOmega}
\end{eqnarray}

In summary, the Langevin SDE of Eq. \ref{LangevinRealspacet}
for the real-space field $\phi_t( \vec x) $
involves the kernel $  \Omega_L(\vec x-\vec y,t)$ of Eq. \ref{KernelOmega}
and the kernel ${\cal D}_L(\vec x- \vec y,t) $ of Eq. \ref{Dkernel}
that governs the correlations of the noise in Eq. \ref{GeneralNoisesRealSpace}.



\subsection{ Functional RG-flows in real-space in terms of the two kernels 
$\Omega_L(\vec x-\vec y,t) $ and ${\cal D}_L( \vec x-\vec y,t) $}

The real-space Langevin SDE of Eq. \ref{LangevinRealspacet}
translates into
the following functional Fokker-Planck RG-flow for the probability distribution ${\cal P}_t[ \phi (.) ]$
of the real-space field configuration $\phi(\vec x)$ at time $t$
 \begin{eqnarray} 
\partial_t {\cal P}_t[ \phi (.) ] =
  \int d^d \vec x \frac{ \partial   }{ \partial \phi(\vec x)} 
\bigg( {\cal P}_t[ \phi (.) ] \int d^d \vec y \Omega_L(\vec x-\vec y,t) \phi(\vec y) \bigg)   
   + \int d^d \vec x 
    \int d^d \vec y {\cal D}_L( \vec x-\vec y,t) \frac{ \partial^2   {\cal P}_t[\phi(.)] }{ \partial \phi(\vec x)\partial \phi(\vec y)} 
     \label{FunctionalFP}
\end{eqnarray}
where the  two real-space translation-invariant kernels $\Omega_L(\vec x-\vec y,t) $ 
and ${\cal D}_L( \vec x-\vec y,t) $ have been discussed in Eqs \ref{KernelOmega}
 and \ref{Dkernel}.
From the perspective of the Wegner-Morris Continuity Eq. \ref{PtdtAdvection},
this corresponds to the choice of the advective velocity
\begin{eqnarray} 
 V_t[\vec x,\phi(.)] && = - \int d^d \vec y \Omega_L(\vec x-\vec y,t) \phi(\vec y)
   - \int d^d \vec y {\cal D}_L( \vec x-\vec y,t) \frac{ \partial \ln  {\cal P}_t[\phi(.)] }{ \partial \phi(\vec y)}
  \label{currentforDiffusion}
\end{eqnarray}
that generalizes the choice of Eq. \ref{vWKscheme}
concerning the single-variable toy-model.

As in the single-variable toy-model (see the summarizing discussion in subsection \ref{subsec_1dflows}),
the linear Fokker-Planck RG-flow of Eq. \ref{FunctionalFP}
can be translated into non-linear RG-flows for the energy ${\cal E}_t[ \phi (.) ] $ or for $\Gamma_t[ \Phi (.) ] $, but into linear RG-flows for the generating functions ${\cal Z}_t[ h (.) ] = {\cal Z}_0 e^{{\cal W}_t[ h (.) ]}$ and 
for ${\cal W}_t[ h (.) ]  $
as we now describe.

Via integration by parts, the Fokker-Planck RG-flow of Eq. \ref{FunctionalFP}
yields the following RG-flow for the exponential of generating function ${\cal W}_t[h(.)] $,

\begin{eqnarray} 
\partial_t  e^{ {\cal W}_t[h(.)] } && =  \int  {\cal D}\phi(.)    e^{ \int  d^d \vec z h(\vec z) \phi(\vec z) }\partial_t  {\cal P}_t[\phi(.) ]
 \nonumber \\
 && =     \int d^d \vec x \int  {\cal D}\phi(.)    e^{ \int  d^d \vec z h(\vec z) \phi(\vec z) }
 \frac{ \partial   }{ \partial \phi(\vec x)} 
\bigg( {\cal P}_t[ \phi (.) ] \int d^d \vec y \Omega_L(\vec x-\vec y,t) \phi(\vec y) \bigg)   
 \nonumber \\
 &&  + \int d^d \vec x     \int d^d \vec y {\cal D}_L( \vec x-\vec y,t)
    \int  {\cal D}\phi(.)    e^{ \int  d^d \vec z h(\vec z) \phi(\vec z) } \frac{ \partial^2   {\cal P}_t[\phi(.)] }{ \partial \phi(\vec x)\partial \phi(\vec y)} 
\nonumber \\
&& 
=  - \int d^d \vec x h(\vec x) \int d^d \vec y \Omega_L(\vec x-\vec y,t)
 \int  {\cal D}\phi(.) \    e^{ \int  d^d \vec z h(\vec z) \phi(\vec z) } 
{\cal P}_t[ \phi (.) ]  \phi(\vec y)
\nonumber \\
&&
 -  \int d^d \vec x h(\vec x)   \int d^d \vec y{\cal D}_L( \vec x-\vec y,t)
  \int  {\cal D}\phi(.) \    e^{ \int  d^d \vec z h(\vec z) \phi(\vec z) } 
 \frac{ \partial   {\cal P}_t[\phi(.)] }{ \partial \phi(\vec y)}
 \nonumber \\
  && 
= -  \int d^d \vec x h(\vec x) \int d^d \vec y \Omega_L(\vec x-\vec y,t)
\frac{\partial}{ \partial h(\vec y) } e^{ {\cal W}_t[h(.)] }  
 +  e^{ {\cal W}_t[h(.)] }  \int d^d \vec x    \int d^d \vec y h(\vec x) {\cal D}_L( \vec x-\vec y,t) h(\vec y) 
   \label{WhtdynFPcalcul}
\end{eqnarray}
that generalizes Eq. \ref{p1FPOUZ} concerning the single-variable toy-model.

The corresponding RG-flow for ${\cal W}_t[h(.)]  $ generalizes Eq. \ref{p1FPOUW}
\begin{eqnarray} 
\partial_t   {\cal W}_t[h(.)]  
= - \int d^d \vec x \int d^d \vec y h(\vec x)  \Omega_L( \vec x-\vec y,t) \frac{\partial {\cal W}_t[h(.)]  }{ \partial h(\vec y) }  
 +  \int d^d \vec x \int d^d \vec y h(\vec x)  {\cal D}_L( \vec x-\vec y,t)  h(\vec y)
   \label{WhtdynFP}
\end{eqnarray}


\subsection{ Special case of the Carosso RG-scheme for the volume $L^d$ described in section \ref{sec_RGVolume} of the main text}

\label{subsec_CarossoRealSpace}

The Carosso RG-scheme of Eq. \ref{LangevinCarosso} involving the Laplacian operator $\Delta$
corresponds to the case
 \begin{eqnarray} 
 \Omega_L(\vec x-\vec y,t)   = - \Delta \delta^{(d)} (\vec x - \vec y) 
   \label{KernelOmegaOppositeLaplacian}
\end{eqnarray}
with its corresponding eigenvalues of Eq. \ref{omega}
 \begin{eqnarray} 
  \omega_L(\vec k)  \equiv  \frac{ 4 \pi^2 }{L^2} \vec k^2
   \label{omegaLapla}
\end{eqnarray}

The special case of Eqs \ref{WhiteNoisesFourierDistri} 
considered in the main text corresponds to Eq. \ref{DL}
\begin{eqnarray} 
 D_L(\vec k, t) = D_L \equiv \frac{D}{2 L^d}  
  \label{DLbis}
\end{eqnarray}
so that the translation-invariant kernel of Eq. \ref{Dkernel} is time-independent and reduces to
 \begin{eqnarray} 
  {\cal D}_L(\vec x-\vec y) 
  &&    
   = \frac{2 D}{L^d}   \sum_{\vec k >0}     \cos \left(  \frac{ 2 \pi }{L} \vec k . (\vec x - \vec y) \right)  
   =  \frac{ D}{L^d}   \sum_{\vec k \ne \vec 0}     e^{i \frac{ 2 \pi }{L} \vec k . (\vec x - \vec y) }
   = D \left[    \sum_{\vec k \in {\mathbb Z}^d}     e^{i \frac{ 2 \pi }{L} \vec k . (\vec x - \vec y) } - \frac{1}{L^d} \right]
    \nonumber \\
  &&       =  D \left[    \delta^{(d)} (\vec x - \vec y)  - \frac{1}{L^d} \right]
          \label{NoiseCorreAvecContrainte}
\end{eqnarray}
The first contribution involving $\delta^{(d)} (\vec x- \vec y) $
 coincides with the expected white-noise expression of Eq. \ref{WhiteNoiseRealSpace},
 while the second term involving the subtraction $\left(- \frac{1}{L^d}\right)$ 
 is produced by
  the global constraint of Eq. \ref{WhiteNoiseRealSpaceConstraint} that imposes in particular
  the vanishing of
  \begin{eqnarray} 
0 && ={\mathbb E}\bigg(  \left[ \int_{L^d} d^d \vec  x \eta_t(\vec x)   \right] 
 \left[ \int_{L^d} d^d \vec  y \eta_{\tau} (\vec y)   \right]  \bigg)
= \int_{[0,L]^d} d^d \vec  x   \int_{[0,L]^d} d^d \vec  y \ {\mathbb E}\bigg(     \eta_t( \vec x)  \eta_{\tau}( \vec y )  \bigg)
\nonumber \\
&& = \delta(t-\tau) 2 \int_{[0,L]^d} d^d \vec  x   \int_{[0,L]^d} d^d \vec  y
  {\cal D}_L(\vec x- \vec y,t)
  \label{WhiteNoiseRealSpaceConstraintSquare}
\end{eqnarray}
that is indeed satisfied by the kernel of Eq. \ref{NoiseCorreAvecContrainte}.


\subsection{ Discussion}

In this Appendix, we have described how the calculations concerning the Carosso RG-scheme 
for the volume $L^d$ described in section \ref{sec_RGVolume} 
can be generalized to time-dependent OU-processes for the Fourier
modes, and we have discussed their real-space interpretations in terms of 
the two translation-invariant kernels 
$\Omega_L(\vec x-\vec y,t) $ and ${\cal D}_L( \vec x-\vec y,t) $.
Further work is needed to see whether these generalizations can be useful 
for some purposes from the RG point of view.


\end{document}